\let\kernelLabel\label
\AddToHook{package/amsmath/before}{\let\label\kernelLabel}
\documentclass[aps,prx,reprint,superscriptaddress,amsmath,amssymb,
  longbibliography]{revtex4-2}
\usepackage{amsthm}
\usepackage{graphicx}
\usepackage{physics}
\usepackage{tikz}
\usetikzlibrary{arrows.meta}

\newtheorem{theorem}{Theorem}
\newtheorem{corollary}[theorem]{Corollary}
\newtheorem{proposition}[theorem]{Proposition}
\newtheorem{lemma}[theorem]{Lemma}

\newcommand{\N}{\mathcal{N}}
\newcommand{\E}{\mathcal{E}}
\newcommand{\D}{\mathcal{D}}
\DeclareMathOperator{\id}{id}
\newcommand{\IO}{\mathrm{IO}}
\newcommand{\All}{\mathrm{all}}

\usepackage[hidelinks]{hyperref}

\newtheorem*{ioallrestated}{Theorem~\ref{thm:io-all-capacity}}
\newtheorem*{additivityrestated}{Proposition~\ref{prop:io-all-additivity-convexity}}
\newtheorem*{iodecoderrestated}{Theorem~\ref{thm:io-decoder-half-bounds}}
\newtheorem*{recoveryrestated}{Theorem~\ref{thm:io-recovery}}
\newtheorem*{capacityrestated}{Corollary~\ref{cor:io-decoder-capacity}}
\newtheorem*{paulirestated}{Theorem~\ref{thm:pauli-capacity}}

\begin{document}
\title{Quantum Capacity with Incoherent Codes}
\author{Oxana Shaya}
\email[Contact author: ]{shayaoxana@gmail.com}
\affiliation{Institute for Information Processing (tnt/L3S),
  Leibniz Universit\"at Hannover, 30167 Hannover, Germany}
\author{Zahra Baghali Khanian}
\affiliation{Perimeter Institute for Theoretical Physics,
  Waterloo, Ontario, Canada}
\affiliation{Institute for Quantum Computing, University of Waterloo,
  Waterloo, Ontario, Canada}
\date{November 24, 2026}
\begin{abstract}
Arbitrarily weak depolarizing noise drives the quantum capacity to zero
in every input basis when the encoder is restricted to incoherent
operations. Even one qubit cannot then be sent reliably: if the output
is replaced by the maximally mixed state with probability
$\varepsilon\le1/2$, the entanglement fidelity is at most
$1-\varepsilon/2$ at every block length. With unrestricted decoding, a
single-letter formula gives the capacity with incoherent encoding as the
maximum Holevo information over ensembles of basis states with identical
environment outputs. For either incoherent or unrestricted encoding,
imposing incoherent decoding reduces the capacity by at most a factor of
two; this factor is optimal. We bound the fidelity of incoherent
decoders and evaluate the capacities with incoherent encoding for
multiqubit Pauli channels.
\end{abstract}

\maketitle

\section{Introduction}
\label{sec:introduction}

Quantum communication protocols usually count communication and shared
entanglement as resources, while treating local operations as free.
Resource theories allow us to reconsider this assumption. The resource
theory of coherence accounts for the coherence required by local quantum
operations~\cite{BCP,SAP,WY,BGMW}. In quantum state merging and state
redistribution, restricting local coherence changes the communication
and entanglement needed to transfer part of a shared
state~\cite{merging,redistribution}. Tan \textit{et al.} construct an
error-correcting protocol that consumes additional coherent ancillary
qubits~\cite{Tan2017}. Kelly \textit{et al.} study how coherence limits
quantum information protection against measurements in monitored
Clifford circuits, and bound the distance of stabilizer codes in terms
of coherence~\cite{Kelly2023}. A different restriction requires
continuous logical symmetries to act transversally on the physical
subsystems, which limits the correction of local
errors~\cite{EastinKnill2009,Faist2020,ZhouLiuJiang2021}.

Different local capabilities also arise when a quantum processor assists
a client with restricted local operations \cite{Wu2020}, or when
heterogeneous networks connect parties with different resource
constraints~\cite{Ganardi2026}. These settings motivate treating
encoding and decoding separately.

Related work quantifies how much classical information can be encoded in
a given quantum state using resource-constrained operations and
recovered after noiseless transmission~\cite{Korzekwa2022,Mondal2026}.

Our question instead concerns asymptotic rates for arbitrary memoryless
channels: how much quantum information can be transmitted when the
encoder, the decoder, or both are restricted to incoherent operations
(IO)? The restrictions refer to fixed reference bases. No external
coherent ancillas are supplied. An IO map admits a Kraus representation
in which each operator sends a reference basis state to a multiple of a
basis state.

The standard quantum capacity is given by the regularized coherent
information and can exceed the optimum for a single channel
use~\cite{Devetak2005,DiVincenzo1998}. With an IO encoder and an
unrestricted decoder, Theorem~\ref{thm:io-all-capacity} gives an exact
single-letter formula. It groups the input basis states by the states
they leave in the environment of the channel, and the capacity is the
largest Holevo information achievable within a single group. Identical
environment outputs allow a classical code to be decoded coherently,
which builds on the link between private classical and quantum
communication~\cite{SchumacherWestmoreland1998,Devetak2005}. The
converse is strong: at every rate above this capacity, the entanglement
fidelity tends to zero exponentially with the number of channel uses.
This capacity is generically zero (Sec.~\ref{sec:zero-capacity}).

Depolarizing noise replaces the channel output by the maximally mixed
state with some probability. An arbitrarily small amount of it reduces
the quantum capacity with an incoherent encoder to zero, for every
finite-dimensional channel and in every input basis, even with
unrestricted decoding. Qubit dephasing illustrates the same
discontinuity. Its computational-basis states still transmit one
classical bit perfectly, and the standard quantum capacity $1-h_2(p)$
approaches one as the phase-flip probability $p$ tends to zero. With an
incoherent encoder in that basis, the quantum capacity is zero for every
$0<p<1/2$, because the two inputs produce different environment states
(Sec.~\ref{sec:pauli}).

The zero-capacity criterion of Sec.~\ref{sec:zero-capacity} explains the
discontinuity. Let $\mathcal R$ be a replacement channel, which outputs
the same state for every input, and consider
$\N_\varepsilon=(1-\varepsilon)\N+\varepsilon\mathcal R$. With
probability $\varepsilon$ the replacement occurs, and the environment
then receives the input state itself. Thus the environment outputs
differ for every pair of distinct basis states when $\varepsilon>0$, and
both capacities with IO encoding vanish
(Corollary~\ref{cor:zero-capacity}). Not even a single qubit can then be
transmitted reliably: for $\varepsilon\le1/2$, every code with an IO
encoder and logical dimension at least two has entanglement fidelity at
most $1-\varepsilon/2$, at every block length. Since
$\|\N_\varepsilon-\N\|_\diamond\le 2\varepsilon$, both capacities are
discontinuous wherever they are positive. The standard quantum capacity
is continuous in the diamond norm~\cite{LeungSmith2009}. Analogous
discontinuities are known for zero-error
capacities~\cite{AhlswedeBjelakovicBocheNoetzel2013}, which require
perfect transmission at finite block length. For IO encoders, the
discontinuity holds for codes whose error tends to zero.

The receiver can use coherence from the channel output for recovery.
Theorem~\ref{thm:io-decoder-half-bounds} shows that an IO decoder
achieves at least half of the capacity attained with unrestricted
decoding, whether the encoder is IO or unrestricted. This factor is
optimal in both cases. We also bound the fidelity attainable with IO
decoding and characterize the two capacities with IO decoding
asymptotically. With an IO encoder, Fourier channels attain the factor
one half: one channel use supplies the coherence that decodes one other
use, and no code does better. Multiqubit Pauli channels lose no rate in
the computational basis.

\section{Communication model}
\label{sec:setting}

\subsection{Incoherent operations}
\label{sec:incoherent-operations}

We consider a finite-dimensional quantum channel $\N_{A\to B}$, that is,
a completely positive, trace-preserving (CPTP) map. Coherence is defined
relative to fixed orthonormal bases~\cite{BCP,SAP}:
$\{\ket{x}:x\in\mathcal X\}$ on $A$ and $\{\ket{y}:y\in\mathcal Y\}$ on
$B$. States diagonal in the reference basis are incoherent; composite
systems use product reference bases. The maximally coherent state of
dimension $d$ is $\ket{\Psi_d}=d^{-1/2}\sum_{j=0}^{d-1}\ket{j}$. In
particular, $\ket{\Psi_2}$ is a maximally coherent qubit.

An incoherent operation ($\IO$) is a quantum channel that admits a Kraus
representation in which every Kraus operator maps each basis state to a
multiple of a basis state~\cite{BCP}. This permits arbitrary
measurements when the measured system is discarded and only the outcome,
recorded in a reference basis state, is kept.
Appendix~\ref{app:incoherent-operations} recalls related classes of free
operations, which we use for comparison.

\subsection{Codes and capacities}
\label{sec:codes-capacities}

An $n$-use code has a CPTP encoder $\E_{A_1\to A^n}$ and decoder
$\D_{B^n\to B_1}$, with logical dimension
$\dim A_1=\dim B_1=M$~\cite[Sec.~24.1]{Wilde}. We call $\E$ and $\D$ the
coding maps. Here $A^n$ and $B^n$ are the channel inputs and outputs.
The logical reference bases are fixed independently of those on $A$ and
$B$, and we identify $\ket{i}_{A_1}$ with $\ket{i}_{B_1}$. The target is
the logical identity channel. Any correction applied at the end is part
of $\D$ and must satisfy the same restriction. Communication is
unassisted, with no preshared entanglement or additional communication
channel. Local ancillary systems are included in the coding maps. No
coherent ancillary states are supplied for implementing an IO map,
although an IO decoder may use coherence already present in the channel
output.

Subspace transmission preserves every pure logical state with fidelity
tending uniformly to one. Entanglement transmission preserves a
maximally entangled state with an inaccessible reference. Their
asymptotic capacities agree for unrestricted
codes~\cite{Barnum2000,KretschmannWerner2004}; we prove the same for IO
encoding, IO decoding, or both
(Appendix~\ref{app:transmission-criteria}). We therefore use
entanglement transmission, with input
\begin{equation}
    \ket{\Phi_M}_{RA_1}
    =\frac{1}{\sqrt{M}}\sum_{i=1}^M\ket{i}_R\ket{i}_{A_1}.
\end{equation}
The reference $R$ is untouched. The decoded state and the entanglement
fidelity are~\cite{Schumacher1996}
\begin{gather}
    \omega_{RB_1}
    =\bigl[\id_R\otimes(\D\circ\N^{\otimes n}\circ\E)\bigr]
       (\dyad{\Phi_M}_{RA_1}),\notag\\
    F_e(\D\circ\N^{\otimes n}\circ\E)
    =\bra{\Phi_M}_{RB_1}\omega_{RB_1}\ket{\Phi_M}_{RB_1}.
    \label{eq:code-entanglement-fidelity}
\end{gather}

A rate $r\ge0$ is achievable if a sequence of $n$-use codes has logical
dimensions $M_n$ with $\liminf_{n\to\infty}n^{-1}\log M_n\ge r$ and
entanglement fidelities tending to one. Rates are measured in qubits per
channel use; the capacity is the supremum of achievable rates.

The superscripts in $Q^{\IO,\All}(\N)$, $Q^{\All,\IO}(\N)$, and
$Q^{\IO,\IO}(\N)$ specify encoder and decoder restrictions,
respectively; $\All$ permits arbitrary CPTP maps. With both
unrestricted, the capacity is $Q(\N)$.

\subsection{Channel dilations and information measures}
\label{sec:channels-coherent-information}

To describe information lost to an environment $E$ unavailable to the
receiver, we use a Stinespring isometry
$V:A\to B\otimes E$~\cite{Wilde}. The channel and its complement are
\begin{align}
    \N_{A\to B}(\rho_A)&=\tr_E\!\left(V\rho_A V^\dagger\right),
    \label{eq:channel-dilation}\\
    \N^c_{A\to E}(\rho_A)&:=\tr_B\!\left(V\rho_A V^\dagger\right).
    \label{eq:complementary-channel}
\end{align}
Whether two inputs give the same environment state does not depend on
the choice of $V$.

All logarithms are base two. For classical random variables, $H(X)$
denotes Shannon entropy, $H(Y|X)$ conditional entropy, and
$I(X;Y)=H(Y)-H(Y|X)$ mutual information. We write
$h_2(p)=-p\log p-(1-p)\log(1-p)$ for the binary entropy. For a quantum
state $\rho$, the von Neumann entropy is $S(\rho)=-\tr(\rho\log\rho)$.
The mutual information of a bipartite state $\omega_{AB}$ is
$I(A;B)_\omega=S(\omega_A)+S(\omega_B)-S(\omega_{AB})$.

The \emph{coherent information} of a bipartite state $\omega_{RB}$ is
$I(R\rangle B)_\omega:=S(\omega_B)-S(\omega_{RB})$. For a channel input
$\rho_A$ with purification $\ket{\psi}_{RA}$, let
$\omega_{RB}=(\id_R\otimes\N)(\dyad{\psi}_{RA})$ and define
$I_c(\rho_A,\N)=I(R\rangle B)_\omega$. The standard quantum capacity is
the regularized coherent
information~\cite{Lloyd1997,Shor2002,Devetak2005},
\begin{equation}
    \begin{aligned}
    Q(\N)&=\lim_{n\to\infty}\frac1n Q^{(1)}(\N^{\otimes n}),\\
    Q^{(1)}(\N)&=\max_{\rho_A}I_c(\rho_A,\N).
    \end{aligned}
    \label{eq:ordinary-quantum-capacity}
\end{equation}

For an input ensemble $\{p_x,\rho_x^A\}$, write $\rho_x^B=\N(\rho_x^A)$
and $\rho_x^E=\N^c(\rho_x^A)$. The receiver's Holevo information is
\begin{equation}
    \chi(p)
    :=S\!\left(\sum_x p_x\rho_x^B\right)
      -\sum_x p_xS(\rho_x^B),
    \label{eq:io-all-holevo}
\end{equation}
and $\chi_E(p)$ is defined by replacing $B$ with $E$. For this ensemble,
$\chi(p)$ bounds how much a measurement at the receiver can reveal about
$x$. Classical rates below $\chi(p)$ are achievable by coding over many
uses and measuring the outputs
jointly~\cite{Holevo1998,SchumacherWestmoreland1997}.

\section{Quantum capacity with incoherent encoding}
\label{sec:io-all}

For each input basis state $\ket{x}$, define the receiver and
environment output states by
\begin{equation}
    \rho_x^B=\N(\dyad{x}),
    \qquad
    \rho_x^E=\N^c(\dyad{x}).
    \label{eq:io-all-basis-outputs}
\end{equation}
Let $\mathcal{P}(\mathcal{X})$ denote the probability distributions on
$\mathcal{X}$. For $p\in\mathcal{P}(\mathcal{X})$, consider the
classical-quantum state
\begin{equation}
    \omega_{XBE}
    =\sum_x p_x\dyad{x}_X\otimes V\dyad{x}_A V^\dagger,
    \label{eq:io-all-cq-state}
\end{equation}
where the classical register $X$ records the chosen basis input. The
receiver's Holevo information is $I(X;B)_\omega=\chi(p)$. For the
environment it is $I(X;E)_\omega=\chi_E(p)$.

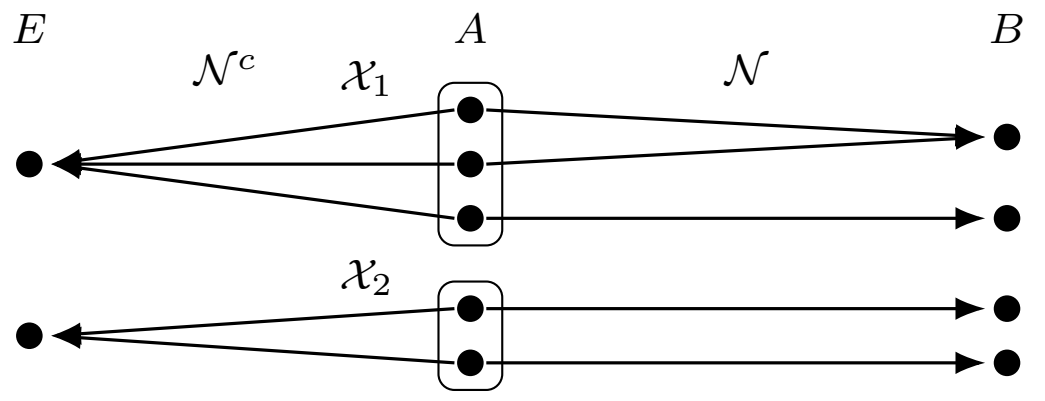
\begin{figure}[t]
    \centering
    \begingroup
\begin{tikzpicture}[
  x=1pt,y=.85pt,font=\footnotesize,text=black,
  line width=.6pt,>=Latex,
  state/.style={circle,draw,fill=black,inner sep=0pt,minimum size=4.4pt}
]
  \node at (34,205) {$E$};
  \node at (117,205) {$A$};
  \node at (218,205) {$B$};
  \draw[rounded corners=3pt,line width=.4pt] (111,157) rectangle (123,193);
  \draw[rounded corners=3pt,line width=.4pt] (111,125) rectangle (123,149);
  \node[anchor=east] at (106,194) {$\mathcal X_1$};
  \node[anchor=east] at (106,150) {$\mathcal X_2$};
  \node at (71,196) {$\N^c$};
  \node at (169,196) {$\N$};
  \foreach \yy in {187,175,163}{
    \draw[->] (114,\yy) -- (38,175);
    \node[state] at (117,\yy) {};
  }
  \foreach \yy in {187,175}{
    \draw[->] (120,\yy) -- (214,181);
  }
  \draw[->] (120,163) -- (214,163);
  \node[state] at (218,181) {};
  \node[state] at (218,163) {};
  \foreach \yy in {143,131}{
    \draw[->] (114,\yy) -- (38,137);
    \draw[->] (120,\yy) -- (214,\yy);
    \node[state] at (117,\yy) {};
    \node[state] at (218,\yy) {};
  }
  \node[state] at (34,175) {};
  \node[state] at (34,137) {};

\end{tikzpicture}
\endgroup
    \caption{Basis inputs belong to the same class exactly when their
    environment output states $\rho_x^E$ coincide. Arrows show the two
    reduced outputs of each input. Receiver states $\rho_x^B$ within a
    class may coincide or differ, as illustrated.}
    \label{fig:capacity-mechanisms}
\end{figure}

We call $I(X;E)_\omega=0$ the privacy condition: the environment learns
nothing about the basis input. Since $X$ is classical, it holds exactly
when $\rho_x^E=\omega_E$ for every $x$ with $p_x>0$. We group basis
inputs with identical environment outputs into \emph{environment
classes} $\mathcal{X}_1,\ldots,\mathcal{X}_\nu$. A distribution
satisfies this privacy condition exactly when it is supported within one
class. We write $\mathcal{P}(\mathcal{X}_c)$ for distributions on a
class, extended by zero outside it (Fig.~\ref{fig:capacity-mechanisms}).

\begin{theorem}[Capacity with incoherent encoding]
\label{thm:io-all-capacity}
For every finite-dimensional quantum channel $\N$,
\begin{equation}
    \begin{aligned}
    Q^{\IO,\All}(\N)
    &=\max_{\substack{p\in\mathcal{P}(\mathcal{X})\\
                     I(X;E)_\omega=0}} I(X;B)_\omega\\
    &=\max_c\max_{p\in\mathcal{P}(\mathcal{X}_c)}\chi(p).
    \end{aligned}
    \label{eq:io-all-capacity}
\end{equation}
\end{theorem}

The inner maximum in the second expression is the classical capacity of
the channel $x\mapsto\rho_x^B$ with inputs restricted to one class.
Quantum transmission additionally requires preserving superpositions of
codewords.

For achievability, we choose a distribution satisfying the privacy
condition. The coding theorem for classical-quantum channels gives basis
codewords that can be distinguished with vanishing error. The privacy
condition ensures that their environment outputs coincide. Uhlmann's
theorem relates their purifications by unitaries on the receiver's
system. Combining these unitaries with a coherent implementation of the
decoding measurement gives a quantum decoder. The coherent recovery
lemma (Lemma~\ref{lem:coherent-recovery}) bounds its entanglement
fidelity from below by the square of the classical decoding success
probability. Thus its entanglement fidelity tends to one. The encoder is
an IO isometry that maps logical basis states to distinct basis
codewords and preserves their superpositions. Such encoders are
physically incoherent and hence also strictly
incoherent~\cite{ChitambarGour2016}, so
Theorem~\ref{thm:io-all-capacity} holds under either stronger encoder
restriction. The formula need not hold for dephasing-covariant
incoherent operations (DIO) or maximally incoherent operations (MIO) as
encoders (Appendix~\ref{app:dio-encoder-separation}).

The converse rests on one identity. For two input basis states,
\begin{equation}
    \norm{\N(\dyad{x}{y})}_1
       =\norm\big{\sqrt{\rho_x^E}\sqrt{\rho_y^E}}_1,
    \label{eq:coherence-fidelity-main}
\end{equation}
where $\norm{X}_1=\tr\sqrt{X^\dagger X}$ is the trace norm
(Appendix~\ref{app:uhlmann}). The right-hand side is the square root of
the fidelity of the two environment outputs. Coherence between two basis
inputs thus reaches the receiver undiminished only if the environment
cannot distinguish them. Each Kraus operator of an IO encoder maps the
logical basis states to multiples of basis codewords, and a reliable
code must preserve the coherence between them. Hence almost all
codewords have the same environment output, and the classical rate bound
applies within one environment class at each channel use
(Appendix~\ref{app:io-all-converse}). The converse is strong: at every
rate above this capacity, the entanglement fidelity tends to zero
exponentially with the number of channel uses
(Proposition~\ref{prop:io-all-strong-converse}).

\subsection{Relation to coherent information}
\label{sec:io-all-interpretation}

The coherent information can be written as
$I_c(\rho,\N)=S(\N(\rho))-S(\N^c(\rho))$
(Appendix~\ref{app:conditional-entropy}). The capacity formula implies
\begin{equation}
    Q^{\IO,\All}(\N)\le Q^{(1)}(\N)\le Q(\N).
    \label{eq:io-all-coherent-upper-bound}
\end{equation}
For any $p\in\mathcal{P}(\mathcal{X}_c)$, all environment outputs
coincide. The joint output $V\ket{x}$ is pure, so
$S(\rho_x^B)=S(\rho_x^E)$. Substituting these equalities into the
coherent information gives~\cite{SchumacherWestmoreland1998}
\begin{equation}
    I_c\!\left(\sum_x p_x\dyad{x},\N\right)
    =I(X;B)_\omega=\chi(p).
    \label{eq:io-all-private-coherent}
\end{equation}
These diagonal input states form a subset of the states over which
$Q^{(1)}(\N)$ is maximized. This proves the first inequality in
Eq.~\eqref{eq:io-all-coherent-upper-bound}. The second holds because
every rate below $Q^{(1)}(\N)$ is
achievable~\cite{Lloyd1997,Shor2002,Devetak2005}.

\subsection{Additivity and convexity}
\label{sec:additivity-convexity}

\begin{proposition}[Additivity and convexity]
\label{prop:io-all-additivity-convexity}
For any two finite-dimensional quantum channels,
\begin{equation}
    Q^{\IO,\All}(\N_1\otimes\N_2)
    =Q^{\IO,\All}(\N_1)+Q^{\IO,\All}(\N_2).
    \label{eq:io-all-additivity}
\end{equation}
For channels $\N_j$ from $A$ to $B$ and probabilities $\lambda_j$,
\begin{equation}
    Q^{\IO,\All}\!\left(\sum_j\lambda_j\N_j\right)
       \le\sum_j\lambda_j Q^{\IO,\All}(\N_j).
    \label{eq:io-all-convexity}
\end{equation}
\end{proposition}

For additivity, the environment classes of a product channel are
products of the individual classes, and the corresponding classical
capacities add. For convexity, a distribution that satisfies the privacy
condition for the mixture also satisfies it for every component with
positive weight. For this fixed distribution, mixing the receiver's
output ensembles cannot increase their Holevo information beyond the
weighted average. The proofs are given in
Appendix~\ref{app:additivity-convexity}.

\subsection{Generic zero capacity}
\label{sec:zero-capacity}

The formula characterizes when incoherent encoding prevents quantum
communication. For every finite-dimensional channel,
$Q^{\IO,\All}(\N)=Q^{\IO,\IO}(\N)=0$ if and only if
\begin{equation}
    \rho_x^E=\rho_y^E\quad\Longrightarrow\quad\rho_x^B=\rho_y^B
    \label{eq:zero-capacity}
\end{equation}
for every pair of input basis states $\ket{x},\ket{y}$
(Appendix~\ref{app:zero-capacity}). The bounds in
Theorem~\ref{thm:io-decoder-half-bounds} below show that the same
criterion applies when the decoder is also IO.

If different basis inputs leave different environment states, every
class contains only one input and the capacity is zero. In particular,
if the complementary channel is injective, both capacities vanish in
every input reference basis.

\begin{corollary}[Arbitrarily weak noise]
\label{cor:zero-capacity}
Let $\N$ be a channel from $A$ to $B$, $\sigma$ a state on $B$, and
$0<\varepsilon\le1$. For every choice of reference bases, the channel
$\N_\varepsilon=(1-\varepsilon)\N+\varepsilon\tr(\cdot)\,\sigma$
satisfies
\[
    Q^{\IO,\All}(\N_\varepsilon)=Q^{\IO,\IO}(\N_\varepsilon)=0.
\]
For $\varepsilon\le1/2$, every code for $\N_\varepsilon$ with an IO
encoder and logical dimension at least two has entanglement fidelity at
most $1-\varepsilon/2$, at every block length.
\end{corollary}

The replacement term vanishes on $\dyad{x}{y}$ for distinct basis
inputs, so $\norm{\N_\varepsilon(\dyad{x}{y})}_1\le1-\varepsilon$. By
Eq.~\eqref{eq:coherence-fidelity-main}, the environment outputs of
distinct basis inputs differ, which gives zero capacity. Two different
codewords differ in at least one position, so the same bound limits the
coherence between them and gives the fidelity bound
(Lemma~\ref{lem:io-classical-codewords},
Appendix~\ref{app:zero-capacity}). For $\sigma=I_B/d_B$ with
$d_B=\dim B$, this is depolarizing noise at the channel output, and the
channels $\N_\varepsilon$ are exactly those with a full-rank Choi
matrix. For fixed finite-dimensional systems $A$ and $B$, these channels
form an open dense subset of the CPTP maps from $A$ to $B$ whose
complement has Lebesgue measure zero. Both capacities with IO encoding
are therefore discontinuous wherever they are positive, although the
fidelity of every fixed code remains continuous
(Appendix~\ref{app:zero-capacity}). For example, erasure and qubit
amplitude-damping channels have zero capacity with IO encoding in every
input basis whenever the erasure or decay probability is nonzero
(Appendix~\ref{app:erasure-amplitude-damping}). Both have positive
standard capacity for noise probabilities below
$1/2$~\cite{BennettDiVincenzoSmolin1997,GiovannettiFazio2005}.

\section{Quantum capacity with incoherent decoding}
\label{sec:io-decoding}

We now require the decoder to be IO. The receiver may still use
coherence present in the channel outputs. The decoder in the
achievability construction for $Q^{\IO,\All}$ uses unitaries that need
not be IO. The quantum Fourier transform channel in
Sec.~\ref{sec:fourier} shows that an IO decoder can fall short of this
rate even when the environment learns nothing about the input.

\begin{theorem}[Capacity bounds with incoherent decoding]
\label{thm:io-decoder-half-bounds}
Restricting the decoder to IO costs at most half the capacity:
$Q^{\All,\IO}(\N)\ge\tfrac12 Q(\N)$ with unrestricted encoding, and
$Q^{\IO,\IO}(\N)\ge\tfrac12 Q^{\IO,\All}(\N)$ with incoherent encoding.
Neither factor $1/2$ can be increased uniformly over finite-dimensional
channels.
\end{theorem}

For the Fourier channel, the factor $1/2$ has a simple origin: with IO
encoding, one use supplies the coherence that decodes one other use, and
no code does better (Sec.~\ref{sec:fourier}). The flagged channels
$\N_d$ of Sec.~\ref{sec:flagged-channels} approach the factor with
unrestricted encoding. The first bound shows that $Q^{\All,\IO}(\N)$ is
positive exactly when $Q(\N)$ is.

\subsection{Entanglement fidelity with incoherent decoding}
\label{sec:io-recovery}

Let $\mathcal T=\N^{\otimes n}\circ\E$ be the encoded channel from $A_1$
to $B^n$, with logical dimension $M$. We consider completely positive
maps $\mathcal A:B^n\to B_1$ admitting Kraus operators $\{A_a\}$ that
satisfy
\begin{equation}
    \begin{aligned}
    \sum_a A_a^\dagger A_a&\le I_{B^n},\\
    \sum_a\norm{A_a\ket{y}}_{\ell_1}^2&\le1
       \quad\text{for every }y,
    \end{aligned}
    \label{eq:admissible-recovery}
\end{equation}
where $y$ runs over the output basis strings and
$\norm{v}_{\ell_1}=\sum_i\abs{\braket{i}{v}}$ uses the logical basis.
The first condition allows the map to decrease trace. If it preserves
trace, the two conditions force every column of every Kraus operator to
have at most one nonzero entry, exactly as required for IO. Allowing
trace loss therefore gives a relaxation of IO decoding. Define
$F_{\ell_1}(\mathcal T)=\sup_{\mathcal A}F_e(\mathcal A\circ\mathcal T)$,
where the supremum is over these maps. For a map that decreases trace,
$F_e$ is the same overlap with $\ket{\Phi_M}$ as in
Eq.~\eqref{eq:code-entanglement-fidelity}, without renormalizing the
output.

\begin{theorem}[Entanglement fidelity with incoherent decoding]
\label{thm:io-recovery}
For every encoded channel $\mathcal T$ of logical dimension $M$,
$\sup_{\D\in\IO}F_e(\D\circ\mathcal T)\le F_{\ell_1}(\mathcal T)$. For
every integer $1\le m\le M$, there is an isometry
$W:\mathbb C^m\to\mathbb C^M$ mapping basis states to distinct logical
basis states such that the restricted channel
$\mathcal T_W(\rho)=\mathcal T(W\rho W^\dagger)$ satisfies
\begin{equation}
    \sup_{\D\in\IO}F_e(\D\circ\mathcal T_W)
       \ge\left(\sqrt{F_{\ell_1}(\mathcal T)}
                   -\sqrt{m/M}\right)_+^2.
    \label{eq:io-decoding-fidelity}
\end{equation}
The second supremum is over IO decoders with $m$-dimensional output, and
$F_e$ uses $\ket{\Phi_m}$. Here $u_+=\max\{u,0\}$.
\end{theorem}

Every IO decoder satisfies Eq.~\eqref{eq:admissible-recovery}, which
gives the upper bound. For the lower bound, we sample: in each column of
a Kraus operator of $\mathcal A$ we keep a single entry, chosen with
probability proportional to its modulus and rescaled so that the
expectation reproduces the column. The samples are incoherent, but a
sampled column $y$ has squared norm $\norm{A_a\ket y}_{\ell_1}^2$
instead of $\norm{A_a\ket y}^2$. The excess comes from pairs of distinct
logical basis states. On a random subcode with $m$ of the $M$ logical
basis states, at most a fraction of about $m/M$ of it remains on
average, and a diagonal rescaling removes it. For the maps in
Eq.~\eqref{eq:admissible-recovery} the resulting loss is the term
$\sqrt{m/M}$ in Eq.~\eqref{eq:io-decoding-fidelity}
(Appendices~\ref{app:admissible-maps} and~\ref{app:fidelity-bounds}).

\begin{corollary}[Capacity with incoherent decoding]
\label{cor:io-decoder-capacity}
For every finite-dimensional quantum channel $\N$, each capacity with IO
decoding is given by
\begin{equation}
    \sup\left\{R\ge0:\lim_{n\to\infty}
         \sup_{\E}F_{\ell_1}(\N^{\otimes n}\circ\E)=1\right\}.
    \label{eq:regularized-io-decoder-capacity}
\end{equation}
The inner supremum is over encoders with $n$ channel inputs and logical
dimension $M\ge2^{nR}$: IO encoders for $Q^{\IO,\IO}(\N)$, and arbitrary
CPTP encoders for $Q^{\All,\IO}(\N)$.
\end{corollary}

For codes at a positive rate, taking $m=\lfloor M/n\rfloor$ in
Theorem~\ref{thm:io-recovery} makes the fidelity tend to one whenever
$F_{\ell_1}\to1$, while reducing the rate by only $O(\log n/n)$. This
proves Corollary~\ref{cor:io-decoder-capacity}
(Appendix~\ref{app:decoder-capacity-formula}). The characterization
replaces the requirement of incoherent Kraus operators by the two
constraints in Eq.~\eqref{eq:admissible-recovery}; the optimization over
encoders and block lengths remains.

Theorem~\ref{thm:io-decoder-half-bounds} follows from the same
construction, applied to the decoder of an arbitrary code. Its excess is
no longer bounded by one, and a second estimate replaces $m/M$ in
Eq.~\eqref{eq:io-decoding-fidelity} by about $m/\sqrt M$. This term
vanishes for $m$ of order $\sqrt M/n$, and keeping only about the square
root of the code dimension halves the asymptotic rate. Restricting a
code to logical basis states leaves an IO encoder incoherent, so the
construction applies to IO and to unrestricted encoders
(Appendix~\ref{app:half-capacity-bounds}).

Let $P(y|i)=\bra y\mathcal T(\dyad i)\ket y$ be the probability of
output $y$ when logical basis input $\ket i$ is sent and the receiver
measures in its reference basis. The constraints in
Eq.~\eqref{eq:admissible-recovery} also imply
(Appendix~\ref{app:decoder-fidelity-upper-bounds})
\begin{equation}
    M\sqrt{F_{\ell_1}(\mathcal T)}
       \le\sum_y\max_i\sqrt{P(y|i)}.
    \label{eq:decoding-probability-bound}
\end{equation}
This bound, and a refinement of it, give the converses for the examples
in Sec.~\ref{sec:examples}.

\subsection{Relation to the standard capacity}
\label{sec:decoder-embedding}

The capacity problem with unrestricted encoding and incoherent decoding
includes the standard quantum capacity problem as a special case. A
maximally coherent state $\ket{\Psi_K}$ suffices to implement any
decoder with a $K$-dimensional output by IO~\cite[Theorem~4]{BGMW}; see
Eq.~\eqref{eq:io-decoder-simulation}. For a channel $\N$ with input
dimension $d_A=\dim A$, consider the channel obtained by appending a
fixed maximally coherent state to its output,
\begin{equation}
    \N'(\rho_A)=\N(\rho_A)\otimes\dyad{\Psi_{d_A}}.
    \label{eq:decoder-capacity-embedding-channel}
\end{equation}
Unrestricted quantum codes with isometric encoders and logical dimension
$M_n\le d_A^n$ attain every rate below $Q(\N)$~\cite[Sec.~24.4]{Wilde}.
After $n$ uses, the appended outputs supply
$\ket{\Psi_{d_A}}^{\otimes n}=\ket{\Psi_{d_A^n}}$. By
Eq.~\eqref{eq:io-decoder-simulation}, with $d=d_A^n\ge M_n$, this
resource suffices to implement any such code's decoder by IO without
changing its entanglement fidelity. The reverse bound follows by
allowing unrestricted decoding. Moreover, $Q(\N')=Q(\N)$ because an
unrestricted receiver can prepare or discard the appended state locally.
Consequently,
\begin{equation}
    Q^{\All,\IO}(\N')=Q(\N')=Q(\N).
    \label{eq:decoder-capacity-embedding}
\end{equation}
Thus a general formula for $Q^{\All,\IO}$ would also determine the
standard quantum capacity of every channel. The equality
$Q^{\All,\IO}=Q$ does not hold for every channel
(Sec.~\ref{sec:flagged-channels}).

\section{Examples}
\label{sec:examples}

We use the computational bases for the examples below.

\subsection{Pauli channels}
\label{sec:pauli}

For multiqubit Pauli noise, symmetries identify the environment classes
explicitly. With IO encoding, restricting the decoder causes no further
loss.

For $m$ qubits, basis states are indexed by binary strings
$x\in\mathbb F_2^m$ of length $m$. We write $0^m$ for the all-zero
string, use $\oplus$ for bitwise addition modulo two, and write
$b\cdot x=\sum_{j=1}^m b_jx_j\pmod 2$ for the binary dot product.
Consider an $m$-qubit Pauli channel with computational reference bases,
\begin{equation}
    \N(\rho)=\sum_{a,b\in\mathbb F_2^m}
       p_{ab}X^aZ^b\rho Z^bX^a,
    \label{eq:pauli-channel}
\end{equation}
where $p_{ab}$ are probabilities, $X$ and $Z$ are the Pauli operators,
$X^a=\bigotimes_{j=1}^mX^{a_j}$, and similarly for $Z^b$. Errors on the
$m$ qubits within one use may be correlated; repeated uses remain
independent.

For the channel in Eq.~\eqref{eq:pauli-channel}, let $T$ be the random
bit-error pattern with $\Pr(T=a)=q_a=\sum_b p_{ab}$. Define the binary
subspace
\[
    L=\bigl(\operatorname{span}\{b\oplus b':
        p_{ab}p_{ab'}>0\text{ for some }a\}\bigr)^\perp,
\]
where the orthogonal complement uses the binary dot product, and write
$T+L$ for the coset of $L$ containing $T$. Let $X$ be uniform on $L$.
Conditional on $X=x$, let $Y$ be the outcome of measuring $\N(\dyad{x})$
in the receiver's reference basis, so that
$\Pr(Y=y\mid X=x)=q_{y\oplus x}$ for $y\in\mathbb F_2^m$.

\begin{theorem}[Pauli channels]
\label{thm:pauli-capacity}
For an $m$-qubit Pauli channel $\N$ with computational reference bases,
and with $L$, $T$, $X$, and $Y$ as defined above,
\begin{equation}
    \begin{aligned}
    Q^{\IO,\IO}(\N)=Q^{\IO,\All}(\N)&=I(X;Y)\\
       &=\dim L-H(T\mid T+L).
    \end{aligned}
    \label{eq:pauli-capacity}
\end{equation}
\end{theorem}

The subspace $L$ is the environment class containing $0^m$ in
Theorem~\ref{thm:io-all-capacity}: for each bit error $a$, all phase
errors $b$ with $p_{ab}>0$ act identically on every basis state
$\ket{x}$ with $x\in L$. Every environment class is a coset of $L$ and
gives the same maximum mutual information, attained by the uniform
input. Thus Eq.~\eqref{eq:pauli-capacity} is determined by the support
of $p_{ab}$ and the bit-error probabilities. The receiver's outputs for
basis inputs are diagonal, and a coherent implementation of classical
decoding attains the rate by IO (Appendix~\ref{app:pauli-capacity}).

For one qubit, bit-flip noise $\mathcal N_p^X$ has $p_{00}=1-p$ and
$p_{10}=p$, whereas dephasing $\mathcal N_p^Z$ has $p_{00}=1-p$ and
$p_{01}=p$; all other probabilities vanish. For $0<p<1/2$, the bit-flip
channel's two basis inputs leave the same environment state and
therefore belong to one environment class, $L=\mathbb F_2$. The receiver
sees a binary symmetric channel, so Theorem~\ref{thm:pauli-capacity}
gives both capacities with incoherent encoding as $I(X;Y)=1-h_2(p)$,
equal to the standard quantum capacity. Dephasing transmits these inputs
perfectly to the receiver but gives different environment outputs. Each
class contains only one input, so $L=\{0\}$ and $I(X;Y)=0$. With only
the decoder restricted, Theorem~\ref{thm:io-decoder-half-bounds}
guarantees a positive rate, and the code in
Appendix~\ref{app:dephasing-io-decoder} attains the full $1-h_2(p)$. The
two channels are related by a Hadamard change of basis, which is not an
incoherent operation. Sufficiently weak additional phase-flip noise
makes the bit-flip channel's environment outputs unequal and both
capacities with incoherent encoding zero while leaving the standard
capacity positive (Appendix~\ref{app:pauli-capacity}).
Table~\ref{tab:channel-capacities} compares these and the following
examples.

\begin{table}[t]
    \caption{Capacities in qubits per channel use, with $0<p<1/2$ and
    $d\ge2$. The standard dephasing capacity follows from
    Ref.~\cite[Appendix~B]{DevetakShor2005}; bit-flip noise is unitarily
    equivalent. The other capacities are derived in
    Appendix~\ref{app:pauli-noise}.}
    \label{tab:channel-capacities}
    \begin{ruledtabular}
    \begin{tabular}{lcccc}
        Channel $\N$ & $Q(\N)$ & $Q^{\IO,\All}(\N)$
                     & $Q^{\All,\IO}(\N)$ & $Q^{\IO,\IO}(\N)$ \\
        \hline
        Dephasing $\mathcal N_p^Z$ & $1-h_2(p)$ & $0$
                     & $1-h_2(p)$ & $0$ \\
        Bit flip $\mathcal N_p^X$ & $1-h_2(p)$ & $1-h_2(p)$
                     & $1-h_2(p)$ & $1-h_2(p)$ \\
        QFT $\mathcal{F}_d$ & $\log d$ & $\log d$
                     & $\log d$ & $\tfrac{1}{2}\log d$ \\
        Flagged $\N_H$ & $1$ & $1$
                     & $\tfrac34$ & $\tfrac34$ \\
    \end{tabular}
    \end{ruledtabular}
\end{table}

\subsection{Quantum Fourier transform channel}
\label{sec:fourier}

For $d\ge2$, the quantum Fourier transform (QFT) channel is
\begin{equation}
    \mathcal{F}_d(\rho)=F_d\rho F_d^\dagger,
    \qquad
    F_d\ket{x}=\frac{1}{\sqrt{d}}\sum_{y=0}^{d-1}
                        e^{2\pi ixy/d}\ket{y}.
    \label{eq:fourier-channel}
\end{equation}
The environment is independent of the input, so all basis states belong
to one class in Theorem~\ref{thm:io-all-capacity}. Their receiver
outputs are orthogonal pure states, and the uniform ensemble gives
$Q^{\IO,\All}(\mathcal F_d)=\log d$. A measurement in the output basis,
however, gives the same uniform distribution for every input label. One
use on $\ket{0}$ supplies $\ket{\Psi_d}$, which lets the receiver
implement $F_d^\dagger$ by IO on the output of a second use carrying the
message. This attains the rate $\tfrac12\log d$. The converse follows
directly from Eq.~\eqref{eq:decoding-probability-bound}: for any IO
encoder, $P(y|i)=d^{-n}$, so $F_e\le F_{\ell_1}\le d^n/M^2$. Hence
$Q^{\IO,\IO}(\mathcal F_d)=\tfrac12\log d$, and the lower capacity bound
in Theorem~\ref{thm:io-decoder-half-bounds} is tight for IO encoding
(Appendix~\ref{app:fourier-unitaries}). With either map unrestricted,
applying $F_d^\dagger$ at that end attains the full $\log d$ rate
(Table~\ref{tab:channel-capacities}). Thus even a channel that loses no
information to an environment can require coherence for decoding at the
full transmission rate.

\subsection{Channels with a classical flag}
\label{sec:flagged-channels}

IO decoding can strictly reduce the capacity even with unrestricted
encoding. An example is the flagged Hadamard channel
\begin{equation}
    \N_H(\rho)=\tfrac12\rho\otimes\dyad{0}
       +\tfrac12H\rho H\otimes\dyad{1},
    \label{eq:flagged-hadamard-channel}
\end{equation}
where $H$ is the Hadamard matrix and the second tensor factor is a flag
register $J$ recording which operation was applied. The identity encoder
is IO, and an unrestricted receiver uses $J$ to reverse the operation.
Hence $Q^{\IO,\All}(\N_H)=Q(\N_H)=1$. With IO decoding,
$Q^{\All,\IO}(\N_H)=Q^{\IO,\IO}(\N_H)=3/4$. The converse uses a
refinement of Eq.~\eqref{eq:decoding-probability-bound}
(Appendix~\ref{app:strict-decoder-loss}). Together with dephasing, this
shows that neither restriction is always the more costly one
(Table~\ref{tab:channel-capacities}).

For $\mathcal F_d$ and $\N_H$, replacing IO decoding by MIO leaves the
capacity unchanged, with IO or with unrestricted encoding. MIO maps need
only send incoherent states to incoherent states
(Appendices~\ref{app:strict-decoder-loss}
and~\ref{app:fourier-unitaries}).

The loss can approach one half even with unrestricted encoding, so the
factor $1/2$ in Theorem~\ref{thm:io-decoder-half-bounds} is optimal in
that case as well. The example uses mutually unbiased bases (MUBs):
orthonormal bases such that any two basis states from different bases
have overlap of modulus $d^{-1/2}$. For each prime power $d$ there are
$d+1$ such bases, given by the standard stabilizer
construction~\cite{WoottersFields1989,Kantor2012}. The flagged MUB
channel $\N_d$ uniformly chooses one of these bases, rotates it into the
reference basis, and records the choice in an orthogonal flag. An
unrestricted receiver can reverse the rotation, so $Q(\N_d)=\log d$. An
entropic uncertainty relation for these bases bounds the rate achievable
with IO decoding:
\begin{equation}
    Q^{\All,\IO}(\N_d)\le\log d-\tfrac12\log\frac{d+1}{2}.
    \label{eq:optimal-decoder-half-factor}
\end{equation}
With Theorem~\ref{thm:io-decoder-half-bounds}, the ratio
$Q^{\All,\IO}(\N_d)/Q(\N_d)$ tends to $1/2$ as $d$ grows
(Appendix~\ref{app:optimal-decoder-factor}).

\section{Discussion and open questions}
\label{sec:discussion}

Arbitrarily weak depolarizing noise can eliminate quantum transmission
with an IO encoder in every input basis. With unrestricted encoding, an
IO decoder always achieves at least half the standard capacity by using
coherence from the channel output.

Evaluating the capacities in Corollary~\ref{cor:io-decoder-capacity}
remains open in general. With unrestricted encoding, the problem
includes the standard quantum capacity problem
(Sec.~\ref{sec:decoder-embedding}). With both maps IO, the capacity is
not additive over different channels: even a channel of zero quantum
capacity can supply coherence that increases the joint rate
(Appendix~\ref{app:nonadditivity}).

For channels with positive $Q^{\IO,\All}$, what determines how much of
that rate remains when both maps are IO? Fourier channels keep half, and
the Pauli channels in Theorem~\ref{thm:pauli-capacity} keep the full
rate. The open problem is to characterize the trade-off between
transmitting quantum information and generating the coherence needed to
decode it. With both coding maps incoherent, one could study achievable
pairs $(R,C)$, where $R$ is the transmission rate and $C$ is the number
of maximally coherent qubits supplied per channel use. Their allocation
between sender and receiver would also be optimized.

The same question extends to other resource theories: what quantum
capacity remains when encoding and decoding are restricted to free
operations? For magic, one can restrict both maps to stabilizer
operations and ask whether optimal communication rates require the
nonstabilizer resources used in universal fault-tolerant
computation~\cite{Veitch2014}. No-go theorems for Gaussian error
correction in bosonic and fermionic systems demonstrate fundamental
limitations imposed by such
restrictions~\cite{NisetFiurasekCerf2009,Tang2026}. For coherence, the
sender and receiver have sharply different requirements: encoding beyond
IO is generically indispensable for positive-rate transmission, whereas
an IO decoder can use channel outputs to achieve at least half the
available capacity.

\begin{acknowledgments}
O.S. thanks Perimeter Institute for Theoretical Physics for its
hospitality during her visit in July 2026. This research was supported
in part by Perimeter Institute for Theoretical Physics. Research at
Perimeter Institute is supported by the Government of Canada through the
Department of Innovation, Science and Economic Development and by the
Province of Ontario through the Ministry of Colleges, Universities,
Research Excellence, and Security.

Z.B.K. acknowledges support from the Ada Lovelace Postdoctoral
Fellowship at Perimeter Institute for Theoretical Physics. During this
work, O.S. was supported by the Federal Ministry of Research, Technology
and Space (BMFTR), Germany, through the QC service center QUICS (grant
no.~13N17418).

Claude Fable (Anthropic) assisted with derivations, literature review,
and manuscript preparation. The authors directed its use and reviewed
its contributions line by line.
\end{acknowledgments}

\section*{Data availability}
This is a purely mathematical work and no data were created or
analyzed in this study.

\clearpage
\onecolumngrid
\appendix
\makeatletter
\newif\ifappendixtoc@
\begingroup
  \renewcommand{\tocname}{Contents of the appendices}%
  \let\main@contentsline\contentsline
  \def\contentsline#1#2#3#4{\ifappendixtoc@\main@contentsline{#1}{#2}{#3}{#4}\fi}%
  \let\main@appendix@toc\appendix@toc
  \def\appendix@toc{\main@appendix@toc\global\appendixtoc@true}%
  \tableofcontents
\endgroup
\makeatother
\section{Technical preliminaries}
\label{app:technical-preliminaries}

We consider a finite-dimensional quantum channel $\N_{A\to B}$, that is,
a completely positive, trace-preserving (CPTP) map. Coherence is defined
relative to fixed orthonormal bases~\cite{BCP,SAP}:
$\{\ket{x}:x\in\mathcal X\}$ on $A$ and $\{\ket{y}:y\in\mathcal Y\}$ on
$B$. States diagonal in the reference basis are incoherent; composite
systems use product reference bases. An incoherent operation ($\IO$) is
a quantum channel that admits a Kraus representation in which every
Kraus operator maps each basis state to a multiple of a basis
state~\cite{BCP}.

\begin{figure}[!htbp]
    \centering
    \resizebox{0.78\linewidth}{!}{\begingroup
\definecolor{setupInk}{HTML}{000000}
\definecolor{setupGray}{HTML}{000000}
\begin{tikzpicture}[
  x=1cm,y=1cm,font=\small,
  >={Latex[length=1.5mm,width=1.1mm]},
  qwire/.style={draw=setupInk,line width=.75pt},
  coding/.style={draw=setupInk,line width=.8pt,fill=white,
    minimum width=1.12cm,minimum height=2.88cm},
  channel/.style={draw=setupInk,line width=.7pt,fill=white,
    minimum width=.80cm,minimum height=.54cm,inner sep=1pt},
  wirelabel/.style={font=\footnotesize,text=setupInk}
]
  \draw[qwire,draw=setupGray]
    (0,1.96) -- (1.12,2.90) -- (9.88,2.90) -- (11,1.96);
  \node[wirelabel,above=3pt] at (1.58,2.90) {$R$};
  \node[wirelabel,above=3pt] at (9.42,2.90) {$R$};
  \node[wirelabel,above=3pt,text=setupGray] at (5.5,2.90) {$\id_R$};
  \node[wirelabel,above left=3pt] at (0,1.96) {$\ket{\Phi_M}_{RA_1}$};
  \node[wirelabel,above right=3pt] at (11,1.96) {$\omega_{RB_1}$};

  \node[coding] (enc) at (2.50,0) {\Large$\E$};
  \node[coding] (dec) at (8.50,0) {\Large$\D$};
  \draw[qwire] (0,1.96) -- (1.10,0) -- (enc.west);
  \draw[qwire] (dec.east) -- (9.90,0) -- (11,1.96);
  \node[wirelabel,above=3pt] at (1.45,0) {$A_1$};
  \node[wirelabel,above=3pt] at (9.55,0) {$B_1$};

  \foreach \yy/\jj in {1.00/1,.24/2,-1.00/n}{
    \node[channel] (N\jj) at (5.50,\yy) {$\N$};
    \draw[qwire,->] (3.06,\yy) --
      node[wirelabel,above=2pt] {$A^{(\jj)}$} (N\jj.west);
    \draw[qwire,->] (N\jj.east) --
      node[wirelabel,above=2pt] {$B^{(\jj)}$} (7.94,\yy);
  }
  \foreach \xx in {4.05,5.50,6.95}{
    \foreach \yy in {-.12,-.27,-.42}{
      \fill[setupInk] (\xx,\yy) circle (.022cm);
    }
  }

\end{tikzpicture}
\endgroup}
    \caption{Entanglement transmission through $n$ uses of $\N$. The
    reference $R$ and logical input $A_1$ start in the maximally
    entangled state $\ket{\Phi_M}_{RA_1}$. The reference is left
    untouched. The encoder $\E$ maps $A_1$ to $A^n$, $\N$ acts
    separately on each input, and the decoder $\D$ acts jointly on $B^n$
    to produce $B_1$. The output $\omega_{RB_1}$ is compared with the
    maximally entangled target $\ket{\Phi_M}_{RB_1}$. For a successful
    sequence of codes, the entanglement fidelity tends to one as $n$
    grows. The encoder, decoder, or both may be restricted to IO.}
    \label{fig:communication-setup}
\end{figure}
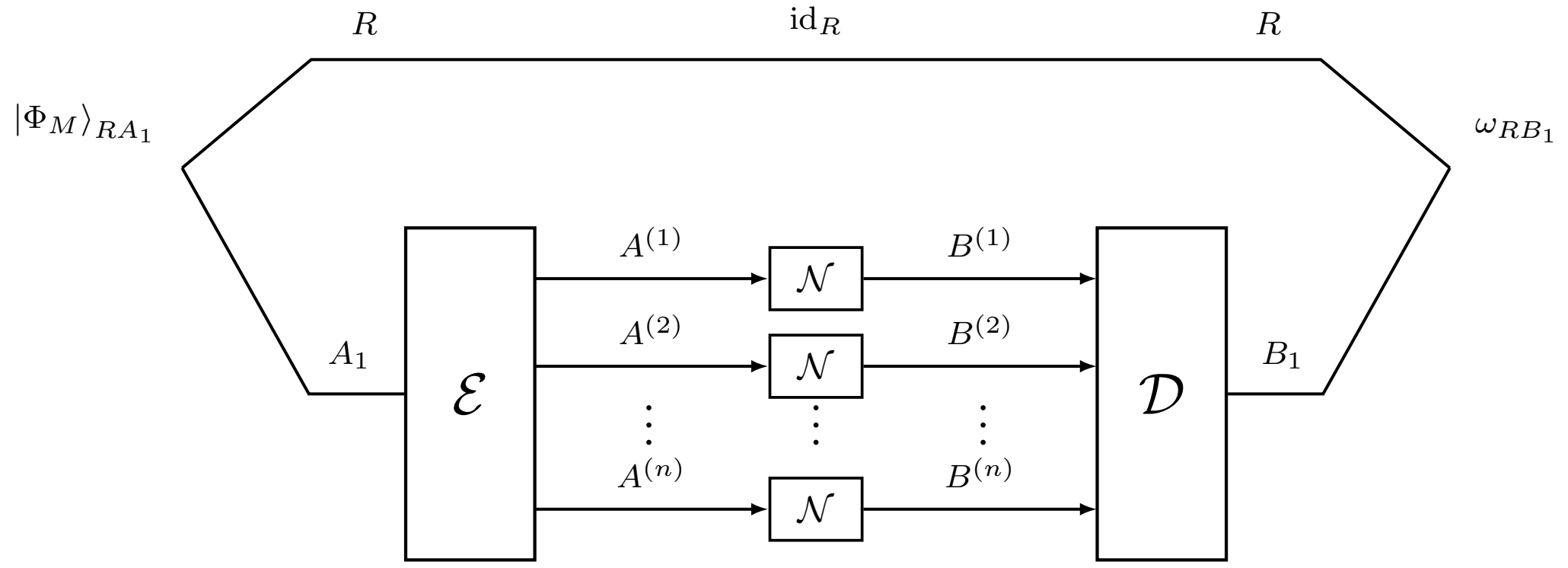

An encoder prepares the inputs to $n$ independent uses of $\N$, and a
decoder acts jointly on their outputs.
Figure~\ref{fig:communication-setup} illustrates this
entanglement-transmission model. To distinguish the individual channel
uses, we write $A^{(k)}$, $B^{(k)}$, and $E^{(k)}$ for the input,
output, and environment of the $k$th use. Thus
$A^n=\bigotimes_{k=1}^n A^{(k)}$, and similarly for $B^n$ and $E^n$.

An $n$-use code has a CPTP encoder $\E_{A_1\to A^n}$ and decoder
$\D_{B^n\to B_1}$, with logical dimension
$\dim A_1=\dim B_1=M$~\cite[Sec.~24.1]{Wilde}. The logical reference
bases are fixed independently of those on $A$ and $B$, and we identify
$\ket{i}_{A_1}$ with $\ket{i}_{B_1}$. Local ancillary systems are
included in the coding maps $\E$ and $\D$. For the effective channel
$\Lambda=\D\circ\N^{\otimes n}\circ\E$, set
\[
    \ket{\Phi_M}_{RA_1}
       =\frac{1}{\sqrt M}\sum_{i=1}^{M}\ket{i}_R\ket{i}_{A_1},
    \qquad
    \omega_{RB_1}=(\id_R\otimes\Lambda)(\dyad{\Phi_M}_{RA_1}).
\]
The reference $R$ is untouched. The entanglement fidelity is
\[
    F_e(\Lambda)=\bra{\Phi_M}_{RB_1}\omega_{RB_1}\ket{\Phi_M}_{RB_1}.
\]

A rate $r\ge0$ is achievable if a sequence of $n$-use codes has logical
dimensions $M_n$ with $\liminf_{n\to\infty}n^{-1}\log M_n\ge r$ and
entanglement fidelities tending to one. Rates are measured in qubits per
channel use; the capacity is the supremum of achievable rates. The
superscripts in $Q^{\IO,\All}(\N)$, $Q^{\All,\IO}(\N)$, and
$Q^{\IO,\IO}(\N)$ specify encoder and decoder restrictions,
respectively; $\All$ permits arbitrary CPTP maps. With both
unrestricted, the capacity is $Q(\N)$.

All logarithms are base two. We follow Refs.~\cite{Wilde,Watrous2018}
for standard properties of quantum channels, fidelity, trace distance,
and entropy.

\subsection{Distance, fidelity, and measurements}
\label{app:distance-fidelity}

For an operator $X$, the trace norm is
$\norm{X}_1=\tr\sqrt{X^\dagger X}$, and the Hilbert--Schmidt norm is
$\norm{X}_2=(\tr X^\dagger X)^{1/2}$. We write $\norm{X}_\infty$ for the
operator norm. The trace distance between states $\rho$ and $\sigma$ is
\begin{equation}
    d(\rho,\sigma)=\frac12\norm{\rho-\sigma}_1.
    \label{eq:trace-distance}
\end{equation}
For positive semidefinite operators $P,Q$, write
$f(P,Q)=\norm{\sqrt P\sqrt Q}_1$ for their root fidelity. For normalized
states, our fidelity convention is $F(\rho,\sigma)=f(\rho,\sigma)^2$.
Quantum channels contract the trace norm, including for non-Hermitian
operators~\cite[Corollary~3.40]{Watrous2018}. In particular, trace
distance cannot increase under a channel. It also bounds the change in
the probability of any measurement outcome:
$\abs{\tr[P(\rho-\sigma)]}\le d(\rho,\sigma)$ for $0\le P\le I$.

For a channel $\Lambda$ between logical systems of dimension $M$,
Schumacher's formula~\cite{Schumacher1996} expresses the entanglement
fidelity in terms of any Kraus representation
$\Lambda(\rho)=\sum_\alpha K_\alpha\rho K_\alpha^\dagger$ as
\begin{equation}
    F_e(\Lambda)
    =
    \frac{1}{M^2}
    \sum_\alpha \abs{\tr K_\alpha}^{2}.
    \label{eq:entanglement-fidelity-kraus}
\end{equation}

A measurement with outcomes $i$ is described by a positive
operator-valued measure (POVM) $\{D_i\}$, where $D_i\ge0$ and
$\sum_iD_i=I$. For input state $\rho_x$, outcome $i$ occurs with
probability $\tr(D_i\rho_x)$. If $X$ records the choice of input from
the ensemble $\{p_x,\rho_x\}$ and $Y$ records the outcome, the Holevo
bound gives
\[
    I(X;Y)\le S\!\left(\sum_xp_x\rho_x\right)-\sum_xp_xS(\rho_x).
\]

\subsection{Dilations and purifications}
\label{app:uhlmann}

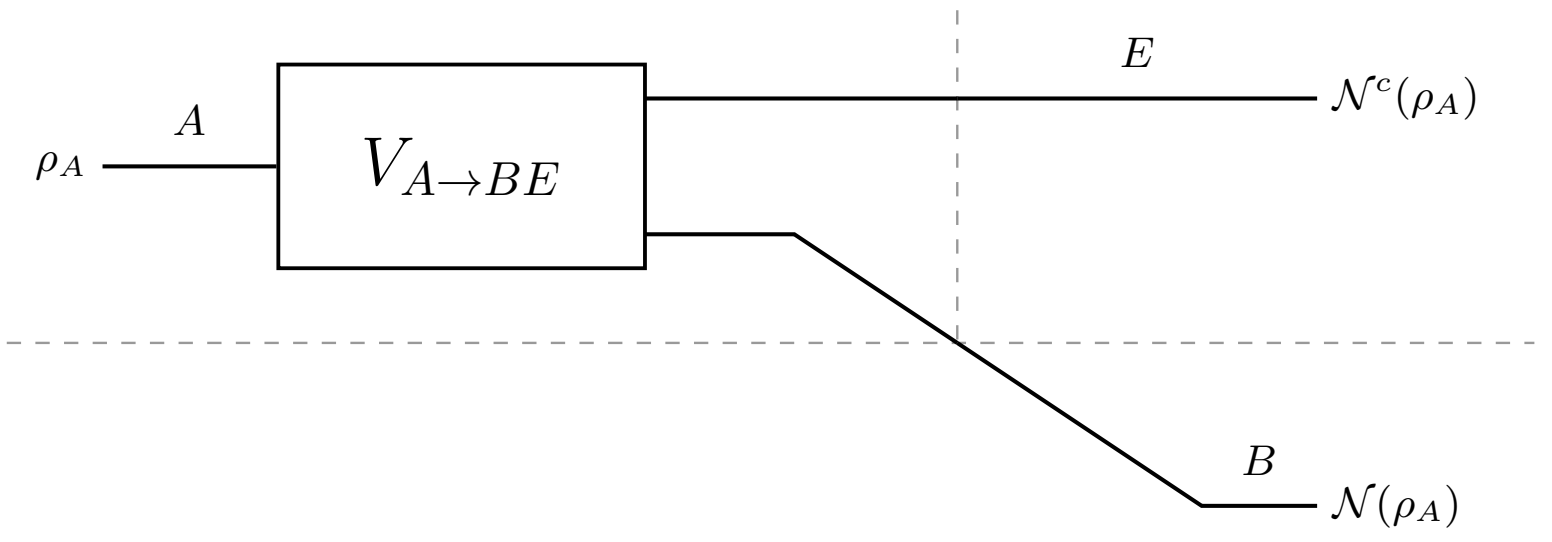
\begin{figure}[!htbp]
    \centering
    \resizebox{0.57\linewidth}{!}{\begingroup
\begin{tikzpicture}[
  x=1cm,y=1cm,font=\small,text=black,
  qwire/.style={draw=black,line width=.85pt},
  dilation/.style={draw=black,line width=.8pt,fill=white,
    minimum width=2.70cm,minimum height=1.50cm,inner sep=5pt},
  wirelabel/.style={font=\footnotesize},
  divider/.style={draw=black!40,dashed,line width=.5pt}
]
  \draw[divider] (0,-1.30) -- (11.25,-1.30);
  \draw[divider] (7.00,1.15) -- (7.00,-1.30);

  \node (input) at (.40,0) {$\rho_A$};
  \node[dilation] (V) at (3.35,0) {\Large$V_{A\to BE}$};
  \draw[qwire] (input.east) --
    node[above=3pt] {$A$} (V.west);

  \draw[qwire] (4.70,.50) -- (7.00,.50);
  \node[anchor=west] (environment) at (9.65,.50)
    {$\N^c(\rho_A)$};
  \draw[qwire] (7.00,.50) --
    node[above=3pt] {$E$} (environment.west);

  \draw[qwire] (4.70,-.50) -- (5.80,-.50) -- (7.00,-1.30);
  \node[anchor=west] (receiver) at (9.65,-2.50)
    {$\N(\rho_A)$};
  \draw[qwire] (7.00,-1.30) -- (8.80,-2.50)
    -- node[above=3pt] {$B$} (receiver.west);
\end{tikzpicture}
\endgroup}
    \caption{Stinespring dilation of one use of $\N_{A\to B}$. The
    isometry $V$ sends $\rho_A$ to the joint state $V\rho_A V^\dagger$
    on $BE$. Tracing out $E$ gives the receiver's state $\N(\rho_A)$,
    and tracing out $B$ gives the environment's state $\N^c(\rho_A)$. In
    general, the receiver and environment are correlated. The circuit
    layout follows Ref.~\cite[Fig.~5.2]{Wilde}.}
    \label{fig:channel-dilation}
\end{figure}

For any Kraus representation
$\N(\rho_A)=\sum_\alpha K_\alpha\rho_A K_\alpha^\dagger$, a Stinespring
isometry is
\begin{equation}
    V=\sum_\alpha K_\alpha\otimes\ket{\alpha}_E,
    \label{eq:kraus-dilation}
\end{equation}
where the states $\ket{\alpha}_E$ of the environment $E$ are
orthonormal. The channel and its complement are
$\N(\rho_A)=\tr_E(V\rho_AV^\dagger)$ and
$\N^c(\rho_A)=\tr_B(V\rho_AV^\dagger)$. Every Stinespring dilation is
obtained from a minimal one by an isometry $W$ on the environment. Since
$W^\dagger W=I$, two environment states satisfy
$W\rho W^\dagger=W\sigma W^\dagger$ if and only if $\rho=\sigma$.
Figure~\ref{fig:channel-dilation} shows how the channel output and the
complementary output are obtained by partial trace.

Uhlmann's theorem relates the root fidelity of two environment states to
the overlap of their purifications. Let $\ket{\phi}_{BE}$ and
$\ket{\psi}_{CE}$ be purifications of states $\rho_E$ and $\sigma_E$,
respectively, with $\dim B=\dim C$. Then
\begin{equation}
    \norm{\sqrt{\rho_E}\sqrt{\sigma_E}}_1
    =
    \max_{U:B\to C}
    \abs{
        \bra{\psi}_{CE}(U\otimes I_E)\ket{\phi}_{BE}
    },
    \label{eq:uhlmann}
\end{equation}
where the maximum is over unitaries from $B$ to $C$. When
$\rho_E=\sigma_E$, the root fidelity is one. A maximizing unitary, with
its phase chosen appropriately, then satisfies
\[
    (U\otimes I_E)\ket{\phi}_{BE}=\ket{\psi}_{CE}.
\]
We use this case in Lemma~\ref{lem:coherent-recovery}.

\paragraph{Comparing environment outputs.}
For two input basis states, Uhlmann's theorem also
gives~\cite[Corollary~3.23]{Watrous2018}
\begin{equation}
    \norm*{\N(\dyad{x}{y})}_1
       =\norm\big{\sqrt{\N^c(\dyad{x})}
                  \sqrt{\N^c(\dyad{y})}}_1
       \le1.
    \label{eq:channel-coherence-fidelity}
\end{equation}
Indeed, for $\ket{\varphi_x}_{BE}=V\ket{x}_A$ and
$\ket{\varphi_y}_{BE}=V\ket{y}_A$, the variational formula for the trace
norm gives
\[
    \begin{aligned}
        \norm*{\N(\dyad{x}{y})}_1
           &=\max_{U_B\text{ unitary}}
              \abs{\tr\!\left[U_B\tr_E
                           \dyad{\varphi_x}{\varphi_y}\right]}\\
           &=\max_{U_B\text{ unitary}}
              \abs*{\bra{\varphi_y}(U_B\otimes I_E)\ket{\varphi_x}}.
    \end{aligned}
\]
This is the root fidelity of the two environment states by
Eq.~\eqref{eq:uhlmann}. The norm in
Eq.~\eqref{eq:channel-coherence-fidelity} equals one exactly when the
two environment states coincide.

\subsection{Information measures}
\label{app:conditional-entropy}

For a bipartite state $\omega_{RB}$, the conditional entropy is
$S(R|B)_\omega=S(\omega_{RB})-S(\omega_B)$. The coherent information of
Sec.~\ref{sec:channels-coherent-information} can therefore be written as
\begin{equation}
    I(R\rangle B)_\omega=-S(R|B)_\omega.
    \label{eq:coherent-information-reference}
\end{equation}

For an input $\rho_A$ purified on $RA$, applying a Stinespring isometry
gives a pure state on $RBE$. Its marginals therefore satisfy
$S(\omega_{RB})=S(\omega_E)$, and the channel coherent information is
\[
    I_c(\rho_A,\N)
       =S\!\left(\N(\rho_A)\right)-S\!\left(\N^c(\rho_A)\right).
\]

For a probability distribution $p=(p_i)$, the R\'enyi entropy of order
$1/2$ is $H_{1/2}(p)=2\log\sum_i\sqrt{p_i}$. If $p$ has at most $M$
outcomes, Cauchy--Schwarz gives $H_{1/2}(p)\le\log M$.

For a classical-quantum state
$\omega_{XE}=\sum_xp_x\dyad{x}_X\otimes\rho_x^E$, the entropy of the
classical register is the Shannon entropy $S(\omega_X)=H(X)$. The states
$\ket{x}_X$ are orthogonal, so
$S(\omega_{XE})=H(X)+\sum_xp_xS(\rho_x^E)$. Substituting these
expressions into $I(X;E)_\omega=S(\omega_X)+S(\omega_E)-S(\omega_{XE})$
cancels the two terms $H(X)$. This gives the Holevo information of the
environment ensemble, $I(X;E)_\omega=S(\omega_E)-\sum_xp_xS(\rho_x^E)$.
The mutual information vanishes exactly when
$\omega_{XE}=\omega_X\otimes\omega_E$. The diagonal blocks corresponding
to $\ket{x}_X$ are $p_x\rho_x^E$ and $p_x\omega_E$, respectively.
Equality holds in every block exactly when $\rho_x^E=\omega_E$ for every
$x$ with $p_x>0$. This is the privacy condition of
Sec.~\ref{sec:io-all}.

\paragraph{Relative entropy and the strong converse.}
\label{app:renyi-divergence}
For a state $\rho$ and a positive definite state $\sigma$, the quantum
relative entropy and the Petz R\'enyi divergence are
\[
    D(\rho\Vert\sigma)=\tr\bigl[\rho(\log\rho-\log\sigma)\bigr],
    \qquad
    D_{1+\alpha}(\rho\Vert\sigma)
       =\frac1\alpha\log\tr\bigl(\rho^{1+\alpha}\sigma^{-\alpha}\bigr).
\]
For singular $\sigma$, these expressions are evaluated on its support
and are infinite unless
$\operatorname{supp}\rho\subseteq\operatorname{supp}\sigma$. For
$0<\alpha\le1$, $D_{1+\alpha}$ cannot increase under a quantum channel,
and it tends to $D$ as $\alpha\downarrow0$. For a finite ensemble of
possible outputs $\{\rho_x\}_x$, the maximum Holevo information also has
the information-radius expression
\[
    \max_p\chi(p)=\inf_{\sigma>0}\max_xD(\rho_x\Vert\sigma),
\]
where the infimum is over positive definite density operators. Thus one
can choose a single comparison state for all outputs, bounding every
relative entropy by the classical capacity plus an arbitrarily small
positive amount. These standard facts~\cite{MosonyiOgawa2017} are used
in the strong converse in Appendix~\ref{app:io-all-converse}.

\subsection{Incoherent operations}
\label{app:incoherent-operations}

In an incoherent Kraus representation, each $K_\alpha$ has at most one
nonzero entry in each column in the reference bases~\cite{WY}. It can
therefore be written as
\begin{equation}
    K_\alpha=\sum_i c_{\alpha i}\dyad{f_\alpha(i)}{i}.
    \label{eq:incoherent-kraus}
\end{equation}
Here $c_{\alpha i}$ is the amplitude for input $\ket{i}$, and
$f_\alpha(i)$ specifies the output basis state when this amplitude is
nonzero. When $c_{\alpha i}=0$, $f_\alpha(i)$ may be chosen arbitrarily.
Different inputs may give the same output basis state.

A maximally incoherent operation (MIO) is a CPTP map that sends every
incoherent state to an incoherent state~\cite{SAP}. By linearity, it is
enough to check this condition on input basis states.

We write $\Delta(\rho)=\sum_i\dyad{i}\rho\dyad{i}$ for complete
dephasing in the reference basis. A subscript specifies the system on
which it acts. A dephasing-covariant incoherent operation (DIO) is a
CPTP map $\Lambda$ satisfying $\Delta\circ\Lambda=\Lambda\circ\Delta$,
with dephasing on the output and input systems, respectively. Every DIO
map is MIO~\cite{ChitambarGour2016}.

A strictly incoherent operation (SIO) is a CPTP map admitting Kraus
operators $K_\alpha$ for which both $K_\alpha$ and $K_\alpha^\dagger$
are incoherent~\cite{Yadin2016}. Equivalently, each Kraus operator has
at most one nonzero entry in each row and each column. SIO maps are DIO,
since they commute with complete dephasing.

\paragraph{Implementing a decoder using coherence.}
A maximally coherent ancillary state
$\ket{\Psi_d}=d^{-1/2}\sum_{j=0}^{d-1}\ket{j}$ allows an IO channel to
implement an arbitrary decoder. For any decoder $\D_{B^n\to B_1}$ with
$\dim B_1=M$ and any integer $d\ge M$, there is an IO channel
$\widetilde\D_{B^nS\to B_1}$, with $\dim S=d$, such that
\begin{equation}
    \widetilde\D\bigl(\rho_{B^n}\otimes\dyad{\Psi_d}_S\bigr)
       =\D(\rho_{B^n})
       \qquad\text{for every state }\rho_{B^n}.
    \label{eq:io-decoder-simulation}
\end{equation}
The case $d=M$ is the simulation result of Ref.~\cite[Theorem~4]{BGMW}.
The following Kraus representation verifies
Eq.~\eqref{eq:io-decoder-simulation} directly for every $d\ge M$. Write
$\D(\rho)=\sum_kK_k\rho K_k^\dagger$ and index the output basis by
$i=0,\ldots,M-1$. For $s=0,\ldots,d-1$, define
\[
    D_{k,s}=\sum_{i=0}^{M-1}\sum_j(K_k)_{ij}
       \ket{i}_{B_1}\bra{j}_{B^n}\otimes\bra{i+s}_S,
\]
where $j$ runs over the input basis and $i+s$ is read modulo $d$. For a
basis input $\ket{j}_{B^n}\ket{t}_S$, at most one term survives: the one
with $i=t-s\pmod d$. Its output is a multiple of $\ket{i}_{B_1}$, or
zero if $i\notin\{0,\ldots,M-1\}$. Thus each $D_{k,s}$ is incoherent.
Since $\sum_s\dyad{i+s}_S=I_S$, orthogonality of the output basis gives
\[
    \sum_{k,s}D_{k,s}^\dagger D_{k,s}
       =\sum_kK_k^\dagger K_k\otimes I_S=I_{B^nS}.
\]
These operators therefore define an IO channel. Moreover,
\[
    D_{k,s}(I_{B^n}\otimes\ket{\Psi_d}_S)=\frac{K_k}{\sqrt d}.
\]
Summing over the $d$ values of $s$ proves
Eq.~\eqref{eq:io-decoder-simulation}.

\section{Proofs for incoherent encoding}
\label{app:io-all-proof}

We use the notation of Sec.~\ref{sec:io-all}.

\begin{ioallrestated}[restated]
For every finite-dimensional quantum channel $\N$,
\begin{equation*}
    Q^{\IO,\All}(\N)
    =\max_{\substack{p\in\mathcal{P}(\mathcal{X})\\
                     I(X;E)_\omega=0}} I(X;B)_\omega
    =\max_c\max_{p\in\mathcal P(\mathcal X_c)}\chi(p).
\end{equation*}
\end{ioallrestated}

For a string $x=(x_1,\ldots,x_n)$, write
$\ket{x}_{A^n}=\bigotimes_{k=1}^n\ket{x_k}_{A^{(k)}}$,
$\rho_{x}^{B^n}=\bigotimes_{k=1}^n\rho_{x_k}^{B^{(k)}}$, and
$\rho_{x}^{E^n}=\bigotimes_{k=1}^n\rho_{x_k}^{E^{(k)}}$.

\begin{figure}[!htbp]
    \centering
    \begin{tikzpicture}[
  x=1cm,y=1cm,font=\small,
  >={Latex[length=1.6mm,width=1.2mm]},
  proofbox/.style={draw=black,line width=.6pt,fill=white,
    text width=5.25cm,minimum height=.9cm,align=center,inner sep=4pt},
  proofarrow/.style={->,draw=black,line width=.65pt},
  reason/.style={font=\footnotesize,align=left,inner sep=1pt}
]
  \node[font=\normalsize\bfseries] at (-3.2,.9) {Achievability};
  \node[font=\normalsize\bfseries] at ( 3.2,.9) {Converse};

  \node[proofbox] (a1) at (-3.2,0)
    {Basis inputs with identical\\environment outputs};
  \node[proofbox] (c1) at ( 3.2,0)
    {Reliable quantum code\\with an IO encoder};
  \node[proofbox] (a2) at (-3.2,-1.7)
    {Classical code on these inputs,\\rate \mbox{$r<\chi(p)$}};
  \node[proofbox] (c2) at ( 3.2,-1.7)
    {Basis codewords with identical\\environment outputs};
  \node[proofbox] (a3) at (-3.2,-3.4)
    {Reliable quantum code\\with an IO encoder};
  \node[proofbox] (c3) at ( 3.2,-3.4)
    {\mbox{$r\le\max_{p:\,\chi_E(p)=0}\chi(p)$}};

  \draw[proofarrow] (a1.south) -- node[reason,right=4pt]
    {Classical coding theorem} (a2.north);
  \draw[proofarrow] (a2.south) -- node[reason,right=4pt]
    {Coherent decoding\\(Lemma~\ref{lem:coherent-recovery})} (a3.north);
  \draw[proofarrow] (c1.south) -- node[reason,right=4pt]
    {Lemma~\ref{lem:io-classical-codewords}; omit a\\vanishing fraction
     of messages} (c2.north);
  \draw[proofarrow] (c2.south) -- node[reason,right=4pt]
    {Classical rate bound} (c3.north);

  \node[draw=black,line width=.6pt,fill=white,
    inner xsep=10pt,inner ysep=6pt] (result) at (0,-5.2) {
    $\displaystyle Q^{\IO,\All}(\N)=\max_{p:\,\chi_E(p)=0}\chi(p)$
  };
  \draw[draw=black,line width=.65pt] (a3.south) -- (-3.2,-4.35) -- (0,-4.35);
  \draw[draw=black,line width=.65pt] (c3.south) -- (3.2,-4.35) -- (0,-4.35);
  \draw[proofarrow] (0,-4.35) -- (result.north);
\end{tikzpicture}%
    \caption{Proof outline for the capacity formula with incoherent
    encoding. Encoding is IO; decoding is unrestricted. A distribution
    $p$ on basis inputs has $\chi_E(p)=0$ exactly when all inputs in its
    support have the same environment output. Achievability turns a
    classical code on such inputs into a quantum code. The converse
    extracts such a classical code from every reliable quantum code.}
    \label{fig:capacity-proof-overview}
\end{figure}
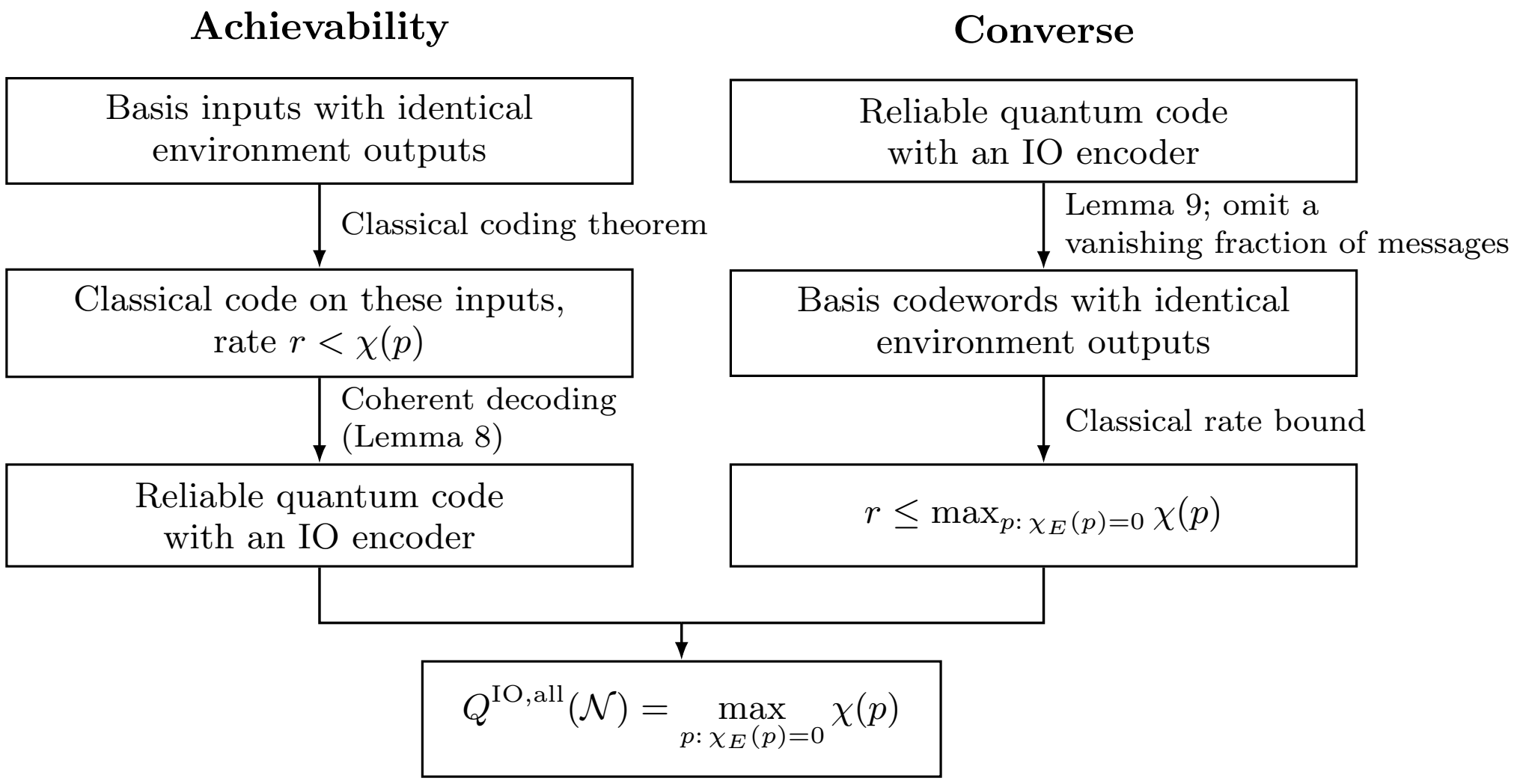

Figure~\ref{fig:capacity-proof-overview} summarizes the two directions
of the proof.

\subsection{Achievability}
\label{app:io-all-achievability}

A decoding measurement, described by a positive operator-valued measure
(POVM), identifies the classical message. For quantum transmission, the
decoder must also preserve superpositions of the logical basis states.
The following lemma implements the measurement coherently and uses
equality of the environment outputs to achieve this. The proof follows
the coherent decoding argument of
Devetak~\cite[proof of Theorem~5]{Devetak2005}, specialized to identical
environment outputs.

\begin{samepage}
\begin{lemma}[Coherent recovery for identical environment outputs]
\label{lem:coherent-recovery}
Let $x_i=(x_{i,1},\ldots,x_{i,n})\in\mathcal{X}^n$, $i=1,\ldots,M$, be
distinct basis strings whose environment outputs $\rho_{x_i}^{E^n}$ are
all equal. Suppose a POVM $\{D_i\}_{i=1}^M$ on $B^n$ has average success
probability
\[
    p_{\mathrm{succ}}=\frac{1}{M}\sum_i
       \tr(D_i\rho_{x_i}^{B^n}),
    \qquad D_i\ge0,\qquad \sum_i D_i=I_{B^n}.
\]
Let $A_1$ and $B_1$ have dimension $M$, and let $\E_{A_1\to A^n}$ be the
channel implemented by the encoding isometry
\[
    \ket{i}_{A_1}\longmapsto\ket{x_i}_{A^n},
    \qquad i=1,\ldots,M.
\]
Then $\E$ is IO, and there is a CPTP decoder $\D_{B^n\to B_1}$ with
\begin{equation}
    F_e(\D\circ\N^{\otimes n}\circ\E)
       \ge p_{\mathrm{succ}}^2.
    \label{eq:coherent-recovery-fidelity}
\end{equation}
\end{lemma}
\end{samepage}

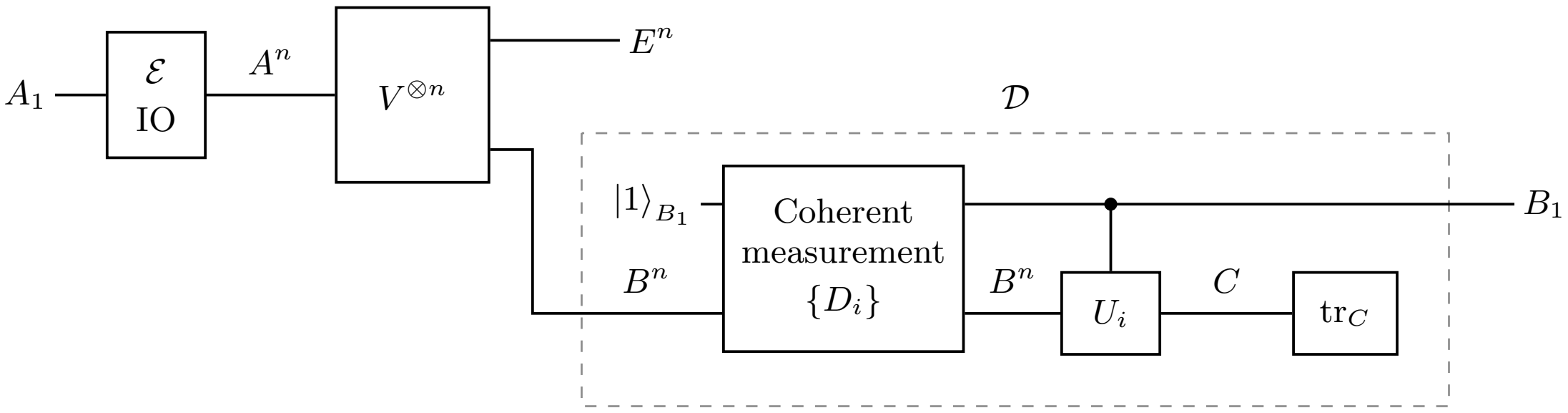
\begin{figure}[htbp]
    \centering
    \resizebox{0.82\linewidth}{!}{\begin{tikzpicture}[
  x=1cm,y=1cm,font=\small,text=black,
  wire/.style={draw=black,line width=.7pt},
  operation/.style={draw=black,fill=white,line width=.7pt,
    align=center,inner sep=4pt},
  system/.style={font=\small,inner sep=2pt},
  boundary/.style={draw=black!45,dashed,line width=.5pt}
]
  \node[system] (input) at (-.1,0) {$A_1$};
  \node[operation,minimum width=.9cm,minimum height=1.15cm]
    (encoder) at (1.1,0) {$\mathcal E$\\[2pt]IO};
  \draw[wire] (input.east) -- (encoder.west);

  \node[operation,minimum width=1.40cm,minimum height=1.60cm]
    (channel) at (3.45,0) {$V^{\otimes n}$};
  \draw[wire] (encoder.east) --
    node[above=3pt,system] {$A^n$} (channel.west);
  \node[system,anchor=west] (environment) at (5.35,.5) {$E^n$};
  \draw[wire] (channel.east |- 0,.5) -- (environment.west);

  \draw[boundary] (5.00,-.35) rectangle (12.95,-2.85);
  \node[above=3pt] at (8.98,-.35) {$\mathcal D$};

  \node[operation,minimum width=2.20cm,minimum height=1.70cm]
    (measurement) at (7.40,-1.50)
    {Coherent\\measurement\\[2pt]$\{D_i\}$};
  \draw[wire] (channel.east |- 0,-.5) -- (4.55,-.5)
    -- (4.55,-2.00)
    -- node[above=3pt,system,pos=.6] {$B^n$}
    (measurement.west |- 0,-2.00);

  \node[system] (ancilla) at (5.65,-1.00) {$\ket{1}_{B_1}$};
  \draw[wire] (ancilla.east) -- (measurement.west |- ancilla.east);

  \coordinate (control) at (9.85,-1.00);
  \node[system,anchor=west] (message) at (13.55,-1.00) {$B_1$};
  \draw[wire] (measurement.east |- control) -- (message.west);

  \node[operation,minimum width=.90cm,minimum height=.75cm]
    (correction) at (9.85,-2.00) {$U_i$};
  \draw[wire] (measurement.east |- correction.west) --
    node[above=3pt,system] {$B^n$} (correction.west);
  \draw[wire] (control) -- (correction.north);
  \fill (control) circle (1.7pt);

  \node[operation,minimum width=.95cm,minimum height=.75cm]
    (discard) at (12.00,-2.00) {$\operatorname{tr}_C$};
  \draw[wire] (correction.east) --
    node[above=3pt,system] {$C$} (discard.west);
\end{tikzpicture}}
    \caption{Quantum transmission with an IO encoder
    (Lemma~\ref{lem:coherent-recovery}), following Devetak's coherent
    decoding argument~\cite{Devetak2005}. The encoder implements
    $\ket{i}_{A_1}\mapsto\ket{x_i}_{A^n}$ for distinct basis codewords
    with identical environment outputs. Here $V$ is a Stinespring
    isometry of $\N$. The goal is to recover the logical quantum state
    in $B_1$. The decoder initializes $B_1$ in $\ket{1}$ and implements
    the decoding POVM $\{D_i\}$ coherently, leaving $B_1$ unmeasured. A
    controlled unitary applies $U_i:B^n\to C$ for each basis state
    $\ket{i}_{B_1}$; $C$ is then traced out. The $U_i$ are chosen by
    Uhlmann's theorem: each maps its corresponding joint state
    immediately after the channel to a common purification.}
    \label{fig:coherent-recovery}
\end{figure}

Figure~\ref{fig:coherent-recovery} shows the encoder and decoder used in
the proof.

\begin{proof}[Proof of Lemma~\ref{lem:coherent-recovery}]
The encoder has the single Kraus operator
$\sum_i\ket{x_i}_{A^n}\bra{i}_{A_1}$. Its columns are orthonormal
because the codewords are distinct basis states, and each column has one
nonzero entry. The encoding is therefore isometric and IO.

For codeword $i$, the Stinespring isometry produces the joint pure state
\[
    \ket{\varphi_i}_{B^nE^n}
       =V^{\otimes n}\ket{x_i}_{A^n}.
\]
All $\ket{\varphi_i}$ purify the same environment state. Choose a
purification $\ket{\eta}_{CE^n}$ on a copy $C$ of $B^n$, so that
$\tr_C\dyad{\eta}_{CE^n}=\rho_{x_i}^{E^n}$ for every $i$. By the
purification property in Appendix~\ref{app:uhlmann}, there are unitaries
$U_i:B^n\to C$ satisfying
\begin{equation}
    (U_i\otimes I_{E^n})\ket{\varphi_i}_{B^nE^n}
       =\ket{\eta}_{CE^n}.
    \label{eq:recovery-purifications}
\end{equation}
Choose an orthonormal basis $\{\ket s_C\}$ and define the decoder by its
Kraus operators
\begin{equation}
    K_s=\sum_i\ket{i}_{B_1}\bra{s}_C U_i\sqrt{D_i},
    \qquad
    \sum_sK_s^\dagger K_s
       =\sum_i\sqrt{D_i}U_i^\dagger U_i\sqrt{D_i}=I_{B^n}.
    \label{eq:coherent-decoder}
\end{equation}
It implements the POVM coherently, recording its outcome in $B_1$
without measuring that register, then applies the corresponding $U_i$
and discards $C$. The displayed identity shows that the resulting
channel is trace preserving.

The factors $\sqrt{D_i}$ can disturb the channel outputs before $U_i$ is
applied. To bound the resulting error, apply Schumacher's formula,
Eq.~\eqref{eq:entanglement-fidelity-kraus}, to the code. Using an
orthonormal basis $\{\ket e_{E^n}\}$ for the environment gives
\begin{align*}
    F_e
       &=\frac1{M^2}\sum_{s,e}\abs{\sum_i
          (\bra s_C\otimes\bra e_{E^n})
          (U_i\sqrt{D_i}\otimes I_{E^n})\ket{\varphi_i}}^2\\
       &=\frac1{M^2}\norm{\sum_i
          (U_i\sqrt{D_i}\otimes I_{E^n})\ket{\varphi_i}}^2.
\end{align*}
The norm is at least the absolute overlap with the pure state
$\ket\eta$. By Eq.~\eqref{eq:recovery-purifications},
\begin{align*}
    \bra\eta(U_i\sqrt{D_i}\otimes I_{E^n})\ket{\varphi_i}
       &=\bra{\varphi_i}(\sqrt{D_i}\otimes I_{E^n})\ket{\varphi_i}\\
       &=\tr(\sqrt{D_i}\rho_{x_i}^{B^n})
        \ge\tr(D_i\rho_{x_i}^{B^n}).
\end{align*}
The inequality follows from $0\le D_i\le I_{B^n}$, which implies
$\sqrt{D_i}\ge D_i$. Summing over $i$ gives
\[
    F_e\ge\frac1{M^2}\abs{\sum_i
          \bra\eta(U_i\sqrt{D_i}\otimes I_{E^n})\ket{\varphi_i}}^2
       \ge\frac1{M^2}\left(\sum_i\tr(D_i\rho_{x_i}^{B^n})\right)^2
       =p_{\mathrm{succ}}^2.
\]
\end{proof}

\begin{proof}[Achievability of the capacity formula]
Fix a class $\mathcal{X}_c$, a distribution
$p\in\mathcal{P}(\mathcal{X}_c)$, and $0<r<\chi(p)$. The
classical-quantum coding
theorem~\cite{Holevo1998,SchumacherWestmoreland1997} gives codes with
strings in $\mathcal{X}_c^n$, rate tending to $r$, and average error
$\varepsilon_n\to0$.

The recovery lemma requires distinct codewords. If several messages use
the same string, add their POVM effects and keep that string only once.
Let $M$ and $M'$ be the original and resulting numbers of codewords. The
sum of the correct decoding probabilities is unchanged, so the new
average success probability satisfies
\[
    p'_{\mathrm{succ}}=\frac{M}{M'}(1-\varepsilon_n)
       \ge1-\varepsilon_n.
\]
Since $p'_{\mathrm{succ}}\le1$, we also have $M'\ge M(1-\varepsilon_n)$,
and the rate still tends to $r$.

Every remaining word belongs to $\mathcal{X}_c^n$. Since the environment
output for one channel use is constant on $\mathcal{X}_c$, all these
words give the same product state on $E^n$.
Lemma~\ref{lem:coherent-recovery} therefore supplies an IO encoder and
an unrestricted decoder with
\[
    F_e\ge(1-\varepsilon_n)^2\longrightarrow1.
\]
Maximizing over $p$ and $c$ proves the lower bound; rate zero is
immediate.
\end{proof}

The same encoders are physically incoherent operations
(PIO)~\cite{ChitambarGour2016}. To implement the isometry
$\ket{i}_{A_1}\mapsto\ket{x_i}_{A^n}$, append $A^n$ in the basis state
$\ket{x_1}$ and apply a permutation of the joint reference basis that
maps
\[
    \ket{i}_{A_1}\ket{x_1}_{A^n}
       \longmapsto\ket{1}_{A_1}\ket{x_i}_{A^n},
       \qquad i=1,\ldots,M.
\]
The prescribed inputs and outputs are distinct basis states, so this map
extends to a permutation of the full basis. Discarding $A_1$ then
implements the encoding isometry. Appending an incoherent state,
applying a basis permutation, and discarding a subsystem are physically
incoherent operations. These are also strictly incoherent. Since the
converse below allows all IO encoders, the capacity formula remains
unchanged for either smaller class of encoders.

\subsection{Converse}
\label{app:io-all-converse}

We first extract basis codewords that the receiver can decode reliably.
Entanglement fidelity close to one also forces almost all of these
codewords to have the same environment output: at every position, their
letters lie in a single environment class. The classical rate bound then
applies within those classes.

\begin{lemma}[Classical codewords from an IO encoder]
\label{lem:io-classical-codewords}
Consider an $n$-use code with an IO encoder $\E_{A_1\to A^n}$ and a
decoder $\D_{B^n\to B_1}$, where $\dim A_1=\dim B_1=M$. There are basis
strings $x_1,\ldots,x_M\in\mathcal X^n$, not necessarily distinct, and
an index $i_0$ such that
\begin{equation}
    F_e(\D\circ\N^{\otimes n}\circ\E)
       \le\frac1M\sum_{j=1}^M\min\Bigl\{\sqrt{s_j},\,
          f(\rho_{x_{i_0}}^{E^n},\rho_{x_j}^{E^n})\Bigr\},
    \qquad
    s_j=\bra j\D(\rho_{x_j}^{B^n})\ket j.
    \label{eq:extracted-classical-code}
\end{equation}
\end{lemma}

Here $s_j$ is the probability of decoding message $j$ correctly in a
classical code: the sender prepares $\ket{x_j}$, and the receiver
applies the original decoder $\D$ and measures $B_1$ in its logical
basis. The fidelities compare the environment outputs with that of one
codeword.

\begin{proof}[Proof of Lemma~\ref{lem:io-classical-codewords}]
Choose an incoherent Kraus representation of the encoder, with
$K_\alpha\ket i=c_{\alpha i}\ket{f_\alpha(i)}$. When $c_{\alpha i}=0$,
choose $f_\alpha(i)$ arbitrarily. Applied to the half $A_1$ of
$\ket{\Phi_M}_{RA_1}$, the operator $K_\alpha$ gives the vector
\[
    (I_R\otimes K_\alpha)\ket{\Phi_M}
       =\frac1{\sqrt M}\sum_i c_{\alpha i}
          \ket i_R\ket{f_\alpha(i)}_{A^n}.
\]
Its squared norm $q_\alpha=M^{-1}\sum_i\abs{c_{\alpha i}}^2$ is the
probability of branch $\alpha$, and $\sum_\alpha q_\alpha=1$ by trace
preservation. The encoded state is the mixture of the normalized
branches with weights $q_\alpha$. The fidelity with a pure target is
linear in the decoded state. It is therefore the average of the
fidelities $F_\alpha$ of the normalized branches,
$F_e=\sum_\alpha q_\alpha F_\alpha$, so some branch of positive
probability has fidelity $F$ at least that of the code.

For this branch, put $x_i=f_\alpha(i)$ and
$c_i=c_{\alpha i}/\sqrt{Mq_\alpha}$. Its normalized input is then
\[
    \ket\psi_{RA^n}=\sum_i c_i\ket i_R\ket{x_i}_{A^n},
    \qquad \sum_i\abs{c_i}^2=1.
\]
Define
$G_{ij}=\bra i\D\bigl(\N^{\otimes n}(\dyad{x_i}{x_j})\bigr)\ket j$.
Expanding $\dyad{\psi}$ and $\ket{\Phi_M}$ in the basis gives
\[
    F=\bra{\Phi_M}(\id_R\otimes\D\circ\N^{\otimes n})
         (\dyad{\psi})\ket{\Phi_M}
     =\frac1M\sum_{i,j}c_i\overline{c_j}G_{ij}.
\]
We use two bounds on $G$. Complete positivity makes $G$ positive
semidefinite: apply $\id_R\otimes(\D\circ\N^{\otimes n})$ to the outer
product of $\sum_i\ket i_R\ket{x_i}_{A^n}$ with itself and take matrix
elements between the states $\ket i_R\ket i_{B_1}$. Consequently,
$\abs{G_{ij}}\le\sqrt{s_i s_j}$, where
$s_i=G_{ii}=\bra i\D(\rho_{x_i}^{B^n})\ket i$ is the probability of
correctly decoding message $i$. Moreover, a matrix entry is at most the
trace norm, and the decoder cannot increase the trace norm. With
Eq.~\eqref{eq:channel-coherence-fidelity} for the channel
$\N^{\otimes n}$, this gives
\[
    \abs{G_{ij}}
       \le\norm*{\D\bigl(\N^{\otimes n}(\dyad{x_i}{x_j})\bigr)}_1
       \le\norm*{\N^{\otimes n}(\dyad{x_i}{x_j})}_1
       =f(\rho_{x_i}^{E^n},\rho_{x_j}^{E^n})=:f_{ij}.
\]
The numbers $m_{ij}=\min\{\sqrt{s_is_j},f_{ij}\}$ are symmetric and
nonnegative, and $\abs{G_{ij}}\le m_{ij}$. Using
$2\abs{c_i}\abs{c_j}\le\abs{c_i}^2+\abs{c_j}^2$ and the symmetry of
$m_{ij}$, we obtain
\[
    F\le\frac1M\sum_{i,j}\abs{c_i}\abs{c_j}m_{ij}
      \le\frac1M\sum_i\abs{c_i}^2\sum_j m_{ij}
      \le\max_i\frac1M\sum_j m_{ij}.
\]
The last step holds because the weights $\abs{c_i}^2$ sum to one. Since
$s_i\le1$, we have $m_{ij}\le\min\{\sqrt{s_j},f_{ij}\}$. An index $i_0$
attaining the maximum proves the claim.
\end{proof}

\begin{proof}[Converse for the capacity formula]
Write $C=\max_c\max_{p\in\mathcal P(\mathcal X_c)}\chi(p)$. There are
finitely many input basis states, and different environment states have
root fidelity strictly below one. Hence
\[
    \varphi=\max_{x,y:\rho_x^E\ne\rho_y^E}f(\rho_x^E,\rho_y^E)<1,
\]
where we set $\varphi=0$ if there is only one environment class.

Consider an $n$-use code with entanglement fidelity at least
$1-\varepsilon$, and take the strings and the index $i_0$ from
Lemma~\ref{lem:io-classical-codewords}. A minimum is at most each of its
two terms, so Eq.~\eqref{eq:extracted-classical-code} gives
\begin{equation}
    \frac1M\sum_j\sqrt{s_j}\ge1-\varepsilon,
    \qquad
    \frac1M\sum_j f(\rho_{x_{i_0}}^{E^n},\rho_{x_j}^{E^n})
       \ge1-\varepsilon.
    \label{eq:app-two-averages}
\end{equation}
For each position $k$, let $\mathcal X_{c_k}$ be the environment class
that contains $x_{i_0,k}$. Select the messages whose codewords use these
classes at every position,
\[
    \mathcal J=\{i:x_{i,k}\in\mathcal X_{c_k}
                         \text{ for every }k\}.
\]
Write $M'=\abs{\mathcal J}$ and $t=1-M'/M$ for the fraction omitted. All
selected words have the same product environment state. For any omitted
word, the root fidelity between its environment output and that of word
$i_0$ is a product over the $n$ positions. At least one factor is at
most $\varphi$, and every other factor is at most one. The second bound
in Eq.~\eqref{eq:app-two-averages} therefore gives
\begin{equation}
    1-\varepsilon
       \le\frac1M\sum_i f(\rho_{x_{i_0}}^{E^n},\rho_{x_i}^{E^n})
       \le\frac{M'+(M-M')\varphi}M
       =1-(1-\varphi)t,
    \qquad t\le\frac{\varepsilon}{1-\varphi}.
    \label{eq:app-class-restriction}
\end{equation}

Since $\sqrt{s_i}\le(1+s_i)/2$, the first bound in
Eq.~\eqref{eq:app-two-averages} gives
$M^{-1}\sum_is_i\ge1-2\varepsilon$. The original classical code
therefore has total decoding error $\sum_i(1-s_i)\le2M\varepsilon$. For
uniformly chosen selected messages, its average error is at most
$\varepsilon'=2\varepsilon/(1-t)$. Decoder outcomes for omitted messages
can be assigned to any selected message without increasing this error.
Relabel the selected words $x_1,\ldots,x_{M'}$ and define
\begin{equation}
    p_k(x)=\frac1{M'}\abs\big{\{i:x_{i,k}=x\}}.
    \label{eq:app-position-distribution}
\end{equation}
This is the input distribution for channel use $k$ under uniform
messages. Its support lies entirely in $\mathcal X_{c_k}$.
Figure~\ref{fig:one-use-ensembles} shows these selected codewords.

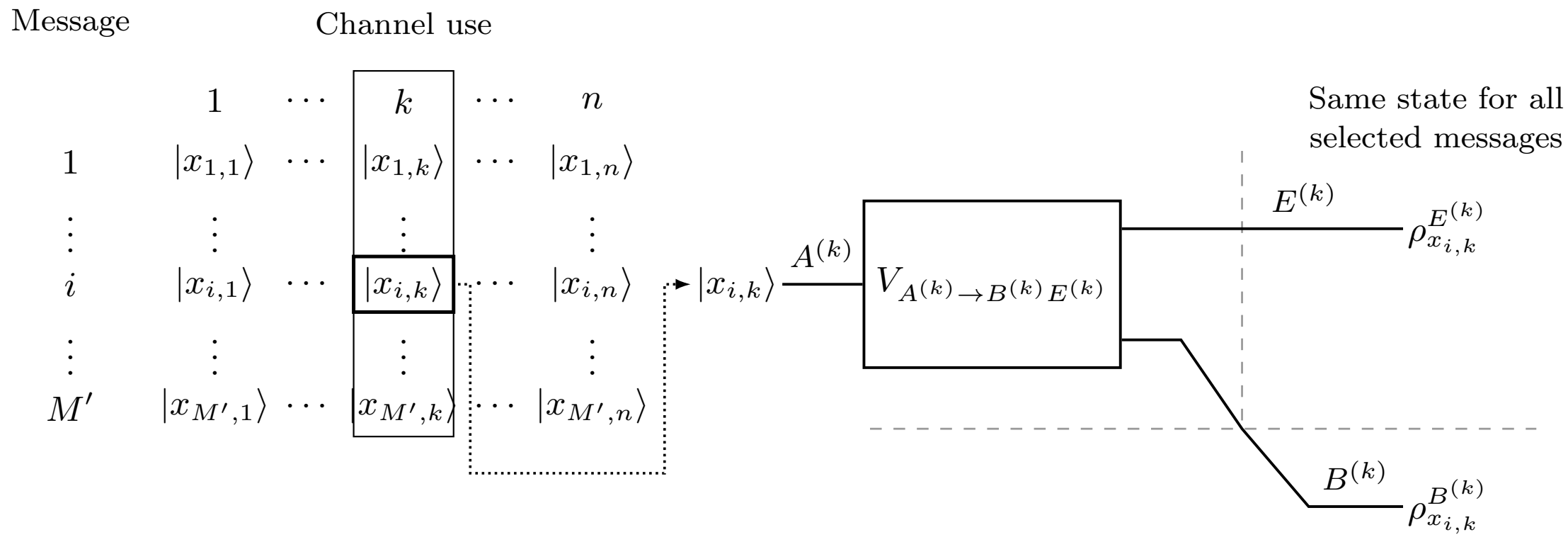
\begin{figure}[!htbp]
    \centering
    \resizebox{0.82\linewidth}{!}{\begin{tikzpicture}[
  x=1cm,y=1cm,
  font=\fontsize{9}{11}\selectfont,
  >={Latex[length=1.4mm,width=1mm]},
  note/.style={font=\fontsize{8}{10}\selectfont,align=center,inner sep=1pt},
  entry/.style={inner sep=1pt,minimum height=.35cm},
  qwire/.style={draw=black,line width=.7pt},
  divider/.style={draw=black!40,dashed,line width=.5pt},
  dilation/.style={draw=black,fill=white,line width=.7pt,
    minimum width=2.2cm,minimum height=1.5cm,inner sep=3pt}
]
  \node[note] at (0,1.80) {Message};
  \node[note] at (3.00,1.80) {Channel use};
  \foreach \xx/\heading in {1.30/1,2.15/\cdots,3.00/k,3.85/\cdots,4.70/n}{
    \node[entry] at (\xx,1.10) {$\heading$};
  }
  \foreach \row/\yy in {1/.55,i/-.55,{M'}/-1.65}{
    \node[entry] at (0,\yy) {$\row$};
    \node[entry] at (1.30,\yy) {$\ket{x_{\row,1}}$};
    \node[entry] at (2.15,\yy) {$\cdots$};
    \node[entry] at (3.00,\yy) {$\ket{x_{\row,k}}$};
    \node[entry] at (3.85,\yy) {$\cdots$};
    \node[entry] at (4.70,\yy) {$\ket{x_{\row,n}}$};
  }
  \foreach \xx in {0,1.30,3.00,4.70}{
    \foreach \yy in {0,-1.10}{
      \node[entry] at (\xx,\yy) {$\vdots$};
    }
  }
  \draw[draw=black,line width=.4pt]
    (2.55,1.37) rectangle (3.45,-1.92);
  \draw[draw=black,line width=.9pt]
    (2.55,-.30) rectangle (3.45,-.80);

  \node[entry] (input) at (6.00,-.55)
    {$\ket{x_{i,k}}$};
  \draw[draw=black,densely dotted,line width=.65pt,->]
    (3.49,-.55) -- (3.60,-.55) -- (3.60,-2.25)
    -- (5.35,-2.25) -- (5.35,-.55) -- (input.west);

  \begin{scope}[xshift=.4cm]
  \draw[divider] (6.80,-1.85) -- (12.80,-1.85);
  \draw[divider] (10.15,.65) -- (10.15,-1.85);
  \node[dilation] (V) at (7.90,-.55)
    {$V_{A^{(k)}\to B^{(k)}E^{(k)}}$};
  \draw[qwire] (input.east) --
    node[above=3pt,note] {$A^{(k)}$} (V.west);

  \node[entry,anchor=west] (environment) at (11.60,-.05)
    {$\rho_{x_{i,k}}^{E^{(k)}}$};
  \draw[qwire] (V.east |- environment.west) --
    node[above=3pt,note,pos=.65] {$E^{(k)}$} (environment.west);

  \node[entry,anchor=west] (receiver) at (11.60,-2.55)
    {$\rho_{x_{i,k}}^{B^{(k)}}$};
  \draw[qwire] (V.east |- 0,-1.05) -- (9.60,-1.05)
    -- (10.15,-1.85) -- (10.75,-2.55)
    -- node[above=3pt,note] {$B^{(k)}$} (receiver.west);

  \node[note] at (11.90,.93)
    {Same state for all\\selected messages};
  \end{scope}
\end{tikzpicture}}
    \caption{The selected basis codewords in the converse. Each row
    corresponds to one of the $M'$ selected messages, and each column to
    a channel use. Under uniform messages, $p_k(x)$ is the fraction of
    column $k$ equal to $x$. All entries of this column belong to
    $\mathcal X_{c_k}$, so their environment outputs coincide. By
    Eq.~\eqref{eq:app-class-restriction}, the fraction of omitted
    messages vanishes with the error.}
    \label{fig:one-use-ensembles}
\end{figure}

Let $J$ be the uniform selected message and $\widehat J$ the decoder's
estimate. Fano's inequality~\cite[Sec.~10.7]{Wilde} and the Holevo
bound~\cite[Theorem~5.49]{Watrous2018} give
\begin{align}
    (1-\varepsilon')\log M'-1
       &\le I(J;\widehat J)\notag\\
       &\le S\!\left(\frac1{M'}\sum_i\rho_{x_i}^{B^n}\right)
                -\frac1{M'}\sum_i S(\rho_{x_i}^{B^n})\notag\\
       &\le\sum_{k=1}^n
          \left[S\!\left(\frac1{M'}\sum_i\rho_{x_{i,k}}^{B^{(k)}}\right)
                 -\frac1{M'}\sum_i S(\rho_{x_{i,k}}^{B^{(k)}})\right]
        =\sum_{k=1}^n\chi(p_k)\le nC.
    \label{eq:app-io-all-rate-bound}
\end{align}
The third inequality uses additivity of entropy for each product output
and subadditivity for their average. Grouping the messages with
$x_{i,k}=x$ gives the equality with $\sum_k\chi(p_k)$. The final bound
follows because each $p_k$ is supported on one class, even though that
class may differ between positions.

For a reliable sequence, $\varepsilon\to0$ implies $t\to0$ and
$\varepsilon'\to0$. Moreover,
\[
    \frac{\log M'}n=\frac{\log M}n+\frac{\log(1-t)}n,
\]
so the selection does not change the asymptotic rate.
Equation~\eqref{eq:app-io-all-rate-bound} then bounds that rate by $C$,
proving the converse.
\end{proof}

In fact, every rate above this capacity forces the entanglement fidelity
to vanish exponentially.

\begin{proposition}[Strong converse for incoherent encoding]
\label{prop:io-all-strong-converse}
Let $R>Q^{\IO,\All}(\N)$. There are constants $\gamma>0$ and $n_0$ such
that every $n$-use code with an IO encoder, an arbitrary decoder, and
$\log M\ge nR$ satisfies
\[
    F_e(\D\circ\N^{\otimes n}\circ\E)\le2^{-\gamma n}
    \qquad(n\ge n_0).
\]
\end{proposition}

\begin{proof}
Set $C=Q^{\IO,\All}(\N)$ and choose $\eta>0$ with $C+2\eta<R$. For each
environment class $\mathcal X_c$, the relative-entropy radius formula in
Appendix~\ref{app:renyi-divergence} gives a full-rank state $\sigma_c$
such that $D(\rho_x^B\Vert\sigma_c)\le C+\eta$ for all
$x\in\mathcal X_c$. Since there are finitely many basis inputs,
continuity in the R\'enyi order gives a common $0<\alpha\le1$ such that
\begin{equation}
    D_{1+\alpha}(\rho_x^B\Vert\sigma_c)\le C+2\eta
       \qquad(x\in\mathcal X_c).
    \label{eq:strong-converse-centers}
\end{equation}
We use these fixed states to bound the success probabilities of a
classical decoding measurement, following the R\'enyi-divergence
argument for classical-quantum channel coding~\cite{MosonyiOgawa2017}.

Take the strings $x_j$, the index $i_0$, and the success probabilities
$s_j$ from Lemma~\ref{lem:io-classical-codewords}. Group the messages by
the environment classes of the letters of their codewords. A pattern
$P=(P_1,\ldots,P_n)\in\{1,\ldots,\nu\}^n$ specifies the classes used at
the $n$ positions, and
\[
    \mathcal I_P=\{j:x_{j,l}\in\mathcal X_{P_l}
                              \text{ for all }l\}
\]
contains the messages with pattern $P$. Repeated codewords cause no
difficulty: the messages keep their distinct logical indices.

Let $T_j=\D^\dagger(\dyad{j})$ be the effects of the decoding POVM and
put $\sigma_P=\bigotimes_{l=1}^n\sigma_{P_l}$. For $j\in\mathcal I_P$,
additivity of the R\'enyi divergence on product states gives
$D_{1+\alpha}(\rho_{x_j}^{B^n}\Vert\sigma_P)\le n(C+2\eta)$. The binary
measurement $\{T_j,I-T_j\}$ accepts these two states with probabilities
$s_j$ and $t_j=\tr(T_j\sigma_P)$. If $T_j=0$, then $s_j=0$. Otherwise
$t_j>0$, because $\sigma_P$ has full rank. The measurement is a quantum
channel with a classical output bit, so the R\'enyi divergence of the
two outcome distributions is at most that of the two states. For
probability distributions $u$ and $v$ on a finite set,
$D_{1+\alpha}(u\Vert v)=\alpha^{-1}\log\sum_\omega
u(\omega)^{1+\alpha}v(\omega)^{-\alpha}$ is the logarithm of a sum of
nonnegative terms. Keeping only the term of the outcome ``accept'' gives
\[
    \frac1\alpha\log\bigl(s_j^{1+\alpha}t_j^{-\alpha}\bigr)
       \le D_{1+\alpha}\bigl((s_j,1-s_j)\big\Vert(t_j,1-t_j)\bigr)
       \le D_{1+\alpha}(\rho_{x_j}^{B^n}\Vert\sigma_P)
       \le n(C+2\eta).
\]
Solving for $s_j$ yields
\[
    s_j\le 2^{\beta n(C+2\eta)}t_j^{\beta},
    \qquad \beta=\frac{\alpha}{1+\alpha}.
\]
The effects $T_j$ sum to the identity, so
$\sum_{j\in\mathcal I_P}t_j\le\tr\sigma_P=1$. Taking square roots and
applying H\"older's inequality with exponents $2/\beta$ and
$(1-\beta/2)^{-1}$ yields
\begin{equation}
    \sum_{j\in\mathcal I_P}\sqrt{s_j}
       \le2^{\beta n(C+2\eta)/2}\sum_{j\in\mathcal I_P}t_j^{\beta/2}
       \le2^{\beta n(C+2\eta)/2}\,\abs{\mathcal I_P}^{1-\beta/2}
       \le M^{1-\beta/2}\,2^{\beta n(C+2\eta)/2}.
    \label{eq:strong-converse-success-sum}
\end{equation}

Let $\varphi<1$ be the largest root fidelity between environment outputs
of basis inputs from different classes, setting $\varphi=0$ if there is
only one class, and fix $0\le k\le n$. If the patterns of $x_j$ and
$x_{i_0}$ differ in more than $k$ positions, multiplicativity of root
fidelity gives
$f(\rho_{x_{i_0}}^{E^n},\rho_{x_j}^{E^n})\le\varphi^{k+1}$. All other
messages belong to one of at most
\[
    V_k=\sum_{l=0}^k\binom nl(\nu-1)^l
\]
patterns. In Eq.~\eqref{eq:extracted-classical-code}, bound the minimum
by the fidelity for the first kind of message and by $\sqrt{s_j}$ for
the second. There are at most $M$ messages of the first kind, and
Eq.~\eqref{eq:strong-converse-success-sum} bounds the sum over each of
the at most $V_k$ patterns of the second kind. Hence
\begin{equation}
    \begin{aligned}
    F_e(\D\circ\N^{\otimes n}\circ\E)
       &\le\frac1M\Bigl[M\varphi^{k+1}
          +V_kM^{1-\beta/2}\,2^{\beta n(C+2\eta)/2}\Bigr]\\
       &=\varphi^{k+1}
          +V_k\,2^{-\frac{\beta}{2}
                             (\log M-n(C+2\eta))}.
    \end{aligned}
    \label{eq:strong-converse-finite-bound}
\end{equation}

Write $b=\frac{\beta}{2}(R-C-2\eta)>0$. If $\varphi=0$, take $k=0$, so
$V_k=1$ and the bound is $2^{-nb}$. Otherwise $\nu\ge2$. Choose a
sufficiently small $0<\delta<(\nu-1)/\nu$ such that
$h_2(\delta)+\delta\log(\nu-1)<b$, and take $k=\lfloor\delta n\rfloor$.
The bound on the size of a Hamming ball gives
\[
    V_k\le2^{n[h_2(\delta)+\delta\log(\nu-1)]}.
\]
Since $\log M\ge nR$, the second term in
Eq.~\eqref{eq:strong-converse-finite-bound} is at most
$2^{-n[b-h_2(\delta)-\delta\log(\nu-1)]}$, and the first is at most
$\varphi^{\delta n}$. Both decay exponentially. Their sum is at most
$2^{-\gamma n}$ for every $\gamma$ below both decay exponents and all
sufficiently large $n$, which proves the proposition.
\end{proof}

\begin{samepage}
\subsection{Generic zero capacity}
\label{app:zero-capacity}

\paragraph{Zero-capacity criterion.}
For every finite-dimensional channel,
$Q^{\IO,\All}(\N)=Q^{\IO,\IO}(\N)=0$ if and only if
\[
    \rho_x^E=\rho_y^E\quad\Longrightarrow\quad\rho_x^B=\rho_y^B
\]
for every pair of input basis states $\ket{x},\ket{y}$.
\end{samepage}

\begin{proof}
Suppose that $\rho_x^E=\rho_y^E$ implies $\rho_x^B=\rho_y^B$ for every
pair of basis inputs. Within every environment class, the receiver's
states then coincide. Hence $\chi(p)=0$ for every distribution supported
on a class, and the capacity formula gives $Q^{\IO,\All}(\N)=0$.

Conversely, if the implication fails, there are two basis inputs with
$\rho_x^E=\rho_y^E$ and $\rho_x^B\ne\rho_y^B$. The distribution
assigning probability $1/2$ to each is supported on one environment
class. Strict concavity of the von Neumann entropy gives $\chi(p)>0$, so
the capacity formula yields $Q^{\IO,\All}(\N)>0$. Since
$\tfrac12 Q^{\IO,\All}(\N)\le Q^{\IO,\IO}(\N)\le Q^{\IO,\All}(\N)$ by
Appendix~\ref{app:half-capacity-bounds}, the two capacities vanish
together. This proves the assertion for an IO decoder as well.
\end{proof}

In particular, an injective complementary channel sends distinct input
states to distinct environment states. The criterion therefore gives
zero capacity with incoherent encoding in every input basis.

\begin{proof}[Proof of Corollary~\ref{cor:zero-capacity}]
The channel $\mathcal R(\rho)=\tr(\rho)\sigma$ sends every input state
to the same output and satisfies $\mathcal R(\dyad{x}{y})=0$ for
distinct basis states of any input reference basis. By
Eq.~\eqref{eq:channel-coherence-fidelity}, the environment outputs of
$\N_\varepsilon=(1-\varepsilon)\N+\varepsilon\mathcal R$ therefore obey
\begin{equation}
    f\bigl(\N_\varepsilon^c(\dyad{x}),
           \N_\varepsilon^c(\dyad{y})\bigr)
       =\norm{\N_\varepsilon(\dyad{x}{y})}_1
       =(1-\varepsilon)\norm{\N(\dyad{x}{y})}_1
       \le1-\varepsilon<1.
    \label{eq:noise-environment-fidelity}
\end{equation}
Every class is a singleton, so both capacities with incoherent encoding
vanish.

For the fidelity bound, let $\varepsilon\le1/2$ and consider an $n$-use
code with an IO encoder and logical dimension $M\ge2$. Take the strings
$x_j$, the index $i_0$, and the success probabilities $s_j$ from
Lemma~\ref{lem:io-classical-codewords}, and let $k$ be the number of
messages $j$ with $x_j=x_{i_0}$. The effects of the decoding POVM sum to
the identity, so these $k$ success probabilities sum to at most one, and
their square roots to at most $\sqrt k$ by the Cauchy--Schwarz
inequality. Every other codeword differs from $x_{i_0}$ in at least one
position. Multiplicativity of root fidelity and
Eq.~\eqref{eq:noise-environment-fidelity} bound its term in
Eq.~\eqref{eq:extracted-classical-code} by $1-\varepsilon$. Hence
\[
    F_e(\D\circ\N_\varepsilon^{\otimes n}\circ\E)
       \le\frac{\sqrt k+(M-k)(1-\varepsilon)}M
       \le1-\varepsilon\Bigl(1-\frac1M\Bigr)
       \le1-\frac\varepsilon2.
\]
The second inequality holds because the numerator does not increase with
$k\ge1$: its increments are
$\sqrt{k+1}-\sqrt k-(1-\varepsilon)\le\sqrt2-\tfrac32<0$.
\end{proof}

\paragraph{Generic channels.}
Let $J(\N)=\sum_{x,y}\dyad{x}{y}_A\otimes\N(\dyad{x}{y})$ be the Choi
matrix of $\N$, and take $\sigma=I_B/d_B$ with $d_B=\dim B$, so that
$J(\mathcal R)=I_{AB}/d_B$. The channels $\N_\varepsilon$ with
$0<\varepsilon\le1$ are then exactly the channels whose Choi matrix has
full rank. Indeed, $J(\N_\varepsilon)\ge\varepsilon I_{AB}/d_B$.
Conversely, if the smallest eigenvalue $\lambda$ of $J(\N)$ is positive,
then $J(\N)-\lambda I_{AB}\ge0$. Put $\varepsilon=\lambda d_B$, which is
at most one because the eigenvalues of $J(\N)$ have mean $1/d_B$. Then
$\N-\varepsilon\mathcal R$ is completely positive, so
$\N=(1-\varepsilon)\N'+\varepsilon\mathcal R$ for some channel $\N'$.

Full rank is an open condition, and $\N_\varepsilon\to\N$ as
$\varepsilon\to0$. These channels therefore form an open dense subset of
the CPTP maps from $A$ to $B$. The Choi matrices of all channels lie in
the real affine space of Hermitian operators $J$ on $AB$ with
$\tr_BJ=I_A$. On this space $\det J$ is a polynomial that does not
vanish identically, since it is positive at $J(\mathcal R)$. Its zero
set has Lebesgue measure zero and contains every channel whose Choi
matrix is singular. Finally,
$\norm{\N_\varepsilon-\N}_\diamond\le2\varepsilon$, so both capacities
are discontinuous at every channel where their value is positive.

\paragraph{Continuity for a fixed code.}
The discontinuity concerns the asymptotic capacity. For a fixed encoder
$\E$, decoder $\D$, and block length $n$,
\[
    \begin{split}
    \bigl|F_e(\D\circ\N^{\otimes n}\circ\E)
    -F_e(\D\circ(\N')^{\otimes n}\circ\E)\bigr|
    \le \frac n2\norm{\N-\N'}_\diamond,
    \end{split}
\]
where $\norm{\cdot}_\diamond$ is the diamond norm. To obtain the bound,
replace one channel use at a time. Each replacement changes the decoded
state by trace distance at most $\tfrac12\norm{\N-\N'}_\diamond$, since
the other channel uses, encoding, and decoding cannot increase this
distance. Entanglement fidelity is a measurement probability, so its
change is bounded by the same distance. This proves continuity at every
fixed block length; the bound grows with $n$ and does not imply
continuity of capacity.

\subsection{Additivity and convexity}
\label{app:additivity-convexity}

\begin{additivityrestated}[restated]
For any two finite-dimensional quantum channels,
\begin{equation*}
    Q^{\IO,\All}(\N_1\otimes\N_2)
    =Q^{\IO,\All}(\N_1)+Q^{\IO,\All}(\N_2).
\end{equation*}
For channels $\N_j$ from $A$ to $B$ and probabilities $\lambda_j$,
\begin{equation*}
    Q^{\IO,\All}\!\left(\sum_j\lambda_j\N_j\right)
       \le\sum_j\lambda_j Q^{\IO,\All}(\N_j).
\end{equation*}
\end{additivityrestated}

\begin{proof}
\emph{Additivity.} For the two channels, denote the outputs of the input
basis states by $\rho_x^{B'},\rho_x^{E'}$ and
$\tau_y^{B''},\tau_y^{E''}$, respectively. With a product Stinespring
dilation, the environment output for the input basis state
$\ket{x}\otimes\ket{y}$ is $\rho_x^{E'}\otimes\tau_y^{E''}$. Two such
product states are equal if and only if their respective marginals are
equal. Thus, writing $\mathcal{X}_c$ and $\mathcal{Y}_d$ for the classes
of the two channels, the classes of their product are precisely
$\mathcal{X}_c\times\mathcal{Y}_d$.

Fix a distribution $p_{xy}$ on $\mathcal{X}_c\times\mathcal{Y}_d$, with
marginals $p_x=\sum_y p_{xy}$ and $q_y=\sum_x p_{xy}$. Write
$\chi_{12}(p_{xy})$, $\chi_1(p_x)$, and $\chi_2(q_y)$ for the Holevo
information of the joint ensemble and its two marginal ensembles,
respectively. Subadditivity of entropy for the average output state and
additivity of entropy for each product output state give
\[
    \chi_{12}(p_{xy})\le\chi_1(p_x)+\chi_2(q_y).
\]
Equality holds for product distributions. The maximum Holevo information
within a product class is therefore the sum of the maxima for its two
individual classes. Maximizing over $c$ and $d$ and applying the
capacity formula proves the claimed additivity.

\medskip\noindent \emph{Convexity.} Let $\N=\sum_j\lambda_j\N_j$,
omitting components with zero weight. Fix a distribution $p$ satisfying
the privacy condition for $\N$, and let $x,y$ lie in its support. Their
environment outputs under $\N$ coincide, so
Eq.~\eqref{eq:channel-coherence-fidelity} and the triangle inequality
give
\[
    1=\norm{\N(\dyad{x}{y})}_1
       \le\sum_j\lambda_j\norm{\N_j(\dyad{x}{y})}_1\le1.
\]
Each norm on the right is at most one and every $\lambda_j$ is positive,
so all these norms equal one. By the same fidelity identity,
$\N_j^c(\dyad{x})=\N_j^c(\dyad{y})$ for every $j$. Thus $p$ satisfies
the privacy condition for each component channel.

Write $\chi(p;\mathcal{M})$ for the Holevo information of
$\{p_x,\mathcal{M}(\dyad{x})\}$. The output ensembles satisfy
\[
    \omega^{(j)}_{XB}=\sum_x p_x\dyad{x}_X\otimes\N_j(\dyad{x}_A),
    \qquad \omega_{XB}=\sum_j\lambda_j\omega^{(j)}_{XB}.
\]
These states have the same $X$ marginal, with entropy $H(X)$. By
concavity of conditional entropy~\cite[Chap.~11]{Wilde}, we obtain
\begin{align*}
    \chi(p;\N)
       &=H(X)-S(X|B)_\omega\\
       &\le\sum_j\lambda_j
                    \bigl[H(X)-S(X|B)_{\omega^{(j)}}\bigr]\\
       &=\sum_j\lambda_j\chi(p;\N_j)
        \le\sum_j\lambda_j Q^{\IO,\All}(\N_j).
\end{align*}
The last inequality holds because $p$ satisfies the privacy condition
for each component channel. Maximizing over the distributions feasible
for $\N$ proves convexity.
\end{proof}

\subsection{Changing the class of allowed encoders}
\label{app:dio-encoder-separation}

Theorem~\ref{thm:io-all-capacity} remains valid for PIO and SIO
encoders, but not for all other choices of free operations. In
particular, the following channel has zero capacity with IO encoding,
yet transmits one qubit perfectly with a DIO encoder and an IO decoder.

Let $A=C\otimes S$ consist of two qubits, and identify the channel
output $B$ with a copy of $C$. Define
\begin{equation}
    U=\dyad{0}_C\otimes I_S+\dyad{1}_C\otimes H_S,
    \qquad
    \N_{A\to B}(\sigma)=\tr_S(U^\dagger\sigma U),
    \label{eq:dio-encoder-channel}
\end{equation}
where $H$ is the Hadamard gate. The unitary $U^\dagger$ is a Stinespring
isometry from $A$ to $B\otimes S$, with $S$ the environment. For the
four input basis states,
\begin{equation}
    \N^c_{A\to S}(\dyad{i}_C\otimes\dyad{j}_S)
       =H^i\dyad{j}H^i,\qquad i,j\in\{0,1\}.
    \label{eq:dio-encoder-environment}
\end{equation}
These are the two computational basis states and the two Hadamard basis
states, all distinct. Each environment class therefore contains a single
input, so Theorem~\ref{thm:io-all-capacity} gives
\[
    Q^{\IO,\All}(\N)=Q^{\IO,\IO}(\N)=0.
\]

Take qubit logical systems $A_1$ and $B_1$, with their basis states
identified with those of $C$ and $B$, respectively. Choose the encoder
\begin{equation}
    \E_{A_1\to A}(\rho)
       =U\left(\rho\otimes\frac{I_S}{2}\right)U^\dagger.
    \label{eq:dio-encoder-map}
\end{equation}
To check its dephasing covariance, write
$\rho=\sum_{i,j=0}^1\rho_{ij}\dyad{i}{j}$. Its blocks on $C$ are
\[
    \E(\rho)=\frac12\sum_{i,j=0}^1
       \rho_{ij}\dyad{i}{j}_C\otimes H^iH^j.
\]
The diagonal blocks are multiples of $I_S$, whereas the other blocks
have zero diagonal on $A$. Consequently,
\[
    \Delta_A(\E(\rho))
       =\frac12\sum_i\rho_{ii}\dyad{i}_C\otimes I_S
       =\E(\Delta_{A_1}(\rho)).
\]
Thus $\E$ is DIO and hence MIO. Direct substitution gives
$\N\circ\E=\id_{A_1\to B}$, so the identity decoder from $B$ to $B_1$
recovers every input state, including its entanglement with a reference.
Applying this code independently on each use achieves one qubit per use.
Since the channel output is a qubit, no higher rate is possible. Thus
its capacity is one with either DIO or MIO encoding, whether the decoder
is IO or unrestricted.

\section{Proofs for incoherent decoding}
\label{app:incoherent-recovery}

This appendix proves Theorems~\ref{thm:io-decoder-half-bounds}
and~\ref{thm:io-recovery} and Corollary~\ref{cor:io-decoder-capacity}.
Fix an encoder $\E_{A_1\to A^n}$ and write
$\mathcal T=\N^{\otimes n}\circ\E$ for the channel from the logical
input $A_1$ to the receiver's system $B^n$. Both logical systems $A_1$
and $B_1$ have dimension $M$, and their basis states with the same index
are identified. We use $i$ for a logical basis index and $y$ for a basis
string of $B^n$.

For a vector $\ket v$ and an operator $X$ on $B_1$, let
\[
    \norm{v}_{\ell_1}=\sum_i\abs{\braket{i}{v}},
    \qquad
    \norm{X}_{\ell_1}=\sum_{i,j}\abs{\bra{i}X\ket{j}}.
\]
Both norms use the logical basis. In particular,
$\norm{\dyad{v}}_{\ell_1}=\norm{v}_{\ell_1}^2$. We reserve $\norm{X}_1$
for the trace norm and use $\norm{v}$ for the Euclidean norm.

\subsection{Admissible maps and incoherent implementation}
\label{app:admissible-maps}

For a channel with Kraus operators $D_a$, trace preservation gives
\[
    \sum_a\norm{D_a\ket y}^2
      =\bra y\sum_aD_a^\dagger D_a\ket y=1.
\]
Every vector satisfies $\norm{v}_{\ell_1}\ge\norm{v}$, with equality
exactly when at most one coordinate is nonzero. Consequently,
$\sum_a\norm{D_a\ket y}_{\ell_1}^2\le1$ for every $y$ if and only if
every $D_a$ has at most one nonzero entry in each column. A channel is
therefore IO precisely when it has such a Kraus representation. We keep
this column condition but allow the trace to decrease.

As in Sec.~\ref{sec:io-recovery}, a finite family $\{A_a:B^n\to B_1\}$
is \emph{admissible} if
\begin{equation}
    \sum_aA_a^\dagger A_a\le I_{B^n},
    \qquad
    \sum_a\norm{A_a\ket y}_{\ell_1}^2\le1
       \quad\text{for every }y.
    \label{eq:sm-admissible-recovery}
\end{equation}
A completely positive map $\mathcal A$ is admissible if it has an
admissible Kraus family. The first inequality makes $\mathcal A$ trace
nonincreasing. The second also implies
\begin{equation}
    \norm{\mathcal A(\dyad y)}_{\ell_1}
       \le\sum_a\norm{A_a\ket y}_{\ell_1}^2\le1,
    \label{eq:sm-basis-output-norm}
\end{equation}
by the triangle inequality for $\norm{\cdot}_{\ell_1}$.

\begin{lemma}[Sampling]
\label{lem:io-sampling}
For every operator $A:B^n\to B_1$, there is a random operator $Z$ taking
finitely many values such that every value has at most one nonzero entry
per column and
\begin{align}
    \mathbb E Z&=A,\notag\\
    \mathbb E[Z^\dagger Z]
       &=A^\dagger A+
          \sum_y\bigl(\norm{A\ket y}_{\ell_1}^2
                       -\norm{A\ket y}^2\bigr)\dyad y.
    \label{eq:sm-sampling-moments}
\end{align}
\end{lemma}

\begin{proof}
Write $A_{iy}=\bra iA\ket y$ and $s_y=\sum_i\abs{A_{iy}}$. For a column
with $s_y>0$, select one row $i$ with probability $\abs{A_{iy}}/s_y$,
and put
\[
    Z_{iy}=s_y\frac{A_{iy}}{\abs{A_{iy}}}.
\]
All other entries of that column are zero. If $s_y=0$, the whole column
is zero. The choices for different columns are independent. Each nonzero
entry $A_{iy}$ is therefore reproduced in expectation, so
$\mathbb E Z=A$.

For $y\ne y'$, independence gives
\[
    \mathbb E\bra yZ^\dagger Z\ket{y'}
      =\sum_i\overline{A_{iy}}A_{iy'}
      =\bra yA^\dagger A\ket{y'}.
\]
The squared norm of column $y$ is always $s_y^2$. Hence the diagonal
entry of $\mathbb E[Z^\dagger Z]$ is $\norm{A\ket y}_{\ell_1}^2$,
whereas that of $A^\dagger A$ is $\norm{A\ket y}^2$. This proves the
second identity.
\end{proof}

For a completely positive map $\mathcal A$ with Kraus operators $A_a$
and an encoded channel $\mathcal T$ with Kraus operators $T_b$, the same
computation as in Eq.~\eqref{eq:entanglement-fidelity-kraus} gives
\begin{equation}
    F_e(\mathcal A\circ\mathcal T)
      =\bra{\Phi_M}(\id_R\otimes\mathcal A\circ\mathcal T)
                       (\dyad{\Phi_M})\ket{\Phi_M}
      =\frac{1}{M^2}\sum_{a,b}\abs{\tr(A_aT_b)}^2.
    \label{eq:sm-cp-entanglement-fidelity}
\end{equation}
This expression is well defined even when $\mathcal A$ decreases the
trace.

Sampling changes the normalization of the Kraus operators. The next
lemma controls this change by keeping only a subset of logical basis
states. A diagonal rescaling then gives an IO decoder whose fidelity
remains close to that of the original map. We allow any
trace-nonincreasing map here: admissible maps give
Corollary~\ref{cor:io-decoder-capacity}, whereas unrestricted decoders
give the lower bounds in Theorem~\ref{thm:io-decoder-half-bounds}.

\begin{lemma}[Incoherent decoding on a subcode]
\label{lem:io-subcode-decoder}
Let $\mathcal T$ be a channel with logical input dimension $M\ge2$ and
finite-dimensional output. Let $\mathcal A$ be a completely positive,
trace-nonincreasing map from its output to an $M$-dimensional logical
system, with Kraus operators $A_a$. Define
\begin{align*}
    \pi_y(i)&=\sum_a\abs{\bra iA_a\ket y}^2,
    &\ell_y&=\sum_i\sqrt{\pi_y(i)},\\
    \kappa_y&=\sum_a\norm{A_a\ket y}_{\ell_1}^2,
    &\bar p(y)&=\bra y\mathcal T(I/M)\ket y.
\end{align*}
Here $\sum_i\pi_y(i)\le1$, and $\bar p$ is the distribution obtained by
measuring the channel output for the maximally mixed logical input. For
$1\le m\le M$, put
\begin{equation}
    \varepsilon_m
      =\sum_y\bar p(y)\min\left\{1,
          \frac{m-1}{M-1}\kappa_y,
          \frac{m-1}{M-1}\ell_y\right\}.
    \label{eq:io-subcode-error}
\end{equation}
There is an isometry $W:\mathbb C^m\to\mathbb C^M$ mapping basis states
to distinct logical basis states and an IO decoder $\D$ with
$m$-dimensional output such that
\begin{equation}
    F_e(\D\circ\mathcal T_W)
      \ge\left(\sqrt{F_e(\mathcal A\circ\mathcal T)}
                  -\sqrt{\varepsilon_m}\right)_+^2,
    \qquad
    \mathcal T_W(\rho)=\mathcal T(W\rho W^\dagger),
    \label{eq:io-subcode-fidelity}
\end{equation}
where $u_+=\max\{u,0\}$.
\end{lemma}

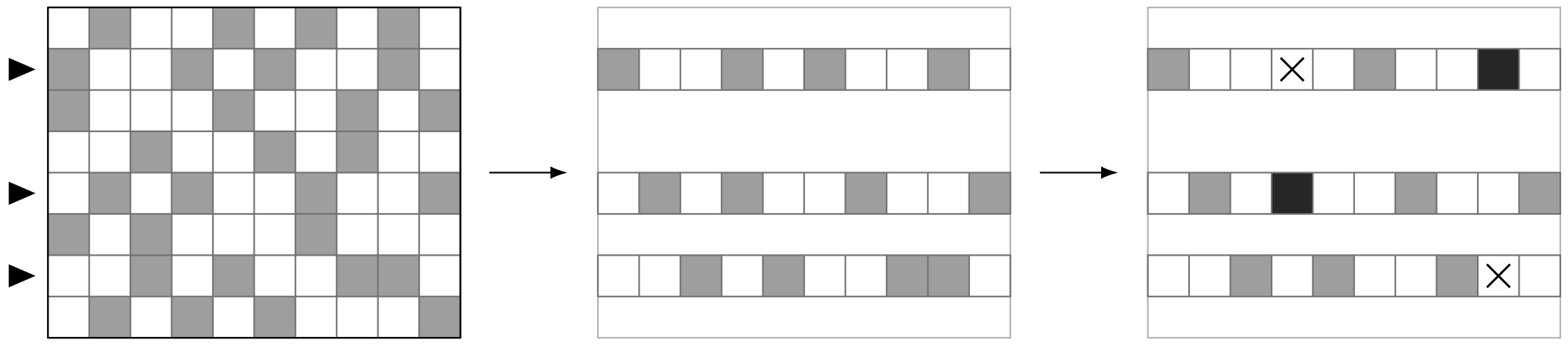
\begin{figure}[!htbp]
    \centering
    \begin{tikzpicture}[x=.36cm,y=.36cm,line width=.4pt,
  >={Latex[length=1.5mm,width=1.1mm]}]
  \def\rowsS{2,5,7}
  \def\inS{1/2,4/2,6/2,9/2,2/5,4/5,7/5,10/5,3/7,5/7,8/7,9/7}
  \def\outS{1/3,1/6,2/1,2/8,3/4,3/6,4/8,5/1,5/3,6/4,6/8,7/1,7/6,8/3,8/4,%
            9/1,10/3,10/8}
  \def\same{1/2,6/2,2/5,7/5,10/5,3/7,5/7,8/7}
  \def\kept{4/5,9/2}
  \def\zero{4/2,9/7}
  \newcommand{\cell}[3]{\fill[#3] (#1-1,-#2) rectangle (#1,-#2+1);}
  \newcommand{\strip}[1]{%
    \draw[black!55] (0,-#1) rectangle (10,-#1+1);
    \foreach \c in {1,...,9} {\draw[black!55] (\c,-#1) -- (\c,-#1+1);}}
  \begin{scope}
    \foreach \c/\r in \inS  {\cell{\c}{\r}{black!38}}
    \foreach \c/\r in \outS {\cell{\c}{\r}{black!38}}
    \draw[black!55,step=.36cm] (0,-8) grid (10,0);
    \draw (0,-8) rectangle (10,0);
    \foreach \r in \rowsS
      {\fill (-.95,-\r+.78) -- (-.3,-\r+.5) -- (-.95,-\r+.22) -- cycle;}
    \node[anchor=north,font=\footnotesize] at (5,-8.4) {(a)};
  \end{scope}
  \draw[->] (10.7,-4) -- (12.6,-4);
  \begin{scope}[xshift=4.8cm]
    \draw[black!30] (0,-8) rectangle (10,0);
    \foreach \c/\r in \inS {\cell{\c}{\r}{black!38}}
    \foreach \r in \rowsS {\strip{\r}}
    \node[anchor=north,font=\footnotesize] at (5,-8.4) {(b)};
  \end{scope}
  \draw[->] (24.05,-4) -- (25.95,-4);
  \begin{scope}[xshift=9.6cm]
    \draw[black!30] (0,-8) rectangle (10,0);
    \foreach \c/\r in \same {\cell{\c}{\r}{black!38}}
    \foreach \c/\r in \kept {\cell{\c}{\r}{black!85}}
    \foreach \r in \rowsS {\strip{\r}}
    \foreach \c/\r in \zero {%
      \draw[line width=.55pt] (\c-.78,-\r+.22) -- (\c-.22,-\r+.78);
      \draw[line width=.55pt] (\c-.78,-\r+.78) -- (\c-.22,-\r+.22);}
    \node[anchor=north,font=\footnotesize] at (5,-8.4) {(c)};
  \end{scope}
\end{tikzpicture}
    \caption{Restriction and sampling in the proof of
    Lemma~\ref{lem:io-subcode-decoder}. Rows are logical basis states,
    columns are output basis strings, and shaded cells are nonzero
    entries. (a) A Kraus operator $A_a$ of the map $\mathcal A$.
    Triangles mark the rows in the random set $S$. (b) The restriction
    $A_{a,S}$ keeps only these rows. Two columns still contain two
    nonzero entries. (c) One value of the random operator $Z_{a,S}$. In
    each of these two columns, one entry is chosen with probability
    proportional to its modulus and rescaled (dark), and the other is
    set to zero (crossed). The remaining columns are unchanged.}
    \label{fig:sampling}
\end{figure}

\begin{proof}
Choose a subset $S\subseteq\{1,\ldots,M\}$ of size $m$ uniformly at
random. Let $W_S:\mathbb C^m\to\mathbb C^M$ be the inclusion that maps
the basis states to the logical basis states with indices in $S$, in
increasing order. Restrict the logical output of each Kraus operator by
setting $A_{a,S}=W_S^\dagger A_a$. For each physical output basis state,
write
\[
    \xi_y(S)=\sum_a\left(\norm{A_{a,S}\ket y}_{\ell_1}^2
                          -\norm{A_{a,S}\ket y}^2\right),
    \qquad
    L_S=\sum_y\frac{\dyad y}{\sqrt{1+\xi_y(S)}}.
\]
The numbers $\xi_y(S)$ are nonnegative. Apply
Lemma~\ref{lem:io-sampling} to each $A_{a,S}$, and denote the resulting
random operator by $Z_{a,S}$ (Fig.~\ref{fig:sampling}). Since
$\sum_aA_{a,S}^\dagger A_{a,S}\le I$, its moment identity gives
\begin{equation}
    \begin{aligned}
    \sum_a\mathbb E[(Z_{a,S}L_S)^\dagger Z_{a,S}L_S]
      &=L_S\left(\sum_aA_{a,S}^\dagger A_{a,S}
                         +\sum_y\xi_y(S)\dyad y\right)L_S\\
      &\le L_S\left(\sum_y\bigl(1+\xi_y(S)\bigr)\dyad y\right)L_S
       =I.
    \end{aligned}
    \label{eq:io-subcode-normalization}
\end{equation}
We use every possible value of $Z_{a,S}L_S$ as a Kraus operator,
multiplied by the square root of its sampling probability. Each operator
is incoherent because the diagonal factor $L_S$ does not introduce any
further nonzero entries in a column. The difference between $I$ and the
left-hand side of Eq.~\eqref{eq:io-subcode-normalization} is positive
semidefinite. Decomposing it as $\sum_j\dyad{v_j}$ allows us to add the
operators $\ket{i_0}\bra{v_j}$ for a fixed logical basis state
$\ket{i_0}$. These operators are also incoherent and complete the family
to an IO channel $\D_S$.

We next bound the change in fidelity caused by $L_S$. For Kraus
operators $T_b$ of $\mathcal T$, define the vector of amplitudes
\[
    v_S(a,b)=\frac1m\tr(W_S^\dagger A_aT_bW_S).
\]
The diagonal rescaling changes this vector by a vector of squared norm
at most
\begin{align}
    &\frac1{m^2}\sum_{a,b}
      \abs{\tr[A_{a,S}(I-L_S)T_bW_S]}^2\notag\\
    &\qquad\le\frac1m\sum_b
       \tr[W_S^\dagger T_b^\dagger(I-L_S)^2T_bW_S]
       =:w_S.
    \label{eq:io-subcode-amplitude-loss}
\end{align}
The inequality uses $\abs{\tr X}^2\le m\tr(X^\dagger X)$ and
$\sum_aA_{a,S}^\dagger A_{a,S}\le I$.

By Eq.~\eqref{eq:sm-cp-entanglement-fidelity}, the fidelity of
$\D_S\circ\mathcal T_{W_S}$ is a sum of nonnegative terms, one for each
Kraus operator of $\D_S$ and each $b$. We drop the terms of the
additional operators $\ket{i_0}\bra{v_j}$. For the sampled operators,
the mean of a squared absolute value is at least the squared absolute
value of the mean, and $\mathbb EZ_{a,S}=A_{a,S}$. Hence
\[
    F_e(\D_S\circ\mathcal T_{W_S})
       \ge\frac1{m^2}\sum_{a,b}
          \mathbb E\abs{\tr(Z_{a,S}L_ST_bW_S)}^2
       \ge\frac1{m^2}\sum_{a,b}
          \abs{\tr(A_{a,S}L_ST_bW_S)}^2.
\]
The right-hand side is the squared norm of the vector with entries
$v_S(a,b)-m^{-1}\tr[A_{a,S}(I-L_S)T_bW_S]$. By
Eq.~\eqref{eq:io-subcode-amplitude-loss} and the triangle inequality,
\begin{equation}
    \sqrt{F_e(\D_S\circ\mathcal T_{W_S})}
       \ge\norm{v_S}-\sqrt{w_S}.
    \label{eq:io-subcode-root-fidelity}
\end{equation}

To average $w_S$, we must also account for the output distribution of
the selected subcode. Put $p(y|i)=\bra y\mathcal T(\dyad i)\ket y$. Then
\[
    w_S=\frac1m\sum_{i\in S}\sum_y p(y|i)
           \left(1-\frac1{\sqrt{1+\xi_y(S)}}\right)^2.
\]
Fix $i,y$ and condition on $i\in S$. Expanding the difference of squared
norms gives
\[
    \xi_y(S)=\sum_a\sum_{\substack{j,k\in S\\j\ne k}}
       \abs{\bra jA_a\ket y}\,\abs{\bra kA_a\ket y}.
\]
Each pair of distinct indices is included with conditional probability
at most $\tau=(m-1)/(M-1)$. Thus
\[
    \mathbb E[\xi_y(S)\mid i\in S]\le\tau\kappa_y.
\]
For a second bound, let $u_j=\sqrt{\pi_y(j)}$ and
$z=\sum_{j\in S\setminus\{i\}}u_j$. The Cauchy--Schwarz inequality in
the Kraus index gives
\[
    \xi_y(S)\le\sum_{\substack{j,k\in S\\j\ne k}}u_ju_k
       \le2u_i z+z^2\le2z+z^2,
\]
because $u_i\le1$. Consequently,
\[
    \left(1-\frac1{\sqrt{1+\xi_y(S)}}\right)^2
       \le\min\{1,\xi_y(S),z\}.
\]
Indeed, $0\le1-(1+\xi)^{-1/2}\le\min\{1,\xi\}$ for every $\xi\ge0$,
which gives the first two bounds. The bound by $z$ follows from
$1+\xi_y(S)\le(1+z)^2$ and $z^2/(1+z)^2\le z$. Each $j\ne i$ lies in $S$
with conditional probability $\tau$, so
$\mathbb E[z\mid i\in S]=\tau(\ell_y-u_i)\le\tau\ell_y$. The expectation
of a minimum is at most the minimum of the expectations, and each $i$
lies in $S$ with probability $m/M$. Conditioning on the transmitted
basis state therefore yields
\begin{align}
    \mathbb E_S w_S
       &=\frac1M\sum_{i,y}p(y|i)\,\mathbb E\!\left[
          \left(1-\frac1{\sqrt{1+\xi_y(S)}}\right)^2
          \,\middle|\,i\in S\right]\notag\\
       &\le\frac1M\sum_{i,y}p(y|i)
                     \min\{1,\tau\kappa_y,\tau\ell_y\}
        =\varepsilon_m.
    \label{eq:io-subcode-average-loss}
\end{align}
The last equality uses $\bar p(y)=M^{-1}\sum_ip(y|i)$.

Finally, each index belongs to $S$ with probability $m/M$, so
\[
    \mathbb E_S v_S(a,b)=\frac1M\tr(A_aT_b).
\]
Convexity of the Euclidean norm and concavity of the square root
therefore give
\[
    \mathbb E_S\norm{v_S}\ge
       \sqrt{F_e(\mathcal A\circ\mathcal T)},
    \qquad
    \mathbb E_S\sqrt{w_S}\le\sqrt{\varepsilon_m}.
\]
Averaging Eq.~\eqref{eq:io-subcode-root-fidelity} shows that at least
one subset $S$ attains the claimed bound.
\end{proof}

\subsection{Entanglement fidelity with incoherent decoding}
\label{app:fidelity-bounds}

We use the definition from Sec.~\ref{sec:io-recovery},
\begin{equation}
    F_{\ell_1}(\mathcal T)
       =\sup_{\mathcal A\ \mathrm{admissible}}
            F_e(\mathcal A\circ\mathcal T).
    \label{eq:sm-recovery-fidelity}
\end{equation}

\begin{recoveryrestated}[restated]
For every encoded channel $\mathcal T$ of logical dimension $M$,
\[
    \sup_{\D\in\IO}F_e(\D\circ\mathcal T)
       \le F_{\ell_1}(\mathcal T).
\]
For every integer $1\le m\le M$, there is an isometry
$W:\mathbb C^m\to\mathbb C^M$ mapping basis states to distinct logical
basis states such that the restricted channel
$\mathcal T_W(\rho)=\mathcal T(W\rho W^\dagger)$ satisfies
\begin{equation}
    \sup_{\D\in\IO}F_e(\D\circ\mathcal T_W)
       \ge\left(\sqrt{F_{\ell_1}(\mathcal T)}
                   -\sqrt{m/M}\right)_+^2.
    \label{eq:sm-subcode-fidelity}
\end{equation}
The second supremum is over IO decoders with $m$-dimensional output, and
$F_e$ uses $\ket{\Phi_m}$. Here $u_+=\max\{u,0\}$.
\end{recoveryrestated}

\begin{proof}
Every IO decoder has an admissible Kraus representation, which proves
the upper bound. For $M=1$, the receiver can discard the channel output
and prepare the unique logical basis state, giving fidelity one.

Suppose $M\ge2$ and fix $m$. For any admissible map $\mathcal A$, the
column condition gives $\kappa_y\le1$. The error in
Lemma~\ref{lem:io-subcode-decoder} therefore obeys
\[
    \varepsilon_m\le\frac{m-1}{M-1}\le\frac mM.
\]
There are finitely many subsets of $m$ logical basis states. For each
subset, choose the inclusion $W$ that preserves their order. Maximizing
over these inclusions, the lemma gives
\[
    \max_W\sup_{\D\in\IO}F_e(\D\circ\mathcal T_W)
       \ge\left(\sqrt{F_e(\mathcal A\circ\mathcal T)}
                   -\sqrt{m/M}\right)_+^2.
\]
The left-hand side does not depend on $\mathcal A$. Taking the supremum
over admissible maps, and using continuity of the right-hand side as a
function of their fidelity, yields Eq.~\eqref{eq:sm-subcode-fidelity}
for at least one of the finitely many inclusions.
\end{proof}

\subsection{Capacity with incoherent decoding}
\label{app:decoder-capacity-formula}

The two bounds in Theorem~\ref{thm:io-recovery} prove
Corollary~\ref{cor:io-decoder-capacity}. For achievability, keeping a
fraction $1/n$ of the logical basis states makes the additional error
vanish while reducing the logarithm of the code dimension by only
$O(\log n)$.

\begin{capacityrestated}[restated]
For every finite-dimensional quantum channel $\N$, each capacity with IO
decoding is given by
\[
    \sup\left\{R\ge0:\lim_{n\to\infty}
         \sup_{\E}F_{\ell_1}(\N^{\otimes n}\circ\E)=1\right\}.
\]
The inner supremum is over encoders with $n$ channel inputs and logical
dimension $M\ge2^{nR}$: IO encoders for $Q^{\IO,\IO}(\N)$, and arbitrary
CPTP encoders for $Q^{\All,\IO}(\N)$.
\end{capacityrestated}

\begin{proof}[Proof of Corollary~\ref{cor:io-decoder-capacity}]
Fix either IO encoders or unrestricted encoders, and let $R_0$ denote
the supremum on the right-hand side of the stated formula. If a rate
$r>0$ is achievable with IO decoding, choose codes attaining that rate
with error tending to zero. For every $R<r$, their dimensions eventually
exceed $2^{nR}$. The upper bound in Theorem~\ref{thm:io-recovery} gives
$F_{\ell_1}\to1$ for these encoders, and hence $r\le R_0$.

Conversely, let $R>0$ be a rate for which the inner supremum tends to
one. Choose encoders $\E_n$ of dimensions $M_n\ge2^{nR}$ such that, for
$\mathcal T_n=\N^{\otimes n}\circ\E_n$, $F_{\ell_1}(\mathcal T_n)\to1$.
Apply Theorem~\ref{thm:io-recovery} with $m_n=\lfloor M_n/n\rfloor$,
which is positive for all sufficiently large $n$. It gives inclusions
$W_n$ for which the restricted channel
$\mathcal T_{n,W_n}(\rho)=\mathcal T_n(W_n\rho W_n^\dagger)$ satisfies
\[
    \sup_{\D\in\IO}F_e(\D\circ\mathcal T_{n,W_n})
       \ge\left(\sqrt{F_{\ell_1}(\mathcal T_n)}
                      -\frac1{\sqrt n}\right)_+^2
       \longrightarrow1,
\]
since $m_n/M_n\le1/n$. Choosing an IO decoder within $1/n$ of this
supremum gives fidelity tending to one for the restricted encoder
$\rho\mapsto\E_n(W_n\rho W_n^\dagger)$. The inclusion $W_n$ maps basis
states to basis states, so these encoders are IO whenever $\E_n$ is IO.
Moreover,
\[
    \frac1n\log m_n
       \ge R-\frac{\log n+1}{n}
\]
for all sufficiently large $n$. Thus every rate below $R$ is achievable
with IO decoding. Taking the supremum over $R$ proves the reverse bound.
Rate zero is achieved with a one-dimensional code, completing the
argument for both encoder classes.
\end{proof}

\subsection{Capacity bounds with incoherent decoding}
\label{app:half-capacity-bounds}

The lower bounds in Theorem~\ref{thm:io-decoder-half-bounds} follow from
Lemma~\ref{lem:io-subcode-decoder}. It converts a decoder for a larger
code into an IO decoder for a subspace spanned by logical basis states.

\begin{iodecoderrestated}[restated]
Restricting the decoder to IO costs at most half the capacity:
$Q^{\All,\IO}(\N)\ge\tfrac12 Q(\N)$ with unrestricted encoding, and
$Q^{\IO,\IO}(\N)\ge\tfrac12 Q^{\IO,\All}(\N)$ with incoherent encoding.
Neither factor $1/2$ can be increased uniformly over finite-dimensional
channels.
\end{iodecoderrestated}

The factor $1/2$ is attained by the Fourier channel with IO encoding
(Appendix~\ref{app:fourier-unitaries}) and approached by the flagged MUB
channels $\N_d$ with unrestricted encoding
(Appendix~\ref{app:optimal-decoder-factor}).

\begin{proof}[Lower bounds in Theorem~\ref{thm:io-decoder-half-bounds}]
Consider either IO encoders or unrestricted encoders. Let $r>0$ be
achievable with unrestricted decoding, and choose a sequence of codes
with logical dimensions $M_n$, entanglement fidelities tending to one,
and $\liminf_n n^{-1}\log M_n\ge r$.

Apply Lemma~\ref{lem:io-subcode-decoder} to the encoded channel
$\mathcal T_n=\N^{\otimes n}\circ\E_n$ and its decoder
$\mathcal A=\D_n$, keeping $m_n=\lceil\sqrt{M_n}/n\rceil$ logical basis
states. For the trace-preserving decoder $\D_n$,
$\pi_y(i)=\bra i\D_n(\dyad y)\ket i$ is the distribution obtained by
measuring its output in the logical basis for input $\ket y$. Thus
$\ell_y=2^{H_{1/2}(\pi_y)/2}$. Since this distribution has at most $M_n$
outcomes, $H_{1/2}(\pi_y)\le\log M_n$, and hence $\ell_y\le\sqrt{M_n}$.
The error term in the lemma is therefore at most
\[
    \frac{(m_n-1)\sqrt{M_n}}{M_n-1}
       \le\frac{M_n}{n(M_n-1)}\le\frac2n
\]
for all sufficiently large $n$. The resulting IO decoders have
entanglement fidelity tending to one, while
\[
    \frac{\log m_n}{n}
       \ge\frac{\log M_n}{2n}-\frac{\log n}{n}.
\]
Thus rate $r/2$ is achievable. The inclusions map basis states to basis
states and are therefore IO, so composing them with the original
encoders preserves the encoder class. Taking the supremum over
achievable $r$ proves both bounds; when the original capacity is zero,
the lower bound is immediate.
\end{proof}

\subsection{Upper bounds for general codes}
\label{app:decoder-fidelity-upper-bounds}

The next lemma allows each Kraus operator of the encoded channel to be
split into two parts. It bounds their contributions to the fidelity
using Hilbert--Schmidt norms for the first part and matrix entries for
the second. This separation gives the uncertainty bound in
Appendix~\ref{app:measurement-uncertainty}, which applies to both $\N_H$
and $\N_d$. Setting the first part to zero gives the simpler bound used
for the Fourier channel in Appendix~\ref{app:fourier-unitaries}.

\begin{lemma}[Upper bound from the Kraus operators]
\label{lem:recovery-amplitude-bound}
Let $\mathcal A:B^n\to B_1$ be a completely positive,
trace-nonincreasing map such that
$\norm{\mathcal A(\dyad y)}_{\ell_1}\le1$ for every $y$. Let $T_b$ be
Kraus operators of a channel $\mathcal T:A_1\to B^n$. For arbitrary
decompositions $T_b=T_b'+T_b''$,
\begin{equation}
    M\sqrt{F_e(\mathcal A\circ\mathcal T)}
       \le\left(M\sum_b\norm{T_b'}_2^2\right)^{1/2}
          +\sum_y\left(\max_i\sum_b
                    \abs{\bra yT_b''\ket i}^2\right)^{1/2}.
    \label{eq:sm-kraus-decomposition-bound}
\end{equation}
\end{lemma}

Every admissible map satisfies these assumptions by
Eq.~\eqref{eq:sm-basis-output-norm}. So does every MIO channel: its
output on a basis state is diagonal and has unit trace.

\begin{proof}
Choose Kraus operators $A_a$ for $\mathcal A$. For a family $X=(X_b)$ of
operators from $A_1$ to $B^n$, set
\[
    q(X)=\left(\sum_{a,b}\abs{\tr(A_aX_b)}^2\right)^{1/2}.
\]
This is the Euclidean norm of a vector depending linearly on $X$, and
therefore satisfies the triangle inequality. Thus
\[
    M\sqrt{F_e(\mathcal A\circ\mathcal T)}
       =q(T)\le q(T')+q(T'').
\]
For the first term, the Cauchy--Schwarz inequality gives
$\abs{\tr Y}^2\le M\tr(Y^\dagger Y)$ for every operator $Y$ on the
logical space. Since $\sum_aA_a^\dagger A_a\le I$, it follows that
\[
    q(T')^2\le M\sum_{a,b}\tr(T_b'^\dagger A_a^\dagger A_aT_b')
       \le M\sum_b\norm{T_b'}_2^2.
\]

For the second term, let $\ket{r_{by}}=(T_b'')^\dagger\ket y$ and
$R_y=\sum_b\dyad{r_{by}}$. Then
$\tr(A_aT_b'')=\sum_y\bra{r_{by}}A_a\ket y$. The triangle inequality for
the Euclidean norm gives
\begin{align*}
    q(T'')
      &\le\sum_y
         \left(\sum_{a,b}\abs{\bra{r_{by}}A_a\ket y}^2\right)^{1/2}\\
      &=\sum_y\left(\tr[\mathcal A(\dyad y)R_y]\right)^{1/2}.
\end{align*}
For matrices $P$ and $R$, expansion of the trace in the basis gives
$\abs{\tr(PR)}\le\norm{P}_{\ell_1}\max_{i,j}\abs{R_{ij}}$. Since
$R_y\ge0$, $\abs{(R_y)_{ij}}^2\le(R_y)_{ii}(R_y)_{jj}$, so its largest
entry in absolute value is a diagonal entry. Consequently,
\begin{align*}
    \tr[\mathcal A(\dyad y)R_y]
       &\le\norm{\mathcal A(\dyad y)}_{\ell_1}
                \max_i\bra iR_y\ket i\\
       &\le\max_i\sum_b\abs{\bra yT_b''\ket i}^2.
\end{align*}
This proves the claimed bound on $q(T'')$.
\end{proof}

For a general encoded channel, set $T_b'=0$ in
Lemma~\ref{lem:recovery-amplitude-bound} and define
\[
    P(y|i)=\bra y\mathcal T(\dyad i)\ket y
       =\sum_b\abs{\bra yT_b\ket i}^2.
\]
This is the probability of output $y$ when logical basis state $\ket i$
is sent and the receiver measures in the reference basis. We obtain
\begin{equation}
    M\sqrt{F_{\ell_1}(\mathcal T)}
       \le\sum_y\max_i\sqrt{P(y|i)}.
    \label{eq:sm-amplitude-bound}
\end{equation}
The same bound holds with $F_{\ell_1}(\mathcal T)$ replaced by
$F_e(\D\circ\mathcal T)$ for any MIO decoder.

\section{Examples}
\label{app:pauli-noise}

This appendix proves the statements on the examples of
Secs.~\ref{sec:zero-capacity} and~\ref{sec:examples}, and shows that
$Q^{\IO,\IO}$ is not additive.

\subsection{Pauli channels}
\label{app:pauli-capacity}

We first show that, for MIO channels with IO encoding, IO decoding
attains the same rate as unrestricted decoding.
Theorem~\ref{thm:pauli-capacity} follows by evaluating this rate for
Pauli channels.

For a basis input $\ket x$, let $P(y|x)=\bra y\N(\dyad x)\ket y$ be the
probability of outcome $Y=y$ when the output is measured in the
reference basis. Write $I(X;Y)_p$ for the mutual information of this
classical channel with input distribution $p$.

\begin{proposition}[Incoherent decoding within an environment class]
\label{prop:io-classical-decoding}
For every finite-dimensional channel $\N$,
\begin{equation}
    Q^{\IO,\IO}(\N)
       \ge\max_c\max_{p\in\mathcal P(\mathcal X_c)}I(X;Y)_p.
    \label{eq:io-classical-decoding}
\end{equation}
If $\N$ is MIO, equality holds and both sides equal $Q^{\IO,\All}(\N)$.
\end{proposition}

\begin{proof}
Fix an environment class $\mathcal X_c$ and a distribution
$p\in\mathcal P(\mathcal X_c)$. At every rate below $I(X;Y)_p$,
classical coding~\cite[Chap.~6]{CsiszarKorner2011} gives distinct basis
codewords $x_i\in\mathcal X_c^n$ and decoding regions $G_i$ that
partition the output basis strings. Their average success probability
tends to one. All codewords have identical environment outputs, so
Lemma~\ref{lem:coherent-recovery} applies with
$D_i=\sum_{y\in G_i}\dyad y_{B^n}$. The quantity
$\tr(D_i\rho_{x_i}^{B^n})$ is exactly the probability that the measured
output string lies in $G_i$. The lemma therefore gives
$F_e\ge p_{\mathrm{succ}}^2\to1$ for the resulting quantum code.

Here the $D_i$ are orthogonal projectors. The Kraus operators in
Eq.~\eqref{eq:coherent-decoder} satisfy
\[
    K_s\ket y_{B^n}
       =\bra s_C U_i\ket y_{B^n}\,\ket i_{B_1}
       \quad(y\in G_i).
\]
Each $K_s$ maps a basis state to a multiple of a basis state, so the
decoder is IO. Maximizing over $p$ and the classes proves
Eq.~\eqref{eq:io-classical-decoding}.

If $\N$ is MIO, each receiver state $\N(\dyad x)$ is diagonal, so
$I(X;Y)_p=\chi(p)$. Theorem~\ref{thm:io-all-capacity} and the lower
bound just proved then give $Q^{\IO,\IO}(\N)=Q^{\IO,\All}(\N)$ and
equality in Eq.~\eqref{eq:io-classical-decoding}.
\end{proof}

We use the notation of Sec.~\ref{sec:pauli} for the Pauli channel of
Eq.~\eqref{eq:pauli-channel}: the probabilities $q_a$, the bit-error
pattern $T$, the subspace $L$, and the random variables $X$ and $Y$.
Here $0^m=(0,\ldots,0)$, all strings have length $m$, and arithmetic is
over $\mathbb F_2$.

\begin{paulirestated}[restated]
For an $m$-qubit Pauli channel $\N$ with computational reference bases,
and with $L$, $T$, $X$, and $Y$ as defined above,
\begin{equation}
    \begin{aligned}
    Q^{\IO,\IO}(\N)=Q^{\IO,\All}(\N)&=I(X;Y)\\
       &=\dim L-H(T\mid T+L).
    \end{aligned}
    \label{eq:pauli-capacity-sm}
\end{equation}
\end{paulirestated}

\begin{proof}
We first identify the environment classes. The Kraus operators
$\sqrt{p_{ab}}\,X^aZ^b$ give the Stinespring isometry
$V=\sum_{a,b}\sqrt{p_{ab}}\,X^aZ^b\otimes\ket{a,b}_E$ of
Eq.~\eqref{eq:kraus-dilation}. Since
$X^aZ^b\ket x=(-1)^{b\cdot x}\ket{x\oplus a}$,
\[
    V\ket x=\sum_{a,b}\sqrt{p_{ab}}\,(-1)^{b\cdot x}
       \ket{x\oplus a}_B\ket{a,b}_E.
\]
The states $\ket{x\oplus a}_B$ with different $a$ are orthonormal.
Tracing out $B$ therefore gives
\[
    \rho_x^E=\sum_a\sum_{b,b'}\sqrt{p_{ab}p_{ab'}}\,
       (-1)^{(b\oplus b')\cdot x}\,\dyad{a,b}{a,b'}_E.
\]
Two basis inputs $x$ and $y$ have the same environment output exactly
when all these matrix entries agree, that is, when
$(b\oplus b')\cdot(x\oplus y)=0$ whenever $p_{ab}p_{ab'}>0$. This says
that $x\oplus y$ is orthogonal to the span in the definition of $L$.
Hence
\begin{equation}
    \rho_x^E=\rho_y^E
       \quad\Longleftrightarrow\quad x\oplus y\in L.
    \label{eq:pauli-environment-classes}
\end{equation}
The environment classes are the cosets $x_0+L=\{x_0\oplus z:z\in L\}$ of
the subspace $L$, and $L$ itself is the class containing $0^m$.

Tracing out $E$ instead gives the receiver's output
\begin{equation}
    \rho_x^B=\sum_a q_a\dyad{x\oplus a}_B.
    \label{eq:pauli-basis-output}
\end{equation}
It is diagonal for every basis input, so $\N$ is MIO. Measuring it in
the reference basis gives the outcome $x\oplus a$ with probability
$q_a$. This is the classical channel $P(y|x)=q_{y\oplus x}$ of
Sec.~\ref{sec:pauli}. Proposition~\ref{prop:io-classical-decoding}
therefore gives
\begin{equation}
    Q^{\IO,\IO}(\N)=Q^{\IO,\All}(\N)
       =\max_c\max_{r\in\mathcal P(\mathcal X_c)}I(X;Y)_r.
    \label{eq:pauli-classical-maximum}
\end{equation}

It remains to evaluate this maximum. Fix a class $x_0+L$ and a
distribution $r$ on it. Let $X_r$ have distribution $r$, independently
of the bit error $T$, and put $Y_r=X_r\oplus T$. Then
$H(Y_r\mid X_r)=H(T)$, so $I(X;Y)_r=I(X_r;Y_r)=H(Y_r)-H(T)$. Recall that
$T+L$ is the coset of $L$ containing $T$. Since $X_r$ lies in $x_0+L$,
the output $Y_r$ determines this coset as $(Y_r\oplus x_0)+L$. For each
value of $T+L$, the output is confined to a coset with
$\abs{L}=2^{\dim L}$ elements. Hence
\[
    H(Y_r)=H(T+L)+H(Y_r\mid T+L)
       \le H(T+L)+\dim L.
\]
When $X_r$ is uniform on the input class, adding any fixed error makes
$Y_r$ uniform on the resulting output coset. Conditional on $T+L$, this
is the same coset for every possible error, so equality holds. Since
$T+L$ is a function of $T$, we have $H(T)=H(T+L)+H(T\mid T+L)$. Every
class therefore gives the same maximum,
\[
    \max_rI(X;Y)_r=\dim L+H(T+L)-H(T)
       =\dim L-H(T\mid T+L),
\]
attained by the uniform distribution on the class. For the class $L$,
this is the distribution of $X$ in the theorem, so the maximum equals
$I(X;Y)$. Together with Eq.~\eqref{eq:pauli-classical-maximum}, this
proves Eq.~\eqref{eq:pauli-capacity-sm}.
\end{proof}

\begin{figure}[!htbp]
    \centering
    \begin{tikzpicture}[
  x=1cm,y=1cm,font=\small,
  >={Latex[length=1.4mm,width=1mm]},
  wire/.style={draw=black,line width=.6pt,->},
  disk/.style={draw=black,line width=.6pt},
  zero/.style={circle,draw=black,fill=black,line width=.5pt,
    inner sep=0pt,minimum size=5pt},
  one/.style={circle,draw=black,fill=white,line width=.6pt,
    inner sep=0pt,minimum size=5pt},
  lbl/.style={inner sep=1pt}
]
  \pgfmathsetmacro{\bp}{0.15}
  \pgfmathsetmacro{\bR}{0.8}
  \def\bdx{2.7}
  \def\bdy{2.65}
  \pgfmathsetmacro{\bz}{\bR*(1-2*\bp)}
  \pgfmathsetmacro{\bx}{\bR*2*sqrt(\bp*(1-\bp))}

  \node[anchor=base] at (0,\bR+.55) {$E$};
  \node[anchor=base] at (\bdx,\bR+.55) {$A$};
  \node[anchor=base] at (2*\bdx,\bR+.55) {$B$};

  \foreach \yy/\name in {0/dephasing,-\bdy/bit flip}{
    \node[rotate=90,anchor=south] at (-\bR-.25,\yy) {\name};
    \foreach \xx in {0,\bdx,2*\bdx}{
      \draw[disk] (\xx,\yy) circle (\bR);
    }
    \draw[wire] (\bdx-\bR-.12,\yy) -- node[above=1pt] {$\N^c$} (\bR+.12,\yy);
    \draw[wire] (\bdx+\bR+.12,\yy) -- node[above=1pt] {$\N$}
      (2*\bdx-\bR-.12,\yy);
    \node[zero] at (\bdx,\yy+\bR) {};
    \node[one]  at (\bdx,\yy-\bR) {};
    \node[lbl,anchor=west] at (\bdx+.28,\yy+\bR+.13) {$\ket{0}$};
    \node[lbl,anchor=west] at (\bdx+.28,\yy-\bR-.13) {$\ket{1}$};
  }

  \node[one]  at (-\bx,\bz) {};
  \node[zero] at (\bx,\bz) {};
  \node[zero] at (2*\bdx,\bR) {};
  \node[one]  at (2*\bdx,-\bR) {};
  \node[anchor=base] at (0,-\bR-.48) {$\rho_0^E\neq\rho_1^E$};

  \begin{scope}[shift={(0,\bz-\bdy)}]
    \fill[white] (0,0) circle (2.5pt);
    \fill[black] (0,2.5pt) arc[start angle=90,end angle=270,radius=2.5pt]
      -- cycle;
    \draw[black,line width=.6pt] (0,0) circle (2.5pt);
  \end{scope}
  \node[zero] at (2*\bdx,\bz-\bdy) {};
  \node[one]  at (2*\bdx,-\bz-\bdy) {};
  \node[anchor=base] at (0,-\bdy-\bR-.48) {$\rho_0^E=\rho_1^E$};
\end{tikzpicture}
    \caption{Basis inputs and reduced outputs in the Bloch $x$--$z$
    plane at $p=0.15$. Filled and open markers correspond to inputs
    $\ket{0}$ and $\ket{1}$; the half-filled marker denotes coincident
    outputs. For dephasing, the environment states are distinct and
    nonorthogonal, giving $Q^{\IO,\All}=0$. For bit-flip noise, the
    environment states coincide, and $Q^{\IO,\All}=1-h_2(p)$.}
    \label{fig:bloch-examples}
\end{figure}
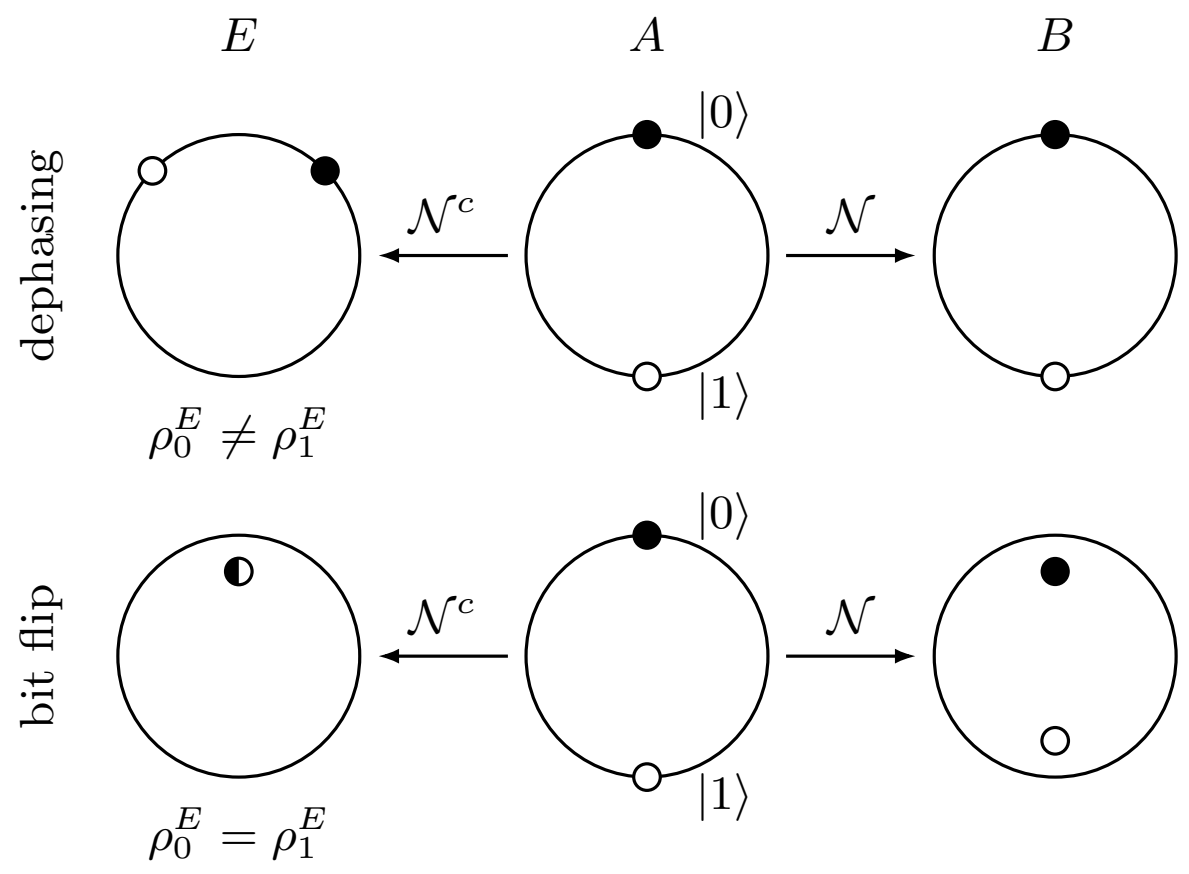

\paragraph{Bit-flip noise and dephasing.}
For bit-flip noise $\mathcal N_p^X$ only $b=0$ occurs, so the two
environment outputs coincide and $L=\mathbb F_2$. For dephasing
$\mathcal N_p^Z$ their root fidelity is $\abs{1-2p}<1$, so they differ
and $L=\{0\}$ (Fig.~\ref{fig:bloch-examples}).

\paragraph{Additional phase-flip noise.}
Adding independent phase-flip noise illustrates the discontinuity
established in Appendix~\ref{app:zero-capacity}. For the composed
channel $\mathcal N_\varepsilon^Z\circ\mathcal N_p^X$ with
$0<\varepsilon<1/2$, Eq.~\eqref{eq:channel-coherence-fidelity} gives the
root fidelity of the environment outputs of $\ket{0}$ and $\ket{1}$ as
\[
    f(\rho_0^E,\rho_1^E)
       =\norm*{(\mathcal N_\varepsilon^Z\circ\mathcal N_p^X)
                       (\dyad{0}{1})}_1
       =1-2\varepsilon<1.
\]
Thus both capacities with incoherent encoding are zero for every such
$\varepsilon$. In contrast, the maximally mixed input gives
\[
    Q(\mathcal N_\varepsilon^Z\circ\mathcal N_p^X)
       \ge I_c(I/2,\mathcal N_\varepsilon^Z\circ\mathcal N_p^X)
       =1-h_2(p)-h_2(\varepsilon).
\]
The receiver output has entropy one, and the environment output has the
four eigenvalues $(1-p)(1-\varepsilon)$, $p(1-\varepsilon)$,
$(1-p)\varepsilon$, and $p\varepsilon$. Its entropy is therefore
$h_2(p)+h_2(\varepsilon)$. For fixed $0<p<1/2$, the standard capacity
remains positive when $\varepsilon$ is sufficiently small.

\subsection{Dephasing with an incoherent decoder}
\label{app:dephasing-io-decoder}

The decoder restriction alone does not reduce the capacity of
$\mathcal N_p^Z$. Its phase errors have the same distribution as
classical bit errors. For any $0<r<1-h_2(p)$, choose binary linear codes
$\mathcal C\subseteq\mathbb F_2^n$ of dimension $k$, with $k/n\to r$,
that correct bit flips of probability
$p$~\cite[Theorems~18.12 and~18.13]{PolyanskiyWu2025}. Let $\mathcal L$
contain the error representative chosen by the syndrome decoder from
each coset of $\mathcal C$. The probability that the actual error lies
in $\mathcal L$ tends to one.

Let the rows of the $k\times n$ matrix $G$ form a basis of $\mathcal C$.
Then $\ker G=\mathcal C^\perp$. For each $u$, the strings $y$ satisfying
$Gy=u$ form a coset of $\mathcal C^\perp$. Regard binary strings as
column vectors and define the encoding isometry
\begin{equation}
    K\ket{u}_{A_1}
       =2^{-(n-k)/2}\sum_{y:Gy=u}\ket{y}_{A^n},
       \qquad u\in\mathbb{F}_2^k.
    \label{eq:dephasing-coset-encoding}
\end{equation}
There are $2^{n-k}$ terms in each sum. The sets of basis states for
distinct $u$ are disjoint, which verifies that $K$ is isometric.

With corresponding input and output bases identified, write
$Z^e=\bigotimes_{j=1}^n Z^{e_j}$ and take
\begin{equation}
    D_e=K^\dagger Z^e,\qquad
    D_e\ket y=2^{-(n-k)/2}(-1)^{e\cdot y}\ket{Gy},
       \quad e\in\mathcal L.
    \label{eq:dephasing-io-decoder}
\end{equation}
Each operator maps a basis state to a multiple of a basis state, so it
is incoherent. To verify trace preservation and recovery, observe that
for $e,e'\in\mathcal L$,
\[
    \bra uK^\dagger Z^{e\oplus e'}K\ket v
       =\delta_{u,v}\,2^{-(n-k)}
          \sum_{y:Gy=u}(-1)^{(e\oplus e')\cdot y}
       =\delta_{u,v}\delta_{e,e'}.
\]
For $e\ne e'$, their difference is outside $\mathcal C$. There is then a
$z\in\mathcal C^\perp$ with $(e\oplus e')\cdot z=1$, and pairing $y$
with $y\oplus z$ cancels the sum. For $e=e'$, the normalization gives
one. Thus the subspaces $Z^e\operatorname{im}K$ are orthogonal. Their
dimensions sum to $2^k\abs{\mathcal L}=2^n$, so they exhaust $B^n$ and
$\sum_e D_e^\dagger D_e=I_{B^n}$. The same identity gives
$D_{e'}Z^eK=\delta_{e,e'}I$, so every error in $\mathcal L$ is corrected
exactly. With $\E(\rho)=K\rho K^\dagger$ and
$\D(\rho)=\sum_e D_e\rho D_e^\dagger$, its entanglement fidelity obeys
\[
    F_e(\D\circ(\mathcal N_p^Z)^{\otimes n}\circ\E)
       \ge\sum_{e\in\mathcal{L}}p^{\abs*{e}}(1-p)^{n-\abs*{e}}
       \longrightarrow1,
\]
where $\abs*{e}$ is the Hamming weight of $e$. Every rate below
$1-h_2(p)$ is therefore attainable with an IO decoder. The standard
quantum capacity gives the matching upper bound, proving
\[
    Q^{\All,\IO}(\mathcal N_p^Z)=Q(\mathcal N_p^Z)=1-h_2(p).
\]

\subsection{Erasure and amplitude damping}
\label{app:erasure-amplitude-damping}

It suffices to show that the complementary channels are injective: then
no two distinct input basis states have the same environment output. The
zero-capacity criterion in Appendix~\ref{app:zero-capacity} gives
$Q^{\IO,\All}=0$, and therefore $Q^{\IO,\IO}=0$.

For the erasure channel with input dimension $d\ge2$ and erasure
probability $p$, the channel and its complement
are~\cite[Secs.~4.7.6 and~5.2]{Wilde}
\begin{equation}
    \mathcal E_p(\rho)=(1-p)\rho\oplus p\dyad{e},
    \qquad
    \mathcal E_p^c(\rho)=p\rho\oplus(1-p)\dyad{e}.
    \label{eq:erasure-complement}
\end{equation}
Here $\oplus$ denotes a direct sum. Each output space contains a copy of
the input space and an orthogonal erasure flag $\ket{e}$. On the copy of
the input space, the environment state has the block $p\rho$. For $p>0$,
this block determines $\rho$.

For qubit amplitude damping $\mathcal A_\gamma$, let $\gamma$ be the
decay probability and write $\rho_{ij}=\bra{i}\rho\ket{j}$ in the energy
basis. Using the standard Kraus operators~\cite[Sec.~4.7.5]{Wilde},
Eq.~\eqref{eq:kraus-dilation} gives
\begin{equation}
    \mathcal A_\gamma^c(\rho)=
    \begin{pmatrix}
        \rho_{00}+(1-\gamma)\rho_{11} & \sqrt\gamma\rho_{01}\\
        \sqrt\gamma\rho_{10} & \gamma\rho_{11}
    \end{pmatrix}.
    \label{eq:amplitude-damping-complement}
\end{equation}
When $\gamma>0$, the lower-right and off-diagonal entries determine
$\rho_{11},\rho_{01},\rho_{10}$; the upper-left entry then determines
$\rho_{00}$. This proves injectivity on all input states. Thus, in every
input reference basis,
\begin{align*}
    Q^{\IO,\IO}(\mathcal E_p)=Q^{\IO,\All}(\mathcal E_p)&=0,
       &&0<p\le1,\\
    Q^{\IO,\IO}(\mathcal A_\gamma)=Q^{\IO,\All}(\mathcal A_\gamma)&=0,
       &&0<\gamma\le1.
\end{align*}
For positive erasure or decay probability below $1/2$, both channels
have positive standard quantum
capacity~\cite{BennettDiVincenzoSmolin1997,GiovannettiFazio2005}.

\subsection{Channels with a classical flag}
\label{app:measurement-uncertainty}

We consider a channel that chooses one of the unitaries
$U_1,\ldots,U_{d_J}$ uniformly at random and records the choice in a
flag register $J$ of dimension $d_J$. For a $d$-dimensional input, the
channel is
\begin{equation}
    \N(\rho)=\frac1{d_J}\sum_{\beta=1}^{d_J}
       U_\beta\rho U_\beta^\dagger\otimes\dyad{\beta}_J.
    \label{eq:flagged-unitary-family}
\end{equation}
The states $\ket\beta_J$ form the reference basis of the flag register,
and the output $BJ$ has the product reference basis. An unrestricted
receiver can read the flag and reverse $U_\beta$; an IO receiver may not
be able to apply its inverse.

To bound the rate with IO decoding, we examine the output probabilities
in the reference basis. Measuring $U_\beta\ket\psi$ in this basis is
equivalent to measuring $\ket\psi$ in the input basis
$\{U_\beta^\dagger\ket y\}_y$. Let $Y_\beta$ denote this measurement
outcome, with probabilities
$\Pr_\psi(Y_\beta=y)=\abs{\bra yU_\beta\ket\psi}^2$, and set
\begin{equation}
    h=\min_{\ket\psi}\frac1{d_J}\sum_{\beta=1}^{d_J} H(Y_\beta)_\psi.
    \label{eq:flagged-unitary-uncertainty}
\end{equation}
The minimum is over normalized pure states. The entropy of a measurement
outcome is concave in the state, which gives the same lower bound for
mixed states.

\begin{proposition}[Converse for channels with a classical flag]
\label{prop:flagged-unitary-converse}
The channel in Eq.~\eqref{eq:flagged-unitary-family} satisfies
\begin{equation}
    Q^{\All,\IO}(\N)\le\log d-\frac12h.
    \label{eq:flagged-unitary-capacity-bound}
\end{equation}
At every rate above this bound, the entanglement fidelity tends to zero
exponentially. This also holds for MIO decoders.
\end{proposition}

\begin{proof}
Let $K_b:A_1\to A^n$ be Kraus operators of an arbitrary encoder $\E$,
with $\dim A_1=M$. For a flag string $s=(s_1,\ldots,s_n)$, put
$U_s=U_{s_1}\otimes\cdots\otimes U_{s_n}$ and
\[
    T_{s,b}=U_sK_b,\qquad
    p_s(y|i)=\bra yU_s\E(\dyad i)U_s^\dagger\ket y
       =\sum_b\abs{\bra yT_{s,b}\ket i}^2.
\]
Here $y\in\{0,\ldots,d-1\}^n$ is an output basis string, and
$\mathbb E_s$ denotes the average over the independent uniform flags.
The encoded states may be mixed and entangled across channel inputs.

We first control the total weight of outcomes with unusually large
probabilities. For $\alpha>0$, define
\[
    g(\alpha)=\max_\rho\frac1{d_J}\sum_{\beta=1}^{d_J}\sum_{y=0}^{d-1}
       \bigl(\bra yU_\beta\rho U_\beta^\dagger\ket y\bigr)^{1+\alpha},
\]
where $\rho$ is a state on one channel input. Measure the inputs
successively in the independently chosen bases. Conditional on earlier
choices and outcomes, the next input is described by a density operator
and its basis choice is still independent and uniform. Its conditional
sum of $(1+\alpha)$-powers, averaged over that choice, is therefore at
most $g(\alpha)$. Summing successively over the last outcome and basis
choice, then the preceding ones, gives
\begin{equation}
    \mathbb E_s\sum_y p_s(y|i)^{1+\alpha}\le g(\alpha)^n.
    \label{eq:flagged-unitary-probability-moment}
\end{equation}

Expanding at $\alpha=0$ yields $\log g(\alpha)=-\alpha h+O(\alpha^2)$.
The remainder is uniform over input states: the second derivative of
$t^{1+\alpha}$ in $\alpha$ is bounded for $0\le t\le1$ and $\alpha\ge0$,
with its value at $t=0$ defined by continuity. For any fixed $\delta>0$,
we can thus choose a fixed $\alpha>0$ such that
$g(\alpha)\le2^{-\alpha(h-\delta/2)}$. Set
\[
    q_n=2^{-n(h-\delta)},\qquad
    e_s=\frac1M\sum_i\sum_{y:p_s(y|i)>q_n}p_s(y|i).
\]
Since $p\le q_n^{-\alpha}p^{1+\alpha}$ when $p>q_n$,
Eq.~\eqref{eq:flagged-unitary-probability-moment} implies
\begin{equation}
    \mathbb E_s e_s\le q_n^{-\alpha}g(\alpha)^n
       \le2^{-n\alpha\delta/2}.
    \label{eq:flagged-unitary-large-probabilities}
\end{equation}

For each flag string, split $T_{s,b}=T'_{s,b}+T''_{s,b}$. The first
matrix contains the entries with $p_s(y|i)>q_n$; the second contains the
others. Use the same choice for every Kraus index $b$. By construction,
\[
    \sum_b\norm{T'_{s,b}}_2^2=Me_s,
    \qquad
    \max_i\sum_b\abs{\bra yT''_{s,b}\ket i}^2\le q_n.
\]
Fix an arbitrary IO decoder $\D$ for the code. Holding its flag input
fixed gives the IO decoder $\D_s(\tau)=\D(\tau\otimes\dyad{s}_{J^n})$.
Lemma~\ref{lem:recovery-amplitude-bound}, applied to this decoder and
the conditional channel with Kraus operators $T_{s,b}$, gives
\[
    \sqrt{F_s}\le\sqrt{e_s}+\frac{d^n\sqrt{q_n}}M,
\]
where $F_s$ is its entanglement fidelity. The first term bounds the
discarded entries; each of the $d^n$ output basis strings contributes at
most $\sqrt{q_n}$ to the second term. The total fidelity is
$F_e=\mathbb E_sF_s$, since the flag probabilities are independent of
the input. The triangle inequality for the root mean square therefore
yields
\begin{equation}
    \sqrt{F_e}\le\sqrt{\mathbb E_s e_s}
          +\frac{d^n\sqrt{q_n}}M
       \le2^{-n\alpha\delta/4}
          +\frac{2^{n(\log d-h/2+\delta/2)}}M.
    \label{eq:flagged-unitary-fidelity-bound}
\end{equation}
At a rate $R>\log d-h/2$, that is, for $M\ge2^{nR}$, choose
$\delta<2(R-\log d+h/2)$. Both terms then decay exponentially. Fixing an
incoherent flag also preserves MIO, and
Lemma~\ref{lem:recovery-amplitude-bound} applies to MIO decoders,
proving the final assertion.
\end{proof}

With both coding maps restricted to IO, a common calculation determines
the capacity of the examples below and of the Fourier channel.

\begin{proposition}[Flagged unitaries with entries of equal modulus]
\label{prop:flagged-basis-capacity}
Let $\N$ be the channel in Eq.~\eqref{eq:flagged-unitary-family} with
$d\ge2$. Suppose that, for every $\beta$, there is an integer
$1\le D_\beta\le d$ such that
\[
    \abs{\bra yU_\beta\ket x}^2\in\{0,D_\beta^{-1}\}
       \qquad\text{for every }x,y.
\]
Then
\begin{equation}
    Q^{\IO,\IO}(\N)
       =\log d-\frac1{2d_J}\sum_{\beta=1}^{d_J}\log D_\beta.
    \label{eq:flagged-basis-capacity}
\end{equation}
Replacing IO by MIO in either or both coding maps leaves this value
unchanged.
\end{proposition}

\begin{proof}
Since the rows and columns of $U_\beta$ have unit norm, each contains
exactly $D_\beta$ nonzero entries. For a flag string
$s=(s_1,\ldots,s_n)$, define
\[
    D(s)=\prod_{k=1}^n D_{s_k},
    \qquad \mu=\frac1{d_J}\sum_{\beta=1}^{d_J}\log D_\beta.
\]
Every row and column of $U_s=\bigotimes_{k=1}^nU_{s_k}$ therefore has
$D(s)$ nonzero entries, each of modulus $D(s)^{-1/2}$. The flags are
independent and uniform, so $n^{-1}\log D(s)\to\mu$ in probability.

For achievability, use the identity encoder on the full input space, of
dimension $M=d^n$. The unrestricted decoder with Kraus operators
$U_s^\dagger(I_{B^n}\otimes\bra s_{J^n})$ reverses the channel exactly.
For Lemma~\ref{lem:io-subcode-decoder}, this gives
\[
    \ell_{s,y}=\sqrt{D(s)},\qquad \bar p(s,y)=d_J^{-n}d^{-n}.
\]
Indeed, $U_s^\dagger\ket y$ has $D(s)$ amplitudes of modulus
$D(s)^{-1/2}$, and the maximally mixed input produces uniform outputs
and flags.

Fix $0<R<\log d-\mu/2$ and apply the lemma with
$m_n=\lceil2^{nR}\rceil$. Its error term is bounded by
\[
    \varepsilon_n=\mathbb E_s\min\!\left\{1,
       \frac{(m_n-1)\sqrt{D(s)}}{d^n-1}\right\}.
\]
Choose $\delta>0$ such that $R<\log d-(\mu+\delta)/2$. Splitting the
average according to whether $\log D(s)\le n(\mu+\delta)$ gives
\[
    \varepsilon_n\le\Pr_s\{\log D(s)>n(\mu+\delta)\}
       +\frac{(m_n-1)2^{n(\mu+\delta)/2}}{d^n-1}
       \longrightarrow0.
\]
The probability tends to zero by the law of large numbers, and the last
term vanishes by the choice of $\delta$. Since the full code has
fidelity one, the lemma supplies IO decoders whose fidelities tend to
one for the selected basis codewords. Their encoders are IO as well,
proving achievability.

For the converse, let $\E$ be a MIO encoder of logical dimension $M$ and
allow a MIO decoder. Each encoded basis input is a diagonal state, so we
can write $\E(\dyad i)=\sum_x r(x|i)\dyad x$ for a probability
distribution $r(\cdot|i)$ on input basis strings. Every matrix entry of
$U_s$ has squared modulus at most $1/D(s)$. Hence
\[
    p_s(y|i)=\sum_x r(x|i)\abs{\bra yU_s\ket x}^2
       \le\frac1{D(s)},
\]
even when the mixture has correlations across channel inputs. Holding
the decoder's flag input fixed preserves MIO.
Equation~\eqref{eq:sm-amplitude-bound} therefore bounds its conditional
entanglement fidelity $F_s$ by
\begin{equation}
    F_s\le\min\!\left\{1,\frac{d^{2n}}{M^2D(s)}\right\},
    \qquad F_e=\mathbb E_sF_s.
    \label{eq:flagged-basis-conditional-fidelity}
\end{equation}
At a rate $R>\log d-\mu/2$, choose $\delta>0$ such that
$R>\log d-(\mu-\delta)/2$. For $M\ge2^{nR}$,
\[
    F_e\le\Pr_s\{\log D(s)<n(\mu-\delta)\}
       +\frac{2^{n(2\log d-\mu+\delta)}}{M^2}
       \longrightarrow0.
\]
This proves the upper bound for MIO coding maps and completes the proof.
\end{proof}

\subsubsection{Flagged Hadamard channel}
\label{app:strict-decoder-loss}

Let $A$, $B$, and $J$ be qubits, with the computational product basis on
$BJ$. The flagged Hadamard channel of Sec.~\ref{sec:flagged-channels} is
\[
    \N_H(\rho)=\frac12\rho\otimes\dyad{0}_J
       +\frac12H\rho H\otimes\dyad{1}_J,
\]
where
$H=2^{-1/2}\bigl(\begin{smallmatrix}1&1\\1&-1\end{smallmatrix}\bigr)$ is
the Hadamard matrix. An unrestricted receiver uses $J$ to reverse the
recorded operation. The identity encoder is IO, so
$Q^{\IO,\All}(\N_H)=Q(\N_H)=1$. With IO decoding, we show that
\begin{equation}
    Q^{\IO,\IO}(\N_H)=Q^{\All,\IO}(\N_H)=\frac34.
    \label{eq:flagged-hadamard-capacity}
\end{equation}

\emph{Converse.} The two measurements in
Proposition~\ref{prop:flagged-unitary-converse} are in the $Z$ and $X$
bases. For a qubit with Bloch coordinates $x,z$, the inequality
$h_2(p)\ge4p(1-p)$ gives
\[
    \frac12\left[h_2\!\left(\frac{1+x}{2}\right)
                 +h_2\!\left(\frac{1+z}{2}\right)\right]
       \ge1-\frac{x^2+z^2}{2}\ge\frac12.
\]
A $Z$-basis state attains equality. Hence $h=1/2$, and the proposition
bounds the capacity by $3/4$, also for MIO decoding.

\emph{Achievability.} The identity and Hadamard branches have $D_1=1$
and $D_2=2$, respectively. Proposition~\ref{prop:flagged-basis-capacity}
applies with $d=2$ and $d_J=2$. It gives $Q^{\IO,\IO}(\N_H)=3/4$, which
also lower-bounds $Q^{\All,\IO}(\N_H)$. This proves
Eq.~\eqref{eq:flagged-hadamard-capacity}. Since the converse also
permits MIO decoders, allowing MIO in place of IO decoding leaves these
two capacities unchanged.

\subsubsection{Flagged MUB channels}
\label{app:optimal-decoder-factor}

For the flagged MUB channels of Sec.~\ref{sec:flagged-channels}, we now
prove the upper bound in Eq.~\eqref{eq:optimal-decoder-half-factor} and
the optimality of the factor $1/2$ in
Theorem~\ref{thm:io-decoder-half-bounds} with unrestricted encoding. Let
$d\ge2$ be a prime power, and choose unitaries $U_0,\ldots,U_d$, with
$U_0=I$, so that $\{U_\beta^\dagger\ket y\}_{y=0}^{d-1}$ form the
standard complete set of $d+1$ stabilizer mutually unbiased bases
(MUBs)~\cite{WoottersFields1989,Kantor2012}. Define
\begin{equation}
    \N_d(\rho)=\frac1{d+1}\sum_{\beta=0}^d
       U_\beta\rho U_\beta^\dagger\otimes\dyad{\beta}_J,
    \label{eq:sm-mub-channel}
\end{equation}
with the product reference basis on $BJ$ and $\dim A=\dim B=d$,
$\dim J=d+1$. An unrestricted receiver uses the flag to reverse the
unitary, so the identity encoder attains
$Q^{\IO,\All}(\N_d)=Q(\N_d)=\log d$.

Every basis with $\beta\ne0$ is unbiased to the reference basis
specified by $U_0=I$. Its unitary therefore has all matrix entries of
modulus $d^{-1/2}$. Thus $D_0=1$ and $D_\beta=d$ for $\beta=1,\ldots,d$.
Proposition~\ref{prop:flagged-basis-capacity} gives the exact value
\begin{equation}
    Q^{\IO,\IO}(\N_d)=\frac{d+2}{2(d+1)}\log d.
    \label{eq:sm-mub-io-capacity}
\end{equation}

Write $Y_\beta$ for the corresponding measurement outcome as above. For
a complete set of mutually unbiased bases, every pure input satisfies
$\sum_{\beta,y}\Pr_\psi(Y_\beta=y)^2=2$~\cite{KlappeneckerRotteler2005}.
Shannon entropy is at least the collision entropy,
$H(Y_\beta)_\psi\ge-\log\sum_y\Pr_\psi(Y_\beta=y)^2$, and $-\log$ is
convex. Jensen's inequality therefore yields
\begin{equation}
    \frac1{d+1}\sum_\beta H(Y_\beta)_\psi
       \ge-\log\!\left[\frac1{d+1}
           \sum_{\beta,y}\Pr_\psi(Y_\beta=y)^2\right]
       =\log\frac{d+1}{2}.
    \label{eq:sm-mub-uncertainty-lower}
\end{equation}
Proposition~\ref{prop:flagged-unitary-converse} therefore gives
\begin{equation}
    Q^{\All,\IO}(\N_d)\le\log d-\frac12\log\frac{d+1}{2},
    \label{eq:sm-mub-decoder-capacity}
\end{equation}
also for MIO decoding. Since $Q^{\All,\IO}(\N_d)\ge Q^{\IO,\IO}(\N_d)$,
we obtain
\[
    \frac{d+2}{2(d+1)}\le\frac{Q^{\All,\IO}(\N_d)}{Q(\N_d)}
       \le\frac12+\frac{\log[2d/(d+1)]}{2\log d}
       \longrightarrow\frac12
\]
as $d$ grows through prime powers. Together with the universal lower
bound in Appendix~\ref{app:half-capacity-bounds}, this proves
\begin{equation}
    \inf_{\N:\,Q(\N)>0}\frac{Q^{\All,\IO}(\N)}{Q(\N)}
       =\frac12,
    \label{eq:sm-optimal-decoder-factor}
\end{equation}
where the infimum is over finite-dimensional channels.

\subsection{Quantum Fourier transform channel}
\label{app:fourier-unitaries}

In the reference bases,
\[
    \bra y F_d\ket x=d^{-1/2}e^{2\pi ixy/d},
    \qquad x,y=0,\ldots,d-1.
\]
If either the encoder or the decoder is unrestricted, it can apply the
inverse Fourier transform, so a single use transmits $\log d$ qubits.
The input dimension gives the matching upper bound.

All entries of $F_d$ have absolute value $d^{-1/2}$.
Proposition~\ref{prop:flagged-basis-capacity} therefore applies to the
single unitary $U_1=F_d$, with $D_1=d$, giving
\[
    Q^{\IO,\IO}(\mathcal F_d)=\tfrac12\log d.
\]
A simple code attains this rate exactly: one use on $\ket0$ outputs
$\ket{\Psi_d}=F_d\ket0$, and by Eq.~\eqref{eq:io-decoder-simulation}
this state lets an IO decoder apply $F_d^\dagger$ to the output of a
second use. The converse in that proposition also allows MIO encoders
and decoders. For $n$ uses, $D(s)=d^n$. For a code of logical dimension
$M$, Eq.~\eqref{eq:flagged-basis-conditional-fidelity} therefore reduces
to
\begin{equation}
    F_e\le\frac{d^n}{M^2}.
    \label{eq:fourier-fidelity-bound}
\end{equation}
It tends to zero exponentially at every rate above $\tfrac12\log d$.
Thus allowing MIO for both coding maps leaves the capacity unchanged.

\subsection{Nonadditivity}
\label{app:nonadditivity}

The coherence supplied by one channel can also assist the decoding of
another, so $Q^{\IO,\IO}$ is not additive over different channels. For
$d\ge2$, let $\mathcal R_d(\rho)=\tr(\rho)\dyad{\Psi_d}$ be the channel
on a $d$-dimensional system that replaces every input by the same
maximally coherent state. It has zero quantum capacity, but
\begin{align*}
    Q^{\IO,\IO}(\mathcal F_d\otimes\mathcal R_d)
       &=\log d\\
       &>Q^{\IO,\IO}(\mathcal F_d)+Q^{\IO,\IO}(\mathcal R_d)\\
       &=\tfrac12\log d.
\end{align*}
To attain $\log d$ per use of the product channel, the encoder sends the
logical state through $\mathcal F_d$ and a fixed basis state through
$\mathcal R_d$. The latter supplies $\ket{\Psi_d}$, which suffices to
implement $F_d^\dagger$ by IO using
Eq.~\eqref{eq:io-decoder-simulation}. The resulting channel is the
identity on the logical system, so it preserves entanglement with a
reference. An unrestricted receiver can prepare the fixed ancillary
state locally, so $Q(\mathcal F_d\otimes\mathcal R_d)=\log d$ gives the
matching upper bound. A channel that carries no input information can
therefore increase the jointly achievable rate by supplying coherence
for decoding.

\section{Equivalence of entanglement and subspace transmission}
\label{app:transmission-criteria}

Entanglement transmission tests a code on a maximally entangled input.
Subspace transmission instead asks the code to preserve every pure state
of the logical input space. For an effective channel $\Lambda$ of
logical dimension $M$, define
\begin{equation}
    F_{\min}(\Lambda)
       =\min_{\braket{\psi}{\psi}=1}
         \bra{\psi}\Lambda(\dyad{\psi})\ket{\psi}.
    \label{eq:minimum-fidelity}
\end{equation}
Corresponding input and output basis states are identified as in the
main text.

For unrestricted codes, the equivalence of these criteria is well
established~\cite{Barnum2000,KretschmannWerner2004}. More recent
formulations select a smaller subspace on which recovery is uniformly
reliable~\cite[Lemma~5 and Corollary~1]{BuscemiDatta2010}
and~\cite[Lemma~24]{BertaChengYao2026}. The selected subspace need not
be spanned by reference basis states, so embedding into it need not be
IO. Here we select a subspace spanned by reference basis states and
control the error throughout that subspace. A theorem on principal
submatrices provides this choice without an asymptotic rate loss. The
inclusion into the original code space and the corresponding restriction
at the output are both IO.

\begin{proposition}[Entanglement and subspace transmission]
\label{prop:transmission-criteria}
Entanglement transmission and subspace transmission have the same
capacity. This remains true when the encoder, the decoder, or both are
restricted to incoherent operations.
\end{proposition}

\begin{proof}
First suppose $F_{\min}(\Lambda)\ge1-\delta$. The fidelity averaged over
pure inputs is at least $1-\delta$ and
satisfies~\cite[Eq.~(3)]{Nielsen2002}
\[
    \int \bra{\psi}\Lambda(\dyad{\psi})\ket{\psi}\,d\psi
       =\frac{MF_e(\Lambda)+1}{M+1},
\]
where $d\psi$ is normalized Haar measure. Thus
$F_e(\Lambda)\ge1-(1+1/M)\delta$.

For the other direction, let $\Lambda$ be the effective channel of an
$n$-use code.

\paragraph{An operator that controls the error.}
Let $K_\alpha$ be Kraus operators of $\Lambda$, and set
\begin{equation}
    a_\alpha=\frac{\tr K_\alpha}{M},
    \qquad
    G=\sum_\alpha(K_\alpha-a_\alpha I)^\dagger
                         (K_\alpha-a_\alpha I).
    \label{eq:criteria-error-operator}
\end{equation}
This is the error operator used in the subspace argument of
Ref.~\cite[Lemma~24]{BertaChengYao2026}. Expanding the products and
using Eq.~\eqref{eq:entanglement-fidelity-kraus} and
$\sum_\alpha K_\alpha^\dagger K_\alpha=I$ gives
\begin{equation}
    \frac{\tr G}{M}=1-F_e(\Lambda),
    \qquad
    0\le G\le2\bigl(1+F_e(\Lambda)\bigr)I\le4I.
    \label{eq:criteria-error-average}
\end{equation}
The operator bound follows from
$(A-B)^\dagger(A-B)\le2A^\dagger A+2B^\dagger B$. Thus successful
entanglement transmission makes the average of the diagonal entries of
$G$ small.

For every pure state $\ket{\psi}$ of the logical system,
\begin{align}
    1-\bra{\psi}\Lambda(\dyad{\psi})\ket{\psi}
       &=\bra{\psi} G\ket{\psi}
          -\sum_\alpha\abs{\bra{\psi} K_\alpha\ket{\psi}-a_\alpha}^2\notag\\
       &\le\bra{\psi} G\ket{\psi}.
    \label{eq:criteria-variance-bound}
\end{align}
The equality follows by expanding the squares and using
$\sum_\alpha\norm{K_\alpha\ket{\psi}}^2=1$. The nonnegative sum gives
the inequality. Thus an operator bound on $G$ controls the fidelity for
every pure input.

\paragraph{Selecting logical basis states.}
Put $\varepsilon=1-F_e(\Lambda)$. Since the diagonal entries of $G$ are
nonnegative and have mean $\varepsilon$, at least $M/2$ of them are at
most $2\varepsilon$. Let $S$ be the set of these indices, and let $G_S$
be the corresponding principal submatrix. Write
\[
    G_S=\operatorname{diag}(G_S)+T,
\]
where $\operatorname{diag}(G_S)$ is the diagonal part of $G_S$ and $T$
contains its off-diagonal entries. The diagonal part is at most
$2\varepsilon I$. The matrix $T$ is Hermitian with zero diagonal, and
$\norm{T}_\infty\le4$: both $G_S$ and $\operatorname{diag}(G_S)$ lie
between $0$ and $4I$.

A theorem of Youssef~\cite[Proposition~5.1]{Youssef2014} states that,
for any $0<\delta<1$, a Hermitian matrix with zero diagonal has a
principal submatrix on at least $c\delta^2$ times its original number of
coordinates, whose operator norm is at most $\delta$ times the original
norm. Here $c>0$ is a universal constant. Apply this theorem to $T$. The
resulting set of basis states defines an inclusion
$W:\mathbb C^m\to\mathbb C^M$ satisfying
\begin{equation}
    m\ge\frac{c\delta^2M}{2},
    \qquad
    W^\dagger GW\le(2\varepsilon+4\delta)I_m.
    \label{eq:criteria-coordinate-restriction}
\end{equation}
The first selection bounds the diagonal part of the error; the theorem
controls the off-diagonal part on a further subset.

\paragraph{Restricting the code.}
Before the original encoder, apply $\E_0(\rho)=W\rho W^\dagger$. After
the original decoder, apply
\begin{equation}
    \D_0(\sigma)=W^\dagger\sigma W
       +\tr\bigl[(I_M-WW^\dagger)\sigma\bigr]\dyad{1}.
    \label{eq:criteria-restricted-decoding}
\end{equation}
The second term sends any component outside the selected subspace to the
fixed logical basis state $\ket{1}$, making $\D_0$ trace preserving. The
encoder has the single Kraus operator $W$. The decoder has Kraus
operators $W^\dagger$ and $\ket{1}\bra{k}$ for the unused basis states
$\ket{k}$. Every one of these operators maps a basis state to a multiple
of a basis state. Thus both maps are IO, and their composition with the
original encoder and decoder preserves the imposed restrictions.

For the new effective channel
$\widetilde\Lambda=\D_0\circ\Lambda\circ\E_0$ and every pure logical
state $\ket{\psi}$, the first term in
Eq.~\eqref{eq:criteria-restricted-decoding} gives
\begin{align*}
    \bra{\psi}\widetilde\Lambda(\dyad{\psi})\ket{\psi}
       &\ge \bra{\psi} W^\dagger
              \Lambda(W\dyad{\psi} W^\dagger)W\ket{\psi}\\
       &\ge1-\bra{\psi} W^\dagger GW\ket{\psi}\\
       &\ge1-2\varepsilon-4\delta.
\end{align*}
The omitted term is nonnegative. The second inequality is
Eq.~\eqref{eq:criteria-variance-bound} applied to the pure state
$W\ket{\psi}$.

Finally, for a sequence of $n$-use codes with positive rate, choose
$\delta=1/n$ for $n\ge2$. Then
\begin{equation}
    F_{\min}(\widetilde\Lambda)
       \ge1-2\bigl(1-F_e(\Lambda)\bigr)-\frac4n,
    \qquad
    \log m=\log M-O(\log n).
    \label{eq:criteria-conversion-error}
\end{equation}
The error tends to zero whenever $F_e(\Lambda)\to1$. The code still uses
$n$ channel uses, and the loss in $\log M$ is sublinear in $n$, so its
asymptotic rate is unchanged. Together with the first direction, this
proves the proposition.
\end{proof}

\bibliography{ref}
\end{document}